\documentclass[12pt,preprint]{aastex}
\usepackage{natbib}
\shorttitle{Variable Stars in Draco}
\shortauthors{Kinemuchi et al.}

\begin{document}

\title{The Variable Stars of the Draco Dwarf Spheroidal Galaxy - Revisited}

\author{K. Kinemuchi}
\affil{Departamento de Astronom\'{i}a, Universidad de Concepci\'{o}n,
  Concepci\'{o}n, Chile \& Department of Astronomy, University of Florida,
  Gainesville, Florida, 32611, USA.} 

\author{H.C. Harris}
\affil{US Naval Observatory - Flagstaff Station, Flagstaff, AZ 86001}

\author{Horace. A. Smith}
\affil{Department of Physics \& Astronomy, Michigan State University, East Lansing, MI 48824}

\author{N. A. Silbermann}
\affil{Spitzer Science Center, Mail Stop 220-6, Pasadena, CA 91125}

\author{L.A. Snyder}
\affil{Department of Physics \& Astronomy, Michigan State University, East Lansing, MI 48824}

\author{A. P. LaCluyz\'{e}}
\affil{Department of Physics \& Astronomy, Michigan State University, East Lansing, MI 48824}

\author{C. L. Clark}
\affil{Department of Physics \& Astronomy, Michigan State University, East Lansing, MI 48824}

\begin{abstract}
We present a CCD survey of variable stars in the Draco dwarf
spheroidal galaxy.  This survey, which has the largest areal coverage
since the original variable star survey by Baade \& Swope, includes
photometry for 270 RR Lyrae stars, 9 anomalous Cepheids, 2 eclipsing
binaries, and 12 slow, irregular red variables, as well as 30
background QSOs.  Twenty-six probable double-mode RR Lyrae stars were
identified. Observed parameters, including mean $V$ and $I$
magnitudes, $V$ amplitudes, and periods, have been derived.
Photometric metallicities of the ab-type RR Lyrae stars were
calculated according to the method of Jurcsik \& Kovacs, yielding a
mean metallicity of $\langle [Fe/H] \rangle = -2.19 \pm 0.03$.  The
well known Oosterhoff intermediate nature of the RR Lyrae stars in
Draco is reconfirmed, although the double-mode RR Lyrae stars with
one exception have properties similar to those found in Oosterhoff
type II globular clusters.  The period-luminosity relation of the
anomalous Cepheids is rediscussed with the addition of the new Draco
anomalous Cepheids.
\end{abstract}


\section{Introduction}
The Draco dwarf spheroidal (dSph) galaxy ($\alpha_{2000.0} = 17^{h} 20^{m} 12.39^{s} $,
$\delta_{2000.0}= +57^{\circ} 54' 55.3''$), a satellite of the Milky Way Galaxy, was first
extensively studied by \citet{Baade:1961} (hereafter known as B\&S).  
They reported discovering over 260 variable stars and obtained photometry
for 138 variables in the central region of Draco, 133 of which
were of RR Lyrae (RRL) type.  Several subsequent studies have 
investigated aspects of the variable star population in Draco.
\citet{Zinn:1976a} reported new observations of the anomalous Cepheids
in Draco.
\citet{Nemec:1985a} reanalyzed the B\&S photometry and produced updated
periods for the B\&S variables.  Both \citet{Nemec:1985a} 
and \citet{Goranskij:1982} reported
new double-mode RRL in Draco.  Recently \citet{Bonanos:2004} provided
a photometric study of Draco which produced light curves for 146 RRL stars, 
four anomalous Cepheids, an SX Phe star, and a field eclipsing
binary.  In this work, we use CCD observations to update the census of
variable stars in Draco.  We cover an area slightly larger than the
full B\&S survey, and we discover new variables with smaller
amplitudes than those found by B\&S.  We provide photometric data,
periods, and light curves for over 320 variable stars.

This paper is organized in the following
manner: Section 2 describes our data acquisition and data
reduction processes.  Section 3 covers our analysis techniques.
Periods, light curves, and classifications of the variable stars 
are presented in Section 4.  A re-discussion of the Oosterhoff
classification of the Draco dwarf spheroidal is presented in Section 5.
Conclusions are summarized in section 6.


\section{Data Acquisition and Reduction}
Our survey of the Draco dSph galaxy was conducted at two
telescopes: the 1.0m at the US Naval Observatory in Flagstaff, AZ.,
and the 2.3m telescope at the Wyoming Infrared Observatory
(WIRO), at Mt. Jelm, Wyoming. Combined, the two datasets cover a time interval of four
years (1993-1996).  Table \ref{jdates} contains the Heliocentric Julian
dates for when the data were observed.

\subsection{USNO observations}
\begin{figure}
\figurenum{1}
\includegraphics[scale=0.8]{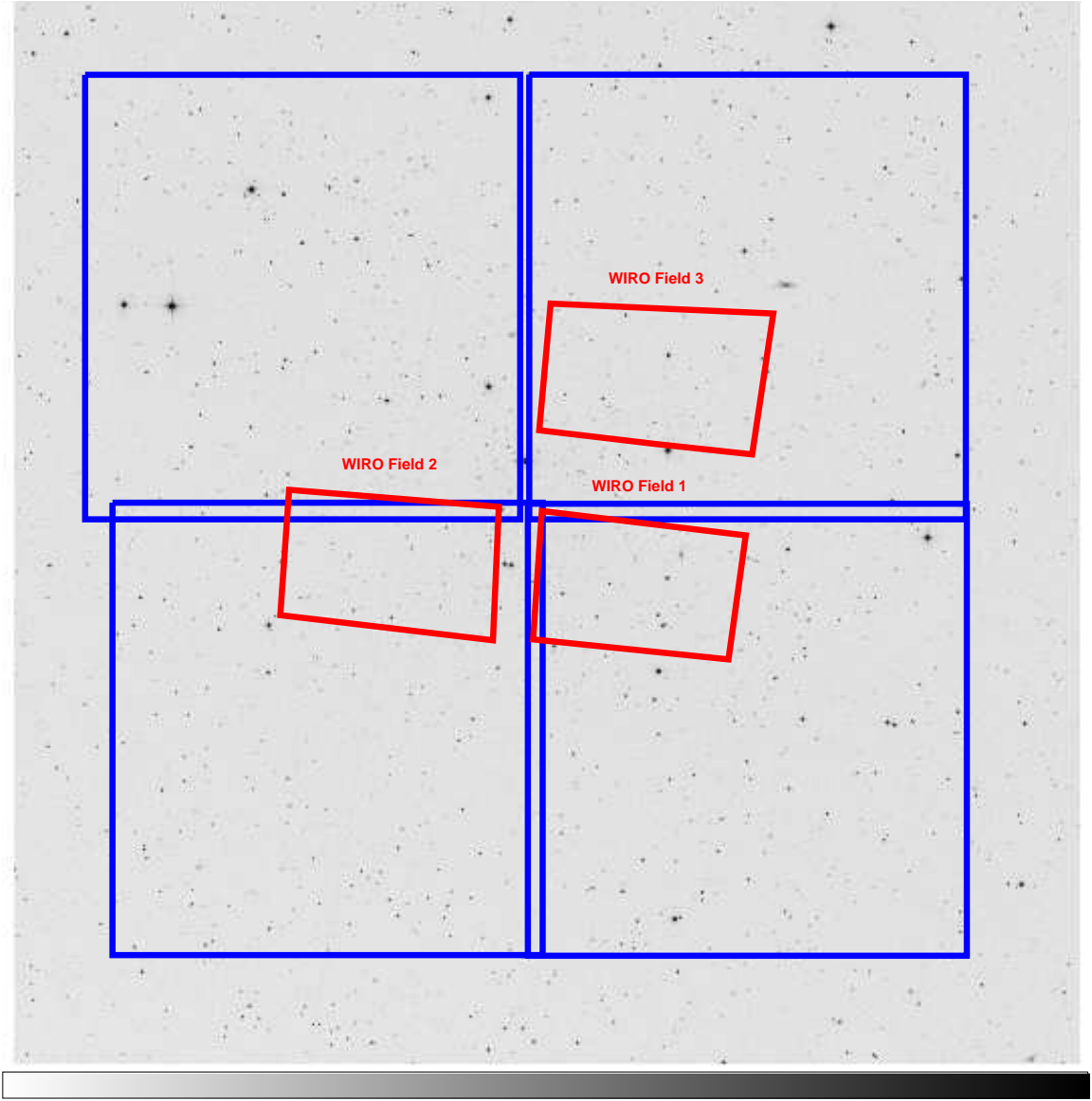}
\caption{The observed fields of the Draco dwarf spheroidal galaxy.  The
  regions outlined in blue are fields observed at USNO while the red boxes
  are the fields observed at WIRO.} 
\label{fields}
\end{figure}

Images of Draco were taken with the 1.0 m telescope of the U.S.
Naval Observatory in Flagstaff, AZ, during the 1995 and 1996
observing seasons.  A Tektronix 2048$\times$2048 CCD was used
with a pixel size of 0.68 arcsec, giving a field size of 23.2 arcmin.
Four fields were observed, each covering one quadrant of Draco,
with 1 arcmin overlap between fields, thus covering approximately
a square region of 45 arcmin size centered on Draco.
The northeast field position was shifted to the east to avoid the
bright star just north of Draco.  Therefore the northeast and
northwest fields did not overlap, and three variable stars (V5, V10,
and V117) were
missed in this narrow gap.  Figure \ref{fields} shows the field placement.
This areal coverage is larger than any other study of the variable
stars in Draco --- it covers about four times more area and more than
two times the variable stars than the study by \citet{Bonanos:2004},
and it provides a useful coverage of about two times more area
than that of B\&S.  The Palomar 200 inch telescope used by B\&S
allowed discovery of some variable stars up to a distance of 24
arcmin from the center of Draco.  However, the degraded image quality
in the outer parts of their field prevented them from measuring
magnitudes or deriving light curves and periods for most variables
beyond an 8 arcmin radius from Draco.  This coverage includes all known
variable stars in Draco from the B\&S study except for two stars they
identified, which were found at large distances from the galaxy (one far east,
V205, and one far west, V333).  Also missing from our study are the three
stars that lie in the gap between the northeast and northwest fields near the
bright star on the north side of Draco.  The first part of Table \ref{omitstars} lists
these stars.

The images were taken with a Johnson $V$ filter throughout the 1995
and 1996 observing seasons, and with a Cousins $I$ filter mostly
during the 1996 season.  The seeing was typically 2$\arcsec$,
and the exposure times were 15-30 min depending on the seeing.
Exposures were taken switching between quadrants, and alternating
filters in 1996, so that each quadrant was observed 1-4 times on a
given night with a given filter.  An effort was made to minimize alias
effects by observing each quadrant over 6-8 hours on several nights,
by observing over three weeks during several months, and by observing
over the full range of months possible each season.  A total of
39-41 $V$ images and 19-20 $I$ images of each quadrant were taken
and are included in the following analysis.  DAOPHOT \citep{Stetson:1987} 
was used to measure all images.  A small radial correction for image
distortion in the corners of each image was applied for the data taken at USNO.  

For the goals of identifying variable stars and measuring accurate
magnitudes and light curves, five sources contribute errors to these
data.  Some stars are crowded or near brighter stars and have erroneous
measurements.  The CCD has a few defects that produce spurious magnitudes
for some stars that occasionally fall on a defect.  The CCD is not physically
flat, so the high center and low corners produce images of stars in
the corners of the field that are not in perfect focus -- together with
astigmatism, the resulting magnitudes have some additional error.
The desire for short exposure times has resulted in images that have
typically 0.03 mag error for each observation for the RR Lyr and other
horizontal-branch stars in Draco.  Finally, the inevitable cosmic rays
occasionally affect a star image.  Therefore, potential variable stars
were examined by eye to decide on real vs. spurious variables.  Table
\ref{omitstars} also lists those stars that B\&S originally marked as
variable candidates but which were found not to be variable in our survey.
The instrumental magnitudes were shifted onto a common system, iteratively
rejecting variable stars, using a method similar to that described by
\citet{Honeycutt:1992}.  

Finally the USNO instrumental magnitudes were transformed to
standard Johnson $V$ and Cousins $I$ magnitudes as follows.
On three photometric nights when Draco images were taken in all
quadrants, \citet{Landolt:1992} standards were also observed and used
to determine transformation coefficients of the form

\begin{equation}
  V = v + C_{0} + C_{1}*(V-I) + C_{2}*Airmass
\end{equation}
\noindent
and similarly for $I$.  On one additional photometric night,
using a different Tektronix 1024x1024 CCD, images were taken
centered on Draco, together with Landolt standards.  Color
coefficients were small, typically 0.01 and 0.03 in $V$ and $I$,
respectively.  These coefficients are presented in Table
\ref{usnotrans} for both $V$ and $I$ bands and per photometric night.  
Three nights were used to determine mean $V$ and $I$ standard magnitudes for a
subset of bright (16-18 mag) nonvariable stars in the Draco images.
The transformation of instrumental magnitudes (after shifting
onto the common system) to standard magnitudes for this subset of bright stars
then was applied to all stars.  A comparison of the resulting standard magnitudes
for nonvariable stars with Stetson's Draco calibration
region\footnote{http://cadcwww.hia.nrc.ca/cadcbin/wdb/astrocat/stetson/query/}
shows good agreement.

The resulting errors in photometry for a single observation are estimated to
be 0.01 from calibration uncertainties, 0.02 from image distortion in the CCD
corners, and photon noise that increases from 0.01 at $V = 18$ to 0.03 at $V
= 20$ to 0.05 at $V = 21$.  After combining frames, the errors in the mean
magnitudes of nonvariable stars at the level of the horizontal branch ($V =
20$) are estimated to be 0.03 in $V$, 0.03 in $I$, and 0.04 in $V-I$.  The
errors in the mean magnitudes of variable stars are generally larger.

\subsection{WIRO Observations}
The USNO dataset was combined with data obtained at WIRO during the
summer quarter observing season of 1993 and 1994.  An RCA $337 \times
527$ pix CCD camera was used, which had a 1.2 ``/pix plate scale.  The
field of view was much smaller compared to the USNO dataset.  The WIRO
fields were $6.4 \times 10.4'$ and overlapped with three quadrants of the USNO
fields.  One WIRO field is roughly 13\% of one USNO field.  Figure
\ref{fields} shows where the WIRO fields are in relation to the USNO
fields.  The data were obtained with Johnson $V$ and Cousins $I$ filters.
From WIRO, a maximum of 28 $V$ and 18 $I$-band images supplemented the USNO
data.  All available data from WIRO was used for light curve and period
analysis of the variable stars.  This brings a maximum of 69 $V$ and 38 $I$
images for stars found in both USNO and WIRO fields.

The WIRO observations were placed on a standard system by using
secondary standards from the USNO analysis.  A total of 45 stars were used for
the calibration, and the dominant source of uncertainty are from the original
calibration done with the USNO dataset for each bandpass (see section 2.1).  Equation \ref{vtran}
and \ref{itran} are the transformation equations for the WIRO dataset.
The coefficients $\alpha_{V}$ and $\alpha_{I}$ were field dependent and were determined
from a weighted mean of differences.  The coefficients $\beta$ and $\gamma$
were obtained from a linear least squares fit between $(V-v_{0})$ and
$(V-I)$.  The standard $V$ magnitude was found through an iterative
process, incorporating the standard $I$ magnitude of that star.  The values of
the transformation coefficients for the WIRO dataset are given in Table \ref{trans}.

\begin{equation}\label{vtran}
V=v_{0} - \alpha_{V} + \beta(v_{0} - I) + \gamma
\end{equation}

\begin{equation}\label{itran}
I = i_{0} - \alpha_{I}
\end{equation}

Photometry was performed on the WIRO data using Stetson's DAOPHOT II
and ALLFRAME stand alone package \citep{Stetson:1987, Stetson:1994}.

\subsection{Variable Star Identifications}
We have kept the original numbering system of B\&S for the first 203
variable stars plus number 204 assigned by \citet{Zinn:1976a}.  All
new variable star identifications, as well as the new long period
variable stars and the QSOs, are an extension of B\&S's system, but
organized by right ascension going east to west.  Our new star
identification, therefore, begins from V205 through V333. 

Stars with high dispersion or high chi-square for
their magnitude were considered to be potential variables and inspected
further.  For the USNO dataset, we used a plot of chi-squared vs. magnitude to
identify potential variables.  We did not use the Welch \& Stetson variability
index \citep{Welch:1993} because it is defined to make use of pairs of images
taken at nearly the same time, and the USNO data generally included unpaired
images in each quadrant each night.  Image differencing might in retrospect be
useful as an additional tool; however, it is more advantageous in fields more
crowded than Draco.

For the WIRO dataset, the variable stars were selected by using a simple
variability index which compared the external to internal uncertainties of the
observations. Our results were then compared to the B\&S catalog and we 
identified the known variable stars.  New variables were found and
classified by their color, period, and location in the color-magnitude diagram
(CMD).  Due to the overlap of the WIRO fields with the USNO fields, the
variable stars found were checked and confirmed between the two datasets. 

A total of 270 RR Lyrae stars, 9 anomalous Cepheids, 12 semi-irregular or
carbon stars, and 2 eclipsing binaries were discovered in this survey.
We were able to recover all of the original B\&S variable stars, and
reclassified 7 stars.  We discuss the variable stars of Draco in more detail
in section 4.

\subsection{Comparison with \citet{Bonanos:2004}}
As discussed in section 2.1, our survey of Draco is nearly four times
larger in area and twice the number of variable stars as were found than
in \citet{Bonanos:2004}'s survey.  Because of a match up error in
preparing their tables, the periods, magnitudes, and the
identifications of 48 stars do not match the RA/DEC star names in
Bonanos' Tables 1 and 2.  We have used corrected versions of the
tables, kindly provided by A. Bonanos, to make the comparison here.
We independently recovered 130 stars that had been found in both the
original B\&S and Bonanos et al. surveys. Those stars with the B04
designation in our Table \ref{properties} have been identified in
Bonanos et al.  They also identified 17 new RRL stars and one new
eclipsing variable; we independently recover all 18 of these stars and
make the same classifications, although we find one star (V289) to be an
RRd that Bonanos et al. classified as RRab.  For this star, we were able to
find a period ($P = 0.6607$ days) close to Bonanos et al. period, but it
produced a noisy light curve with our data.  Our solution produces a tighter
light curve for our photometry.  Bonanos et al. identify
9 red variables with small amplitudes near the tip of the giant
branch.  We find four of them to vary, and find the other five not to
vary significantly in our data, so we omit them from our tables.
Finally, the SX Phe star Bonanos et al. found was too faint for our
survey and was not included in our analysis.  For most RRL stars, we
find excellent agreement with the periods and RRab/RRc classifications
with Bonanos et al.  The typical difference between Bonanos et al.'s
and our periods for the RRL stars is 0.00002 days.  For a few stars,
we find a different alias period.  The greater number of nights
covered by our observations usually make alias problems less important
in our analysis, so we prefer our period solutions.


\section{Data Analysis}
\begin{figure}
\figurenum{2}
\includegraphics[scale=0.8]{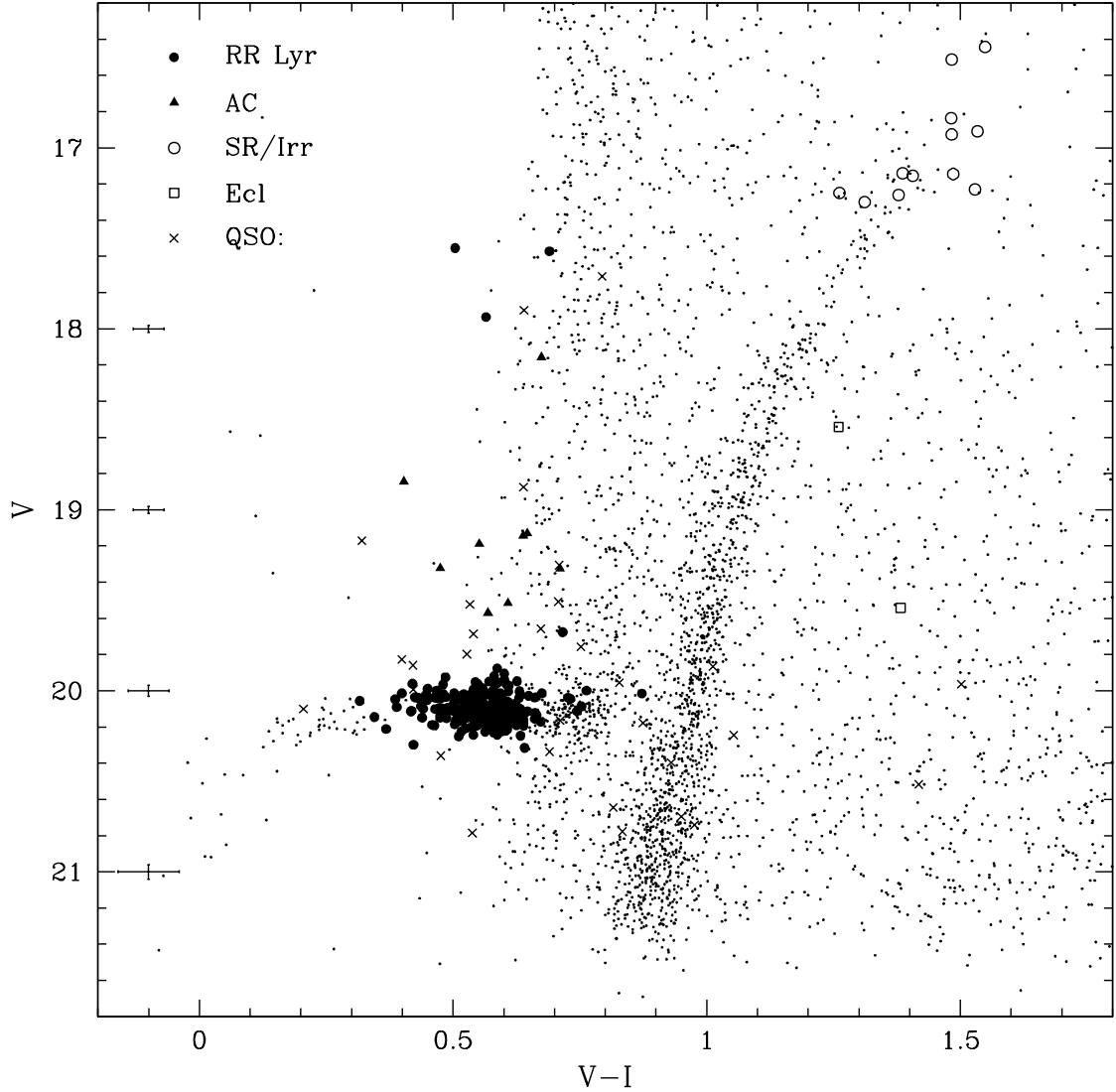}
\caption{Color-magnitude diagram of the Draco dwarf spheroidal
  galaxy.  Variable stars (RR Lyrae, Cepheids, eclipsing binaries, and
  semi-regular) are identified in the figure.  Background QSOs are
  included in this diagram.  Representative error bars for nonvariable stars
  are shown at the left edge of the figure.}
\label{cmd}
\end{figure}

\begin{figure}
\figurenum{3}
\includegraphics[scale=0.8]{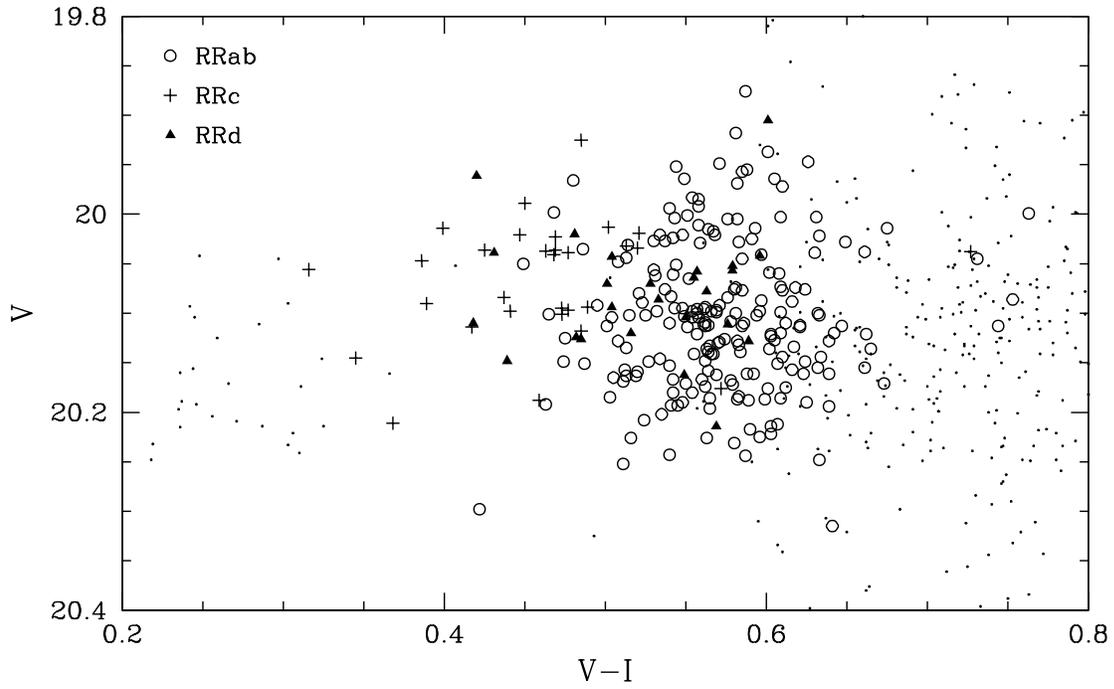}
\caption{A closer view of the horizontal branch.  RRab stars are the open circles,
  RRc stars are the plus signs, and the RRd stars are the filled triangles.}
\label{hbcmd}
\end{figure}

Once the datasets from USNO and WIRO were independently reduced, the
data were combined.  This increased the number of epochs for 103
variable stars.  Using our combined datasets, we present a robust CMD of the
Draco dSph galaxy down to a limiting magnitude of $V = 21$ in figure
\ref{cmd}.  In this updated CMD, we have identified pulsating and eclipsing
stars as well as background QSOs in the Draco field.  Our census has yielded
279 stars that are either of the RR Lyrae or Cepheid type of pulsating
variable star.  We have found 12 variable stars which were not RRL, anomalous
Cepheids, or eclipsing stars, but belong to other types, either slow,
semi-regular, red, or other objects.  There appears to be 30 background QSOs
found in our coverage of the Draco galaxy.  The rest of the stars plotted in
figure \ref{cmd} are non-variable stars (approximately 4700 stars).  There is
also contamination of field stars from the Milky Way, and thus, a
likelihood of field RRL in our survey.  We address this possibility in section
4.1.  Figure \ref{hbcmd} is a close up view of the horizontal branch region of
the CMD.  Here we identify the individual RRL Bailey types as well as the
non-variable stars.  We note that there is a large scatter of nearly 0.4
magnitudes for the RRL. 

The subsequent analysis was done in four steps: 1) period
searching, 2) amplitudes and mean magnitudes calculation, 3)
Fourier decomposition of the light curves, and 4) deriving distances from the
RRL population.  The Fourier decomposition work is discussed in detail in
section 4.2. 

For the full dataset, we anticipated minimizing any period alias
solutions, specifically any yearly aliases.  Our primary period
searching method was the date compensated discrete Fourier transform
(DCDFT) program \citep{Ferraz:1981, Foster:1995}.  This program
was particularly useful for datasets that have a patchy distribution
of data points (i.e. the observations were more or less annual).  The
actual DCDFT program is part of the CLEANEST program
\citep{Foster:1995}.  An updated version of this program is available
through Peranso \footnote{www.peranso.com}.  As a check for the
period solutions, the IRAF version of the phase dispersion
minimization statistic (PDM) \citep{Stellingwerf:1978} was used as
well as the Supersmoother routine \citep{Reimann:1994}.  Overall our
periods are good to about 0.00001 to 0.00003 days.  To obtain the amplitudes
of the $V$ and $I$ variable star data, we use a spline fit to the phased light
curve. 


\section{Variable Star Census}
\subsection{RR Lyrae Stars}
\begin{figure}
\figurenum{4}
\includegraphics{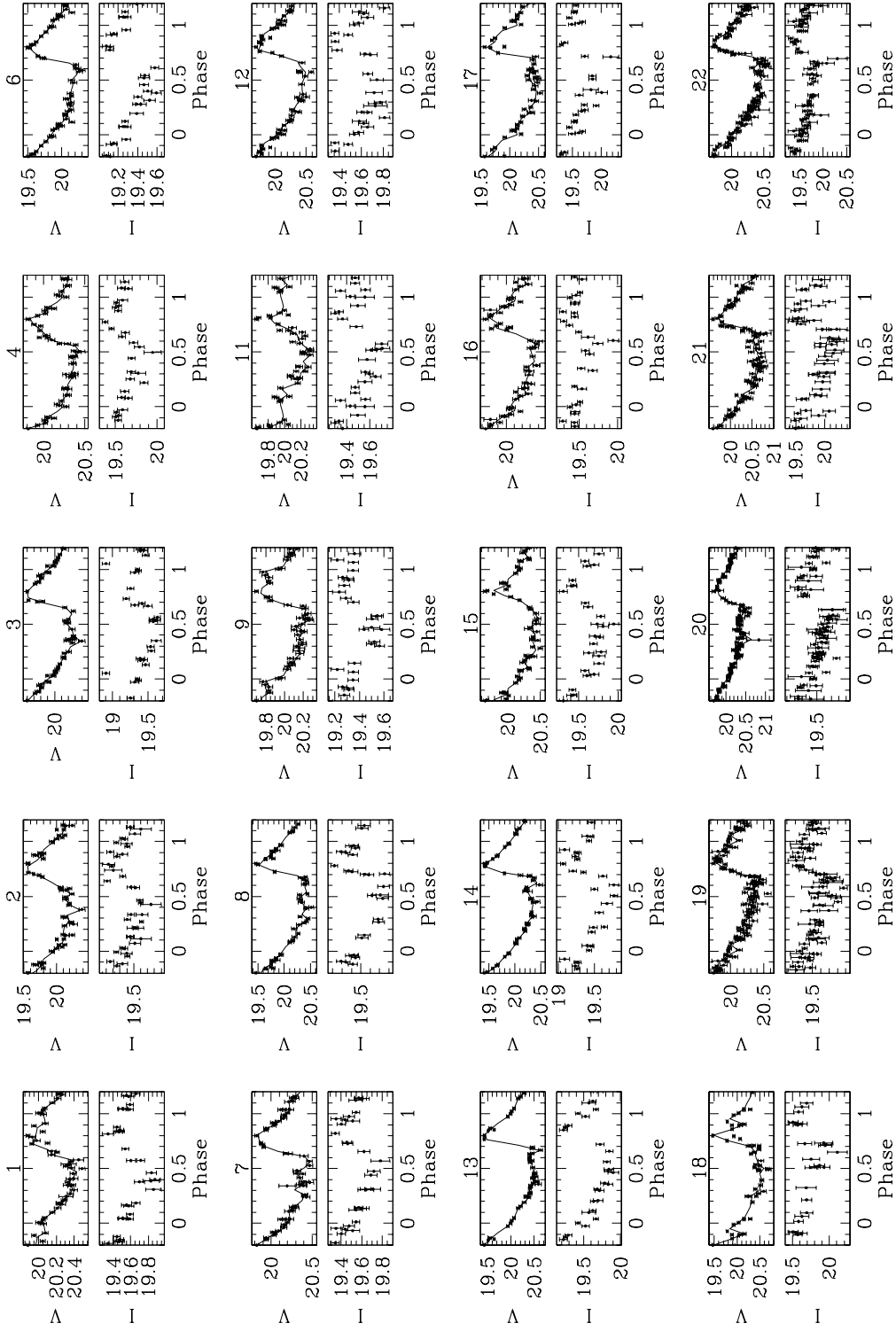}
\caption{Phased light curves of the Draco RRL population.  We present a sample
  of these light curves here.  The full figure is provided in the online version of this paper.}
\label{ltc}
\end{figure}

\begin{figure}
\figurenum{5}
\includegraphics[scale=0.7,angle=-90]{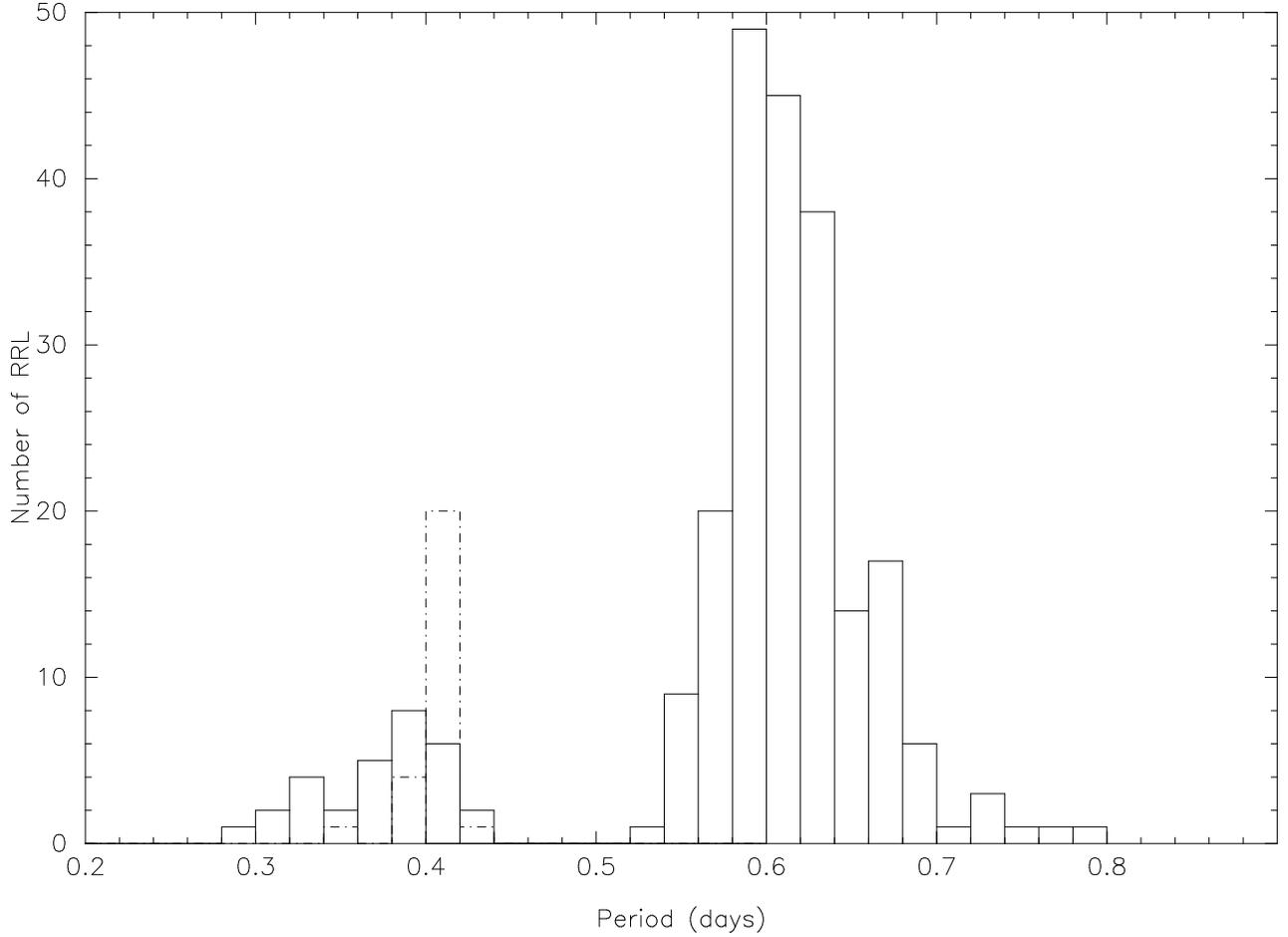}
\caption{Period distribution of all Draco RR Lyrae stars.  The dash-dot
  histogram is of the double-mode RR Lyraes. Average periods for each Bailey
  type of RRL: $\langle P_{ab} \rangle = 0.615d$, $\langle P_{c} \rangle =
  0.375d$, $\langle P_{d} \rangle = 0.407d$.} 
\label{rrlhist}
\end{figure}

Figure \ref{ltc} shows the phased $V$ and $I$ light curves.  The $V$ light
curves have our best spline fit included to aid the eye.  Fourier series fits
to our light curves were not used because they often give biassed results at
rapidly changing phases (rising and maximum light) if few data points are
available to constrain the fit.  With typically 40 $V$ observations, some
stars in our data have few points at these phases.  Table
\ref{properties} lists the RRL positions (RA and DEC J2000.0), the period
solutions (column 4), the $V$ amplitude (column 5), the intensity-weighted
mean magnitudes in $V$ and $I$ (columns 6 and 7), and the type of RRL with
additional notes (column 8).  We find in our survey 270 RR Lyrae stars, of
which 214 are RRab, 30 RRc, and 26 RRd stars.  Of these 81 are new RRL
compared to the B\&S study.  Including these new RRL stars, we find the
average period of the RRab stars to be $\langle P_{ab} \rangle = 0.615 \pm
0.003$ days and for the RRc stars an average period $\langle P_{c} \rangle =
0.375 \pm 0.006$ days.  In Figure \ref{rrlhist}, we show the period
distribution of the RRL stars.  The average period for the RRd stars
is $\langle P_{d} \rangle = 0.407 \pm 0.002$ days.  As originally
noted by B\&S, the mean period of the RRab stars is Oosterhoff
intermediate.  The Oosterhoff properties of the Draco dwarf system are
discussed more fully in section 5. 

Foreground RRL have been found in our survey.  Using the surface
density for RRL in the SA57 field \citep{Kinman:1994}, and assuming a
halo space density of $R^{-3.5}$, we calculated the volume and RRL per
magnitude along our line of sight.  From the calculation, we expected
0.9 field RRL in the line of sight, but in actuality we find 3 field
RRL (V327, V321, and V276).  One of these stars (V327) was previously
discovered by \citet{Wehlau:1986}.  The distribution of stars per magnitude
peaked around $V=17$, thus we should see field RRL around this
magnitude.  The three field RRL are flagged in the main RRL properties
table, Table \ref{properties}.

\subsubsection{Double-Mode RR Lyrae Stars}
\citet{Goranskij:1982} used the photometry of \citet{Baade:1961} to identify
three RR Lyrae stars in Draco that were pulsating simultaneously in the first
overtone and fundamental radial modes.  Also using the \citet{Baade:1961}
observations, \citet{Nemec:1985a} identified seven more of these stars
\citep[RRd variables in Nemec's nomenclature, or RR01 stars in the
nomenclature of][]{Clement:2001}.  \citet{Bonanos:2004} redetermined periods
for six of the RRd stars found by \citet{Nemec:1985a}.  

We carried out a search for double-mode behavior among the RR Lyrae stars
that had light curves that did not seem to be adequately described by a
single period.  Using the CLEANest routine \citep{Foster:1995} to prewhiten the
V-band observations, we removed the primary frequency and its first four harmonics.
A search was then undertaken for evidence of a significant secondary
period.  If a secondary period seemed possible, the CLEANest routine was
used to simultaneously fit the primary and secondary periods and their
first four harmonics.  Although higher order harmonics and cross frequency
terms have been detected in the light curves of double-mode RR Lyrae stars,
the current set of observations is not sufficient to identify them.
For suspected RRd stars, results from the CLEANest routine were verified
using the Period04 program \citep{Lenz:2005}.

\begin{figure}
\figurenum{6}
\includegraphics[scale=1.0]{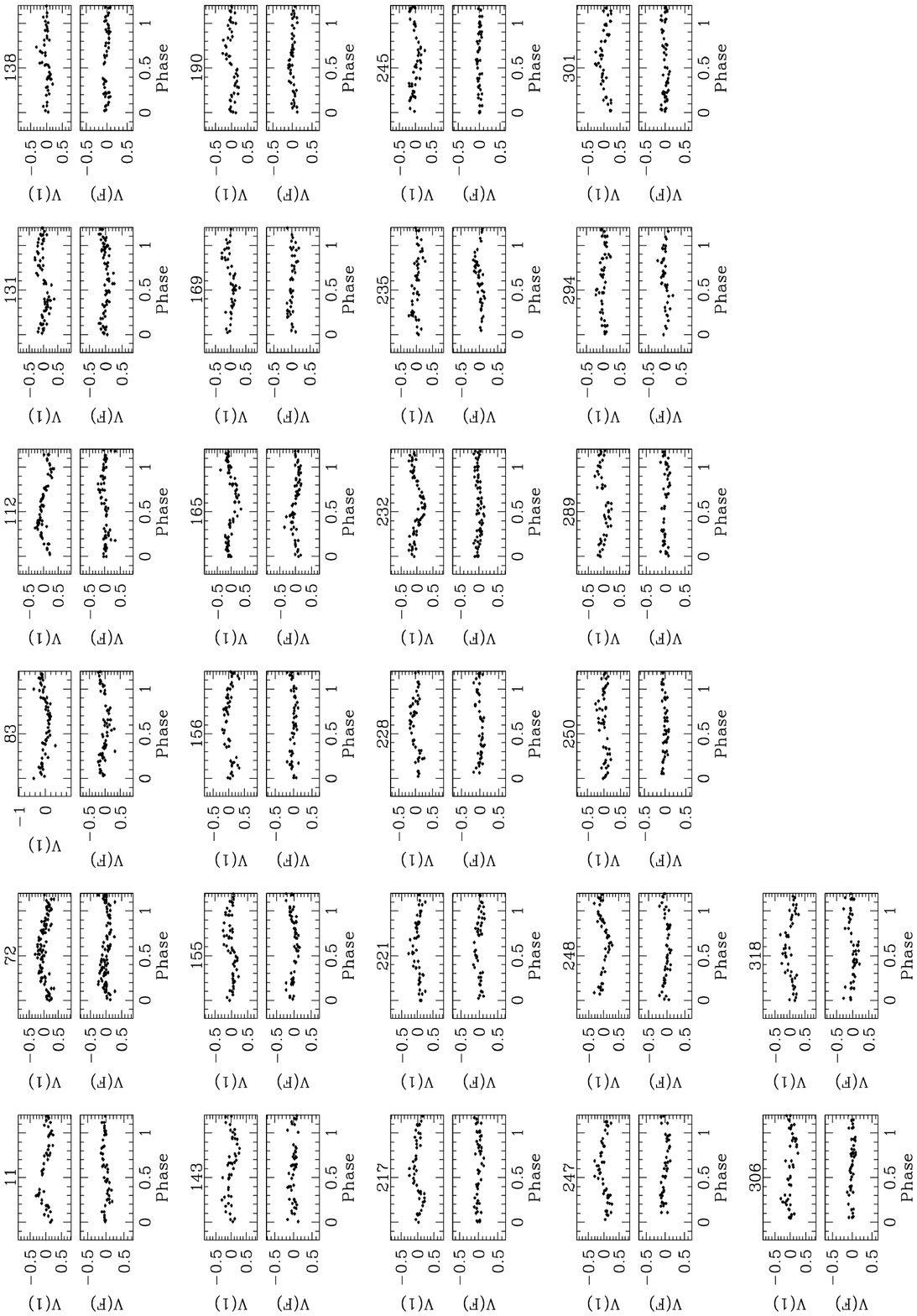}
\caption{Deconvolved light curves of the Draco RRd stars.  Upper plots are the
first overtone pulsation while the lower is of the fundamental mode.}
\label{rrdltc}
\end{figure}

By this means we found all ten of the RRd stars identified by \citet{Goranskij:1982}
and \citet{Nemec:1985a}.  In addition, we have identified 16 probable RRd
variables, giving a total of 26.  The first overtone mode was the dominant
mode in each case.  First overtone mode periods, fundamental mode periods,
and period ratios for probable RRd stars are shown in Table \ref{rrd}.  The listed
uncertainties are the formal errors given by the CLEANest program.  Results
for stars with asterisks are more uncertain, usually because of the possibility
of a period alias for the fundamental mode period. Deconvolved first overtone
and fundamental mode period light curves for the RRd stars are shown in
Figure \ref{rrdltc}.

\begin{figure}
\figurenum{7}
\includegraphics{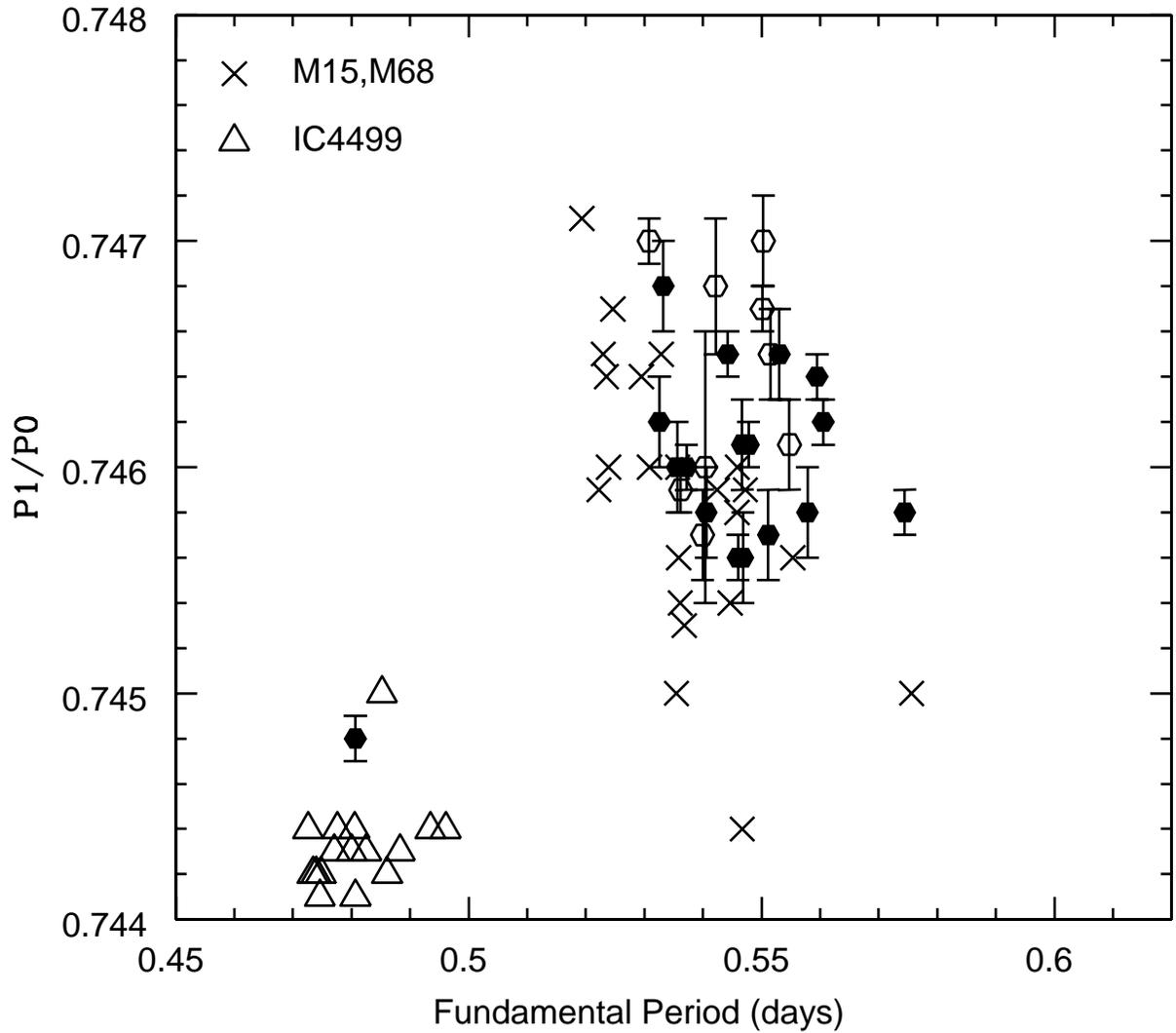}
\caption{Petersen diagram of Draco RRd stars.  The Draco RRd stars with
  uncertainties in the period are plotted as open circles.  For comparison,
  RRd stars from the Oosterhoff type II clusters M15 (\citet{Nemec:1985b};
  \citet{Purdue:1995}) and M68 \citep{Walker:1994} and Oosterhoff type I
  cluster IC 4499 \citep{Walker:1996} are included.}
\label{petersen}
\end{figure}

In plotting the Petersen diagram \citep{Petersen:1973} of period ratio versus 
fundamental period, \citet{Nemec:1985a} discovered that V165 had a
position in this diagram similar to those seen among
RRd stars in Oosterhoff type I globular clusters, but that all of the other
stars had properties similar to those of RRd stars in Oosterhoff type II
clusters.  Figure \ref{petersen} shows the Petersen diagram for all 26
probable RRd stars.  RRd stars whose locations in this diagram are
somewhat uncertain (the asterisked stars in Table \ref{rrd})
are plotted as open points.  For comparison, the locations of RRd stars in
the Oosterhoff type I globular cluster IC 4499 \citep{Walker:1996} and the
Oosterhoff type II globular clusters M15 and M68 \citep{Nemec:1985b,
  Purdue:1995, Walker:1994} are also plotted.  V165 still remains the only RRd star
with properties similar to those of RRd stars in Oosterhoff type I clusters.

\begin{figure}
\figurenum{8}
\includegraphics{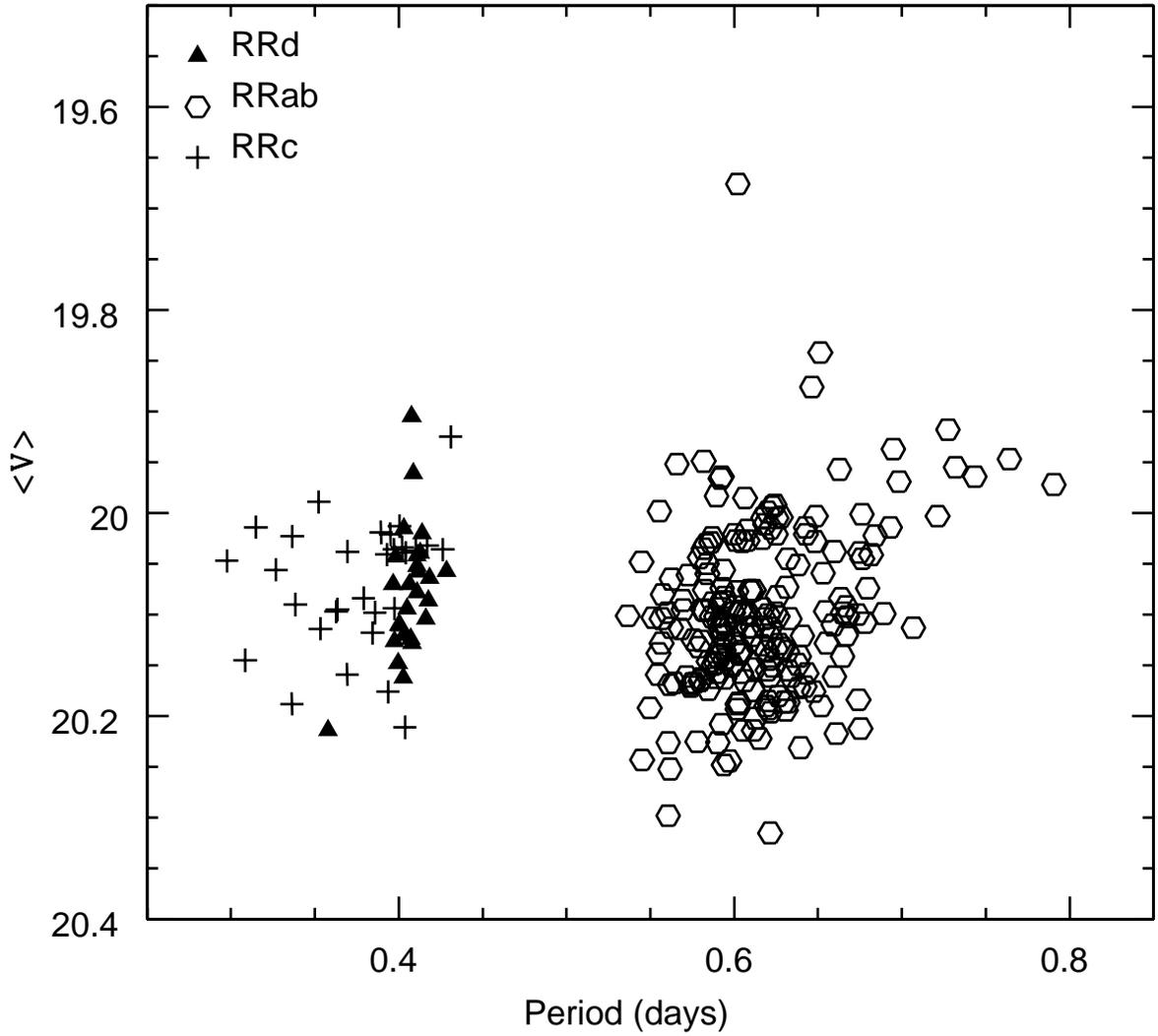}
\caption{The intensity weighted mean $V$ magnitude is plotted against
  period for all RRL stars in our study.  Open circles are the RRab, plus signs are the RRc, and the filled triangles are the RRd stars.}
\label{pv}
\end{figure}

Figure \ref{pv} plots the luminosity weighted mean V magnitude against the
primary period for all of the Draco RR Lyrae stars.  V165, the sole Oosterhoff
type I RRd star, is also the faintest RRd star.  This is at least qualitatively
consistent with other findings that RR Lyrae stars in Oosterhoff type I
clusters are less luminous than those in Oosterhoff type II systems 
\citep[e.g.,][]{Sandage:1958, Sandage:1981}.

\subsubsection{Blazhko Effect}
The Blazhko effect is a second order modulation most evident in the
shape of the RRL light curve.  The maximum light phase can be
depressed by the Blazhko effect.  This effect is also periodic -- on
the order of tens to hundreds of days.  What causes the Blazhko effect
is not clearly known, but there are several proposed explanations 
\citep[see][]{Kolenberg:2006,Stothers:2006}. 

We do not have enough observations to determine Blazhko periods for
those RRL stars in our sample that show the Blazhko effect.  We
can, however, identify as Blazhko effect candidates those RRL
stars that have unusually large scatter in their light curves and which do not
seem to be RRd stars.  We list these Blazhko candidates in Table
\ref{properties} by noting ``Bl'' in the last column.  Stars V26,
V33, V34, V35, V37, V39, V41, V68, V75, V96, V123, V129, V147, V150, V160,
V184, and V196 have already been identified as possible Blazhko variables by
\citet{Nemec:1985a} and \citet{Bonanos:2004}.  The mean period of the
Blazhko effect candidates among the RRab stars is $\langle P_{Bl}
\rangle = 0.603 \pm 0.006$ days.

\subsection{Fourier Decomposition}
The Fourier decomposition of the light curves was done only on the $V$
data. Using Simon's MINFIT program \citep{Simon:1979, Simon:1982}, a
cosine series up to 8th order was fit to the light curves:

\begin{equation}
m = A_{0} + \sum A_{i} \cos (i\omega (t-t_{0}) + \phi_{0})
\qquad\mbox{where i = 1, 2 ...}
\end{equation}

\noindent
Once the amplitude ($A_{i}$) and phase ($\phi_{i}$) terms were
obtained, the Fourier parameters, $R_{ji}$ and $\phi_{ji}$, were
calculated up to the 4th order.

We applied the \citet{Jurcsik:1996} photometric metallicity relation
using the Fourier decomposition parameter $\phi_{31}$ and the period
(their equation 3).  The Jurcsik \& Kovacs method works best when RRab
light curves are fully sampled and where photometric uncertainties are
relatively small.  The light curves for individual RRab stars in our
sample do not always meet these criteria.  To test the quality of the
RRab light curve for this method, a compatibility test called the
$D_{M}$ deviation parameter, is calculated.  This deviation parameter
is determined from a comparison of the observed and predicted Fourier
parameters.  An updated version of this test is provided in
\citet{Kovacs:1998}.  In order for a star to be a good candidate for
the Jurcsik \& Kovacs method, the $D_{M}$ parameter criterion must be
met.  For our RRab sample, we chose $D_{M} < 3.0$ (as recommended by
Jurcsik \& Kovacs) and $D_{M} < 5.0$ \citep[as recommended
by][]{Clement:1999}.  Stars that have passed the criteria are listed in Table
\ref{met} with asterisks.  Table \ref{met} also lists the Fourier
decomposition parameters and photometric metallicities of the Draco RRab
stars.  All photometric metallicities on are the metallicity scale of the
Jurcsik \& Kovacs method \citep{Jurcsik:1995}.

\begin{figure}
\figurenum{9}
\includegraphics{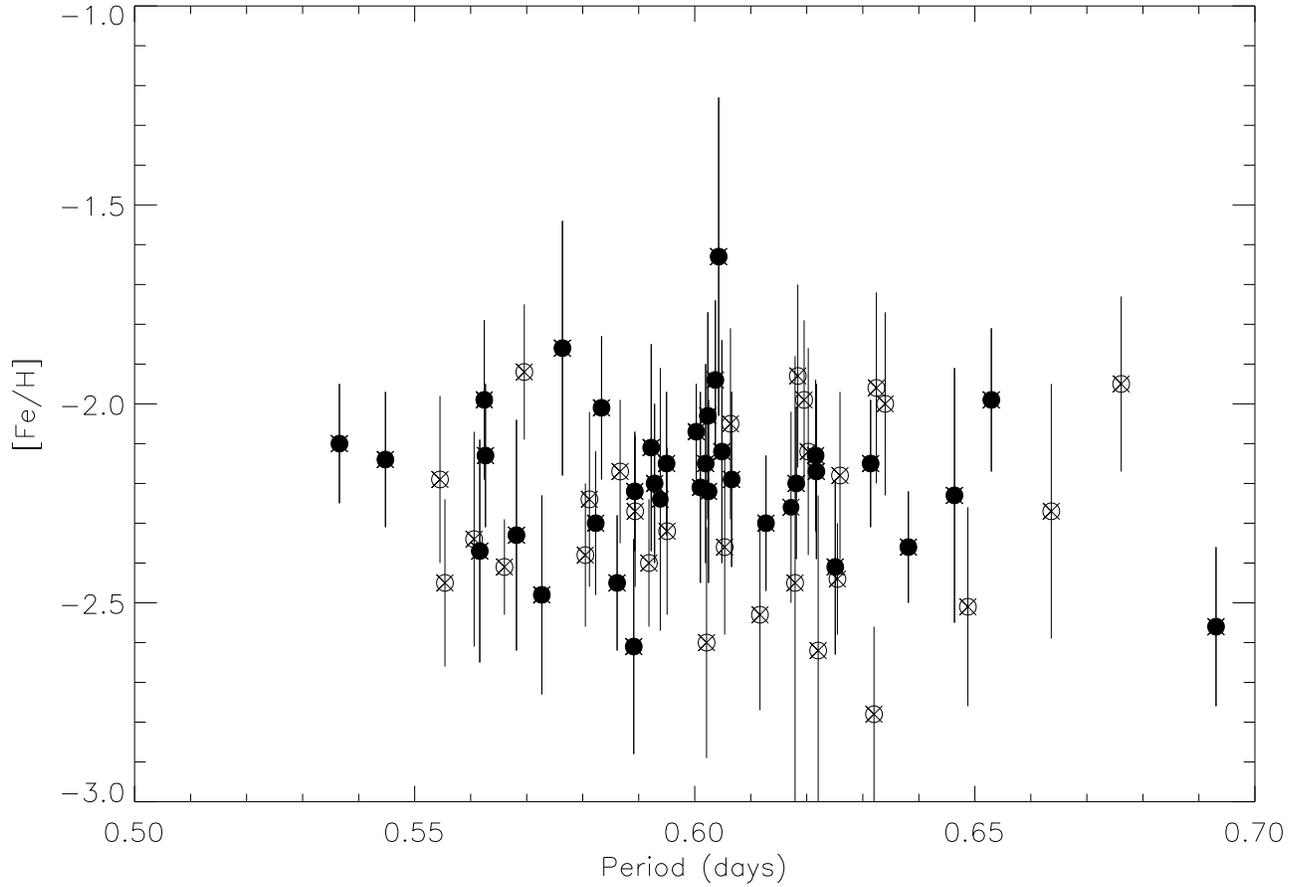}
\caption{Metallicity distribution with respect to period for 63 Draco
  RRab stars.  [Fe/H] values were determined via the empirical method
  described in \citet{Jurcsik:1996}. The filled points correspond to $D_{M} <
  3.0$ and the open points are for $3.0 < D_{M} < 5.0$.}
\label{fehz}
\end{figure}

The [Fe/H] values derived from the Jurcsik \& Kovacs method may in
this case be more useful in deriving a mean [Fe/H] value for Draco
than in the determination of metallicities for individual stars.  It
is quite likely that some of the outlying [Fe/H] values in Table
\ref{met}, at both the high and low end do not really reflect the
metallicities of the stars for which they are derived. The average
[Fe/H] for Draco, as determined by the photometric metallicities of
the RRab stars, is $\langle [Fe/H] \rangle = -2.19 \pm 0.03 $, if we
assume the stars are not undergoing the Blazhko effect (see section
4.1.1) and have passed the $D_{M} < 3.0$ criterion.  For the case
where $D_{M} < 5.0$, and assuming no Blazhko, the average metallicity
of Draco is $\langle [Fe/H] \rangle = -2.23 \pm 0.03$.  Figure
\ref{fehz} shows the metallicity distribution of the RRab stars that
have passed the $D_{M} < 5.0$ criterion with respect to period.

Using Stromgren photometry, \citet{Faria:2007} recently obtained a
mean $[Fe/H]$ of $-1.74$ for Draco and field red giant branch stars, with most
stars falling within the limits $-1.5 > [Fe/H] > -2.0$.  That result
is broadly consistent with the earlier results of
\citet{Shetrone:2001b} and \citet{Zinn:1978}, although Shetrone et
al. did find one red giant star as metal poor as $[Fe/H] = -2.97$.
\citet{Faria:2007} calibrated their derived metallicities to the work of
\citet{Hilker:2000}, which analyzed the red giants of three globular clusters
and spanned a metallicity range of $-2.0$ to 0.0 dex.  Therefore, we must be
cautious when comparing out metallicity results to those of other studies
since there are dependencies to various metallicity calibrations.  However,
there is a suggestion that the average metallicity of the Draco RRab stars is
lower than that of the Draco red giant stars.  The reality of this difference
in metallicity is uncertain due to the nature of the different calibration
methods.  If this difference is real, then presumably the red horizontal
branch stars in Draco would have to be more metal-rich on average than the
RRab stars.

\subsection{RRL distance for Draco}

Since RRL are excellent distance indicators, we calculate the distance to the
Draco dwarf galaxy.  We use the metal-poor ($[Fe/H] < 1.5$) relation from
\citet{Cacciari:2003} (their equation 4).  As with Bonanos et al.'s work, we
use an $E(B-V) = 0.027$ from the \citet{Schlegel:1998} reddening maps, and the
corrections for the extinction as suggested by the work of
\citet{Cardelli:1989}, thus, $A_{V} = 0.091$.  From our sample of RRL stars,
the intensity-weighted mean V magnitude is $\langle V \rangle = 20.10 \pm
0.04$ ($\sigma_{RMS} = 0.08$).  For this value, we omitted the magnitudes of
the foreground RRL (V276, V321, and V327) and V176 since it is blended with a
bright star.  The uncertainties given for this mean
magnitude accounts for the calibration errors, image distortion, and photon
noise (see Section 2.1).  The value of $\langle V(RR) \rangle = 20.10 \pm
0.04$ in this paper is brighter than those of \citet{Bonanos:2004}: $\langle
V(RR) \rangle = 20.18 \pm 0.02$, of \citet{Aparicio:2001}: $ \langle V(HB)
\rangle = 20.2 \pm 0.1$, and of \citet{Bellazzini:2002}: $\langle V(HB)
\rangle = 20.28 \pm 0.10$, with a 2-sigma difference from the most precise
value of Bonanos et al.

If we assume a metallicity for Draco from our Fourier decomposition analysis,
$[Fe/H] = -2.19 \pm 0.03$, and using the \citet{Cacciari:2003} relation, our
resultant absolute magnitude is $\langle M_{V} \rangle = 0.43 \pm 0.13$.
Therefore, using the present mean $V$ magnitude of the RRL stars and
accounting for the extinction, we derive a dereddened distance modulus to
Draco of $\mu_{0} = 19.58$ or $D = 82.4 \pm 5.8$ kpc.  However, if we assume a
different metallicity for Draco, our distance changes slightly.
\citet{Shetrone:2001b} obtained a mean metallicity of $[Fe/H] = -2.00 \pm
0.21$ from high resolution spectroscopy of Draco red giants, whereas
\citet{Faria:2007} found $[Fe/H] = -1.74$.  If we assume the metallicity
values of $-2.00$ and $-1.74$, and using the same Cacciari \& Clementini
relation and the present RRL mean $V$ magnitude, the resultant distances are
81.2 and 79.8 kpc, respectively.  \citet{Pritzl:2002a} arrived at a distance
to Draco independently from the anomalous Cepheid stars (see section 4.4).
Their value is $\mu_{0} = 19.49$ or $D = 79.1$ kpc, but using a reddening
value of $E(B-V) = 0.03$.  Within our uncertainties, we agree with all these
distance values from different Draco studies.

\subsection{Anomalous Cepheids}
In our study of the Draco dwarf galaxy, we increase the number of
known anomalous Cepheids (AC) to nine.  \citet{Baade:1961} had
identified what appeared to be five overly bright RR Lyrae stars in
their original survey.  \citet{Norris:1975}, followed by
\citet{Zinn:1976a} first classified these variables as AC stars (V134,
V141, V157, V194, and V204).  \citet{Nemec:1988a} reidentified the five
stars in Draco as AC, based on a reanalysis of
B\&S's photographic survey.  These five AC stars were
confirmed in our study.  We have been able to add four new AC's: V31,
V230, V282, and V312, to the census.  Table \ref{tableac} lists all
the photometrically derived parameters.

\begin{figure}
\figurenum{10}
\includegraphics[angle=-90]{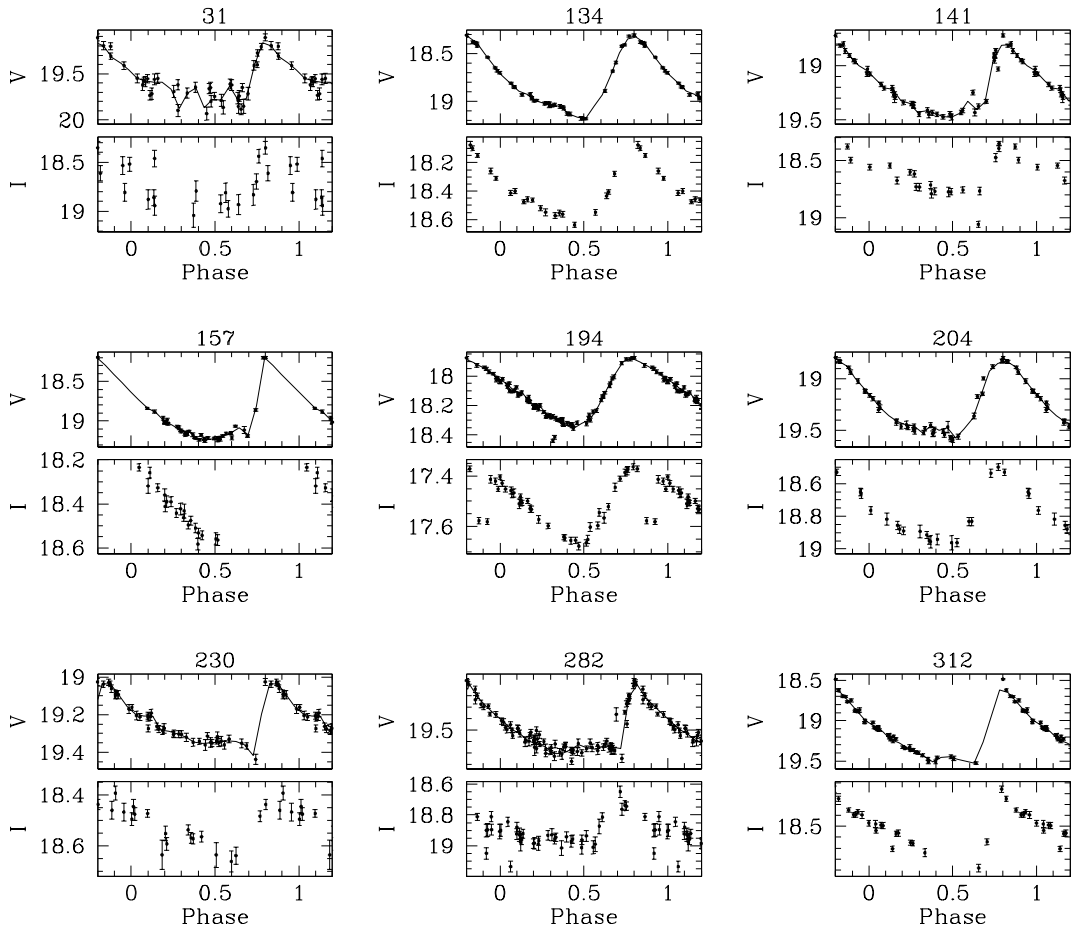}
\caption{Phased light curves of the Draco anomalous Cepheids.}
\label{acltc}
\end{figure}

Of the new anomalous Cepheids, one star, V31, has been reclassified.
Originally, it was identified by B\&S as an RRL variable star based on
eye estimates only.  However, it lies only 13" from a bright BD star.
The $I$ and the $V-I$ color are particularly uncertain because of
scattered light from the nearby bright red star.  Our CCD
data show that it is significantly brighter than other Draco RRL
stars, so we believe it is a new AC.  The bright star is saturated in our data
and contributes significant scattered light around V31.  Nevertheless, after
doing careful background subtraction, the estimated errors in our photometry
are 0.1 in $V$ and 0.2 in $I$, leaving it 0.5 mag brighter than the horizontal
branch.  The other AC stars do not have companions visible in our data.
Futhermore, most are either sufficiently bright or have large
amplitudes that they cannot be an RRL star made brighter by an unresolved
companion.  However for V31, V230, and V282, we cannot exclude this
possibility of RRL-plus-unresolved companion.  In Figure \ref{acltc} we
present the light curves of all the AC found in this survey with a spline fit
added to aid the eye. 

Generally, these variable stars are brighter than the RRL population
by 0.5 (for shorter period, $P \sim 0.3$ days) to 2 magnitudes (longer
period, $P \sim 2.0$ days).  These stars are also more massive than
the RRL, typically 1.0-2.0 $M_{\sun}$ \citep[][and references
therein]{Pritzl:2002a}, and must be relatively metal-poor in order for
the progenitor stars to reach the instability strip. Anomalous
Cepheids have been found in all the known dwarf spheroidal galaxies of
the Local Group, however, they are not typically found in the Galactic
globular clusters.  The exceptions are V19 in NGC 5466
\citep{Zinn:1976b} and two candidates in $\omega$ Cen
\citep{Wallerstein:1984}.  XZ Ceti is a well known field AC.  The
origins of these stars still remains unsolved, but the leading
theories suggest that they are either 1) intermediate aged stars ($t <
5$ Gyrs) or 2) primordial binary systems undergoing mass transfer.
These mechanisms provide alternative origins for the blue straggler
populations that have been speculated to be the progenitor stars of
the AC.

Recently, \citet{Momany:2007} investigated the frequency of blue
straggler stars in the Local Group dSph population, compared to the
frequency of such stars in Galactic globular clusters, open clusters,
and the field.  They find that, in general, the blue straggler
frequency is higher in dSph galaxies than in globular clusters.  If
the blue straggler stars are progenitors of the ACs, then this higher
frequency is consistent with a higher frequency of ACs among the dSph
systems.  It is noteworthy, too, that some mechanisms for creating blue
stragglers by mass transfer may not operate in systems of low stellar
density, such as the dSph.  For example it has been suggested that
collisional binary systems might create blue straggler stars,
but such collisions would be infrequent in dSph systems \citep{Momany:2007}.
Thus, to consider the blue straggler star frequency, one must only
consider those stellar formation mechanisms that will be likely in
a dSph environment if one wishes to correlate with the number of AC
stars found.

Anomalous Cepheids of dSph galaxies have also been used to create a
period-luminosity (P-L) relation.  Recent work by \citet{Dall'Ora:2003},
\citet{Pritzl:2002a}, and \citet{Nemec:1994} have presented empirical
anomalous Cepheid P-L relations associated with the pulsational mode.  Both
empirical and theoretical P-L relations have shown that they are not parallel
\citep{Pritzl:2002a, Bono:1997}.  However, there is still some question as to
whether the two apparent P-L relations are real, due to distinct fundamental
and first-overtone mode relations, or whether the results might instead be
interpreted as a single P-L relation with large scatter.  That scatter might
be a reflection of the range of AC masses as well as the finite width of the
instability strip.

\begin{figure}
\figurenum{11}
\includegraphics[scale=0.7, angle=-90]{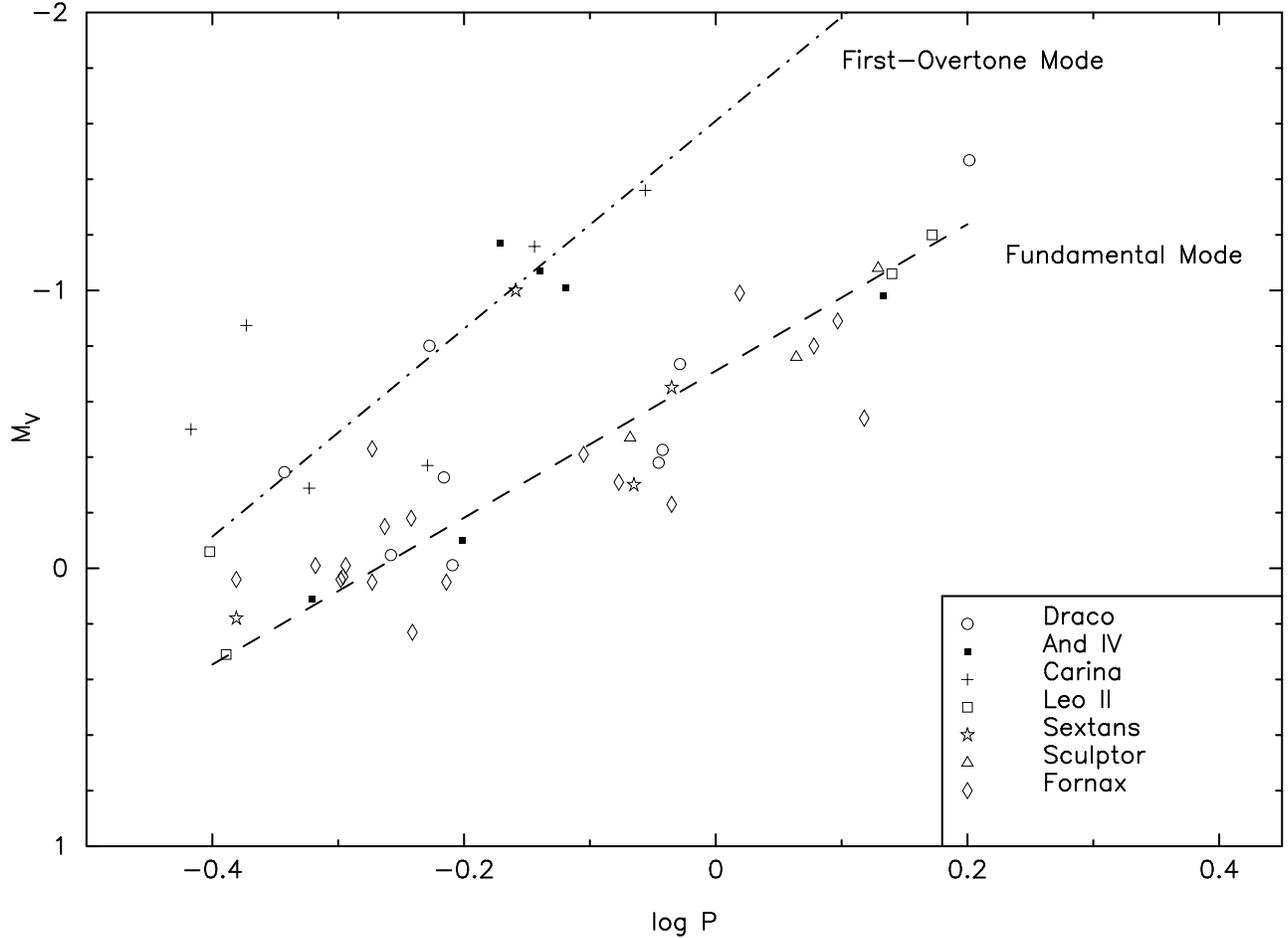}
\caption{Draco anomalous Cepheids with respect to other anomalous
  Cepheids found in dSph galaxies.  Information of anomalous Cepheids
  of other dwarf galaxies are from \citet{Pritzl:2002a,
  Dall'Ora:2003}.  Period-luminosity relations for
  the fundamental and first overtone pulsational modes from
  \citet{Pritzl:2002a} are included.}
\label{acpl}
\end{figure}

For the Draco AC sample, we applied the empirical P-L relations of
\citet{Pritzl:2002a} to see whether the location of the additional Draco stars
would support the reality of two distinct P-L relations.
We have calculated absolute magnitudes for the Draco AC
stars assuming a distance modulus of $(m-M)_{0} = 19.49$ and an
$E(B-V) = 0.03$ \citep{Pritzl:2002a} in order to incorporate our results with
their empirical P-L relations.  Figure \ref{acpl} shows
the location of the Draco AC stars with respect to the AC stars found
in other Local Group dwarf galaxies.  We see
that most of the Draco ACs (V31, V141, V157, V194, V230, V282, and
V312) fall along the P-L relation for stars pulsating in the
fundamental mode, but two, V134 and V204, fall closer to the first
overtone mode P-L relation.  As discussed in \citet{Pritzl:2002a}, it
is difficult to assign the pulsational mode in this manner, especially
if phase coverage is not complete.  We find that to be the case for
the Draco ACs as well.  Two possible first overtone pulsators, V134
and V204, have light curves showing only modest asymmetry.  Among RR
Lyrae stars that is a sign of RRc or first overtone mode pulsation.
However, the light curves for the supposed fundamental mode pulsator,
V194, seems similar.  Thus, we can only indicate that while there is
evidence in Figure \ref{acpl} for two distinct AC P-L relations, the
actual situation is still uncertain.  For example, a range of masses
among the ACs might influence the positions of the Draco AC within the
P-L diagram, and it perhaps cannot be entirely excluded that a single
P-L relation with scatter could account for the observations.

\subsection{Other Variable Stars}
Three categories of variable stars other than RR Lyraes
and Cepheids appear in our data:  two eclipsing binaries,
30 ``bluish long-period variables'', 12 red semi-regular or irregular
variables, and carbon stars have been found and are listed in Tables
\ref{longper} and \ref{qso}.  The following subsections discusses each of
these types of stars.

\subsubsection{Eclipsing Binary Stars}
\begin{figure}
\figurenum{12}
\includegraphics[scale=1.0]{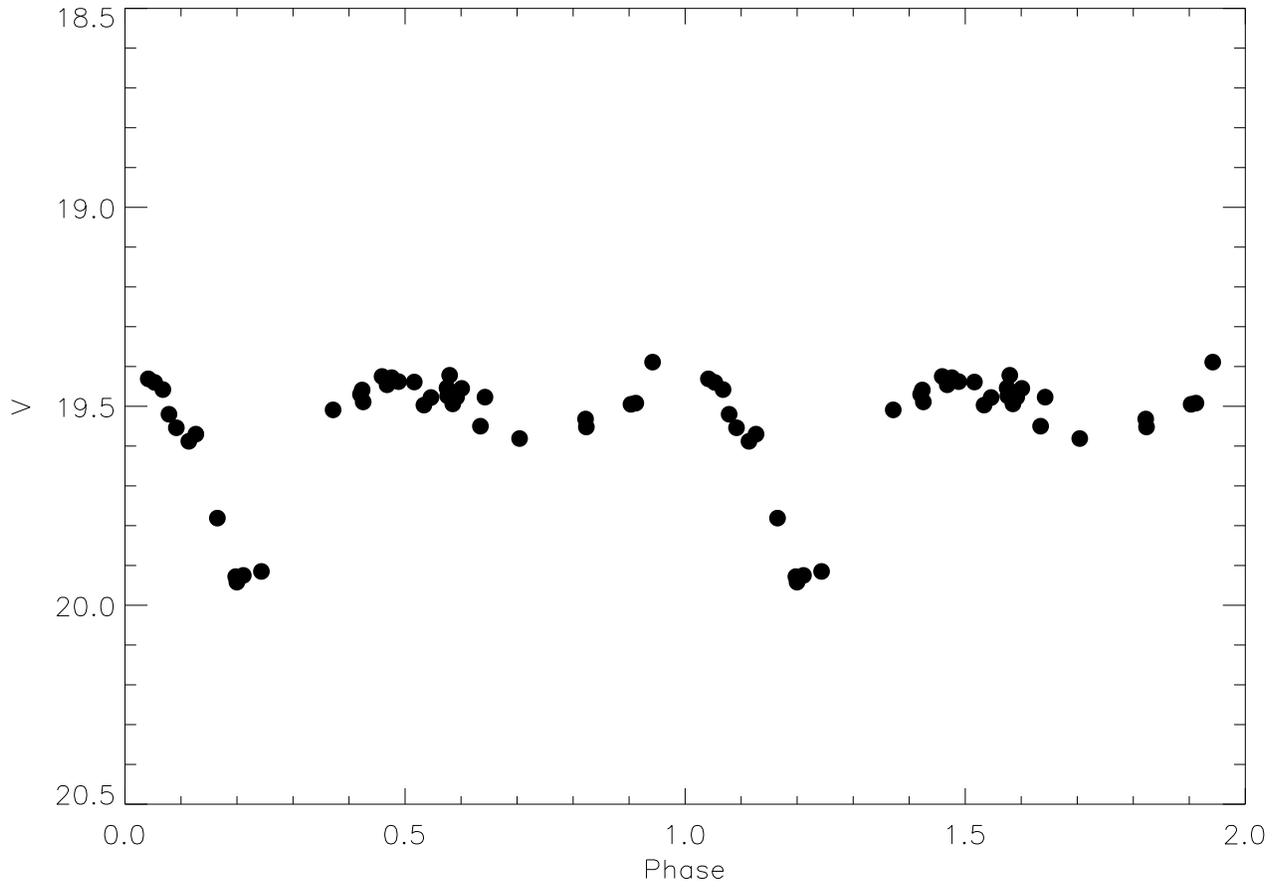}
\caption{Field eclipsing binary star found in
  \citet{Bonanos:2004} (their ID:J171906.2+574120.9) and in this work
  (our ID: V296). Period is 0.2435 days, which agrees well with their
  derived period.}
\label{eclip}
\end{figure}

A field eclipsing binary star (V296) was found in the survey completed by
\citet{Bonanos:2004}, which we have recovered in our work.  We agree
with their period solution for this star with a period of 0.2435
days.  Figure \ref{eclip} shows the light curve of V296 phased to this
period.  Additionally, we have also found another possible eclipsing
binary star with a small amplitude change.  This star, V256, has few
faint observations, and our period result is somewhat uncertain. In Table
\ref{longper}, we provide two plausible period solutions.  However, to
truly confirm the nature of this eclipsing binary, a careful follow up
will be needed to arrive at the correct period.  

\subsubsection{Carbon Stars}
A population of stars redward of the red giant branch (RGB) have been
often identified as carbon stars \citep{Aaronson:1983}.  There are
six carbon stars known in Draco \citep[C1-C3][]{Aaronson:1982};
\citep[C4][]{Azzopardi:1986}; \citep[C5][]{Armandroff:1995};
\citep[C6][]{Shetrone:2001a}.  We find the stars C1, C2, and C5 to be variable
with $V$ amplitudes close to 0.2 mag.  Stars C3, C4, and C6 do not appear to
vary during two seasons of observations at USNO.  Shetrone et al. also
reported C2 as a definite variable, and C5 as a possible variable.

The unusual nature of star C1 was noted by \citet{Aaronson:1982} and
by \citet{Margon:2002} from their independent study of the star in
a spectrum from the Sloan Digital Sky Survey.  The strong emission
lines of hydrogen and helium, the blue continuum flux, and the X-ray
emission indicate it is a symbiotic carbon star like UV Aur.
Therefore, its photometric variability is not surprising.  It also has
a variable radial velocity \citep{Olszewski:1995} that is likely
caused by orbital motion and may be independent of its variable
brightness.  The other carbon stars, including the two that we find to
be variable, have not shown variable velocities.

\subsubsection{Long Period Variables and QSOs}

\begin{figure}
\figurenum{13}
\includegraphics[angle=-90]{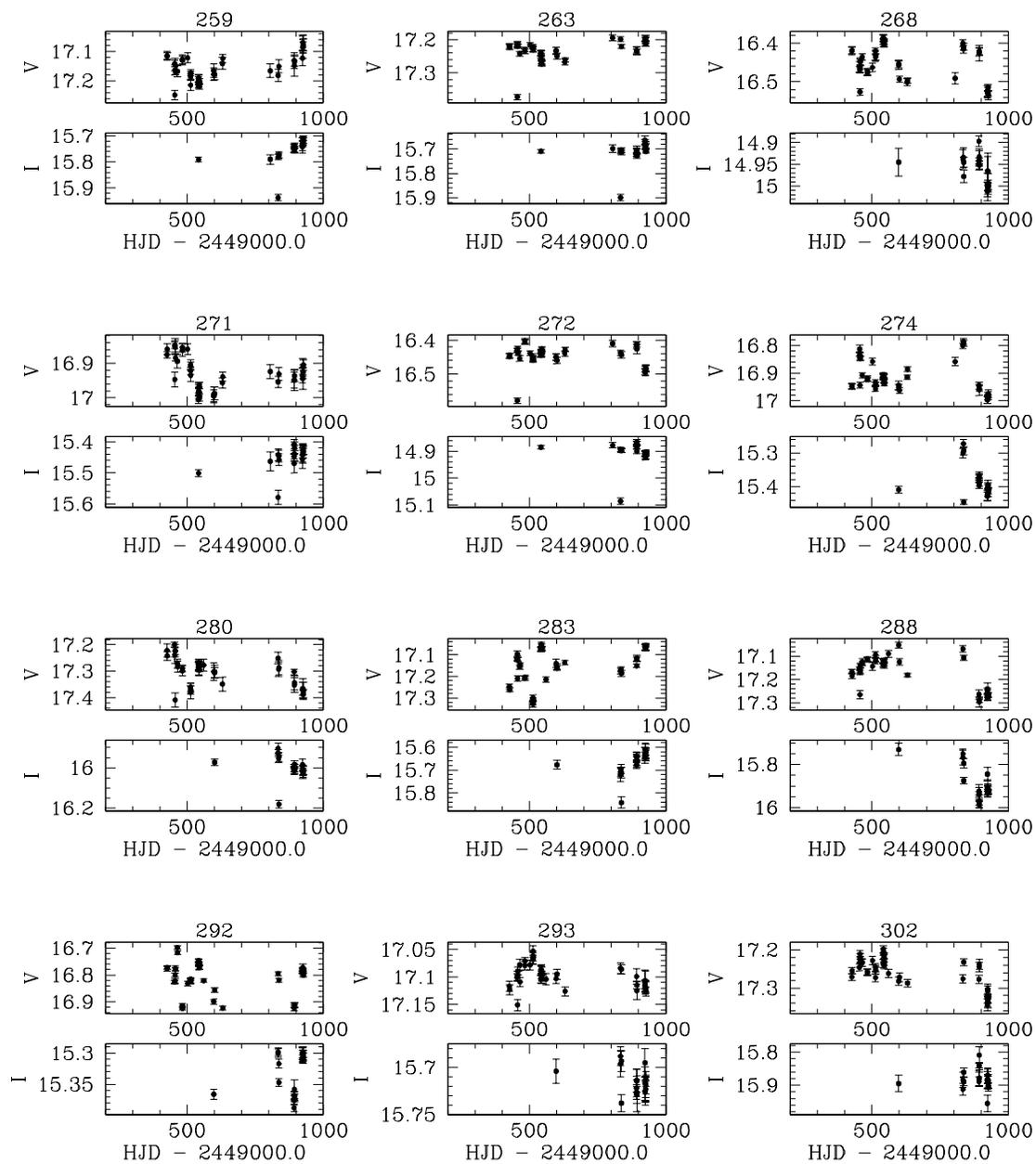}
\caption{Time series data of the long period variable stars found in
  our survey.  The x-axis is the heliocentric Julian date
  (HJD-2449000.0).}
\label{srltc}
\end{figure}

The characteristics of the bluish long-period variables are
slow variability, no apparent period, amplitudes typically 0.25 mag,
colors blueward of the Draco giant branch, and no clear concentration
toward Draco.  These characteristics suggest that most of them are
background QSOs, and this hypothesis has been supported by available
spectroscopy.  The red semi-regular variables have colors and
magnitudes placing them near the tip of the red giant branch (RGB)
in Draco, and they all have radial velocities and/or proper motions
showing they are members of Draco.  Figure \ref{srltc} shows our
time series photometric data of the red long period variable stars.
Spline fits were not included since we assume the coverage of the full
variation was not obtained through the time coverage of our dataset. Also, due
to the approximately 40 $V$ observations, we cannot provide robust estimations
in the amplitudes of these long period variables.  In Table \ref{longper} we
list mean magnitudes rather than intensity-weighted mean magnitudes due to our
spotty phase coverage and to the low amplitudes of these objects.

B\&S remarked on the lack of red variables found in Draco.
\citet{Bonanos:2004} showed there are variables among the stars
near the tip of the giant branch, as is also shown in Figure \ref{cmd}.
Our 12 red variables are mostly of low amplitude, and the amplitudes
must have been just below the threshold for detection by B\&S.
We now know that in metal-poor systems like Draco, high-amplitude
red variables like Miras are absent, and low amplitude semi-regular
or irregular variables are more common \citep[e.g.][]{Harris:1987}.

Distinguishing between background QSOs and red variables in Draco
is not always obvious, however, because QSOs sometimes can be red,
and some Draco variables might be bluish and without regular periods.
For example, UU~Her and RV~Tauri stars are found in globular
clusters and could be confused here with our limited data.
Therefore, spectroscopy is useful to insure accurate classification:
22 bluish long-period variables are confirmed as QSOs with spectra
taken with the WIYN telescope and described in a separate paper
(Harris \& Munn 2008, in preparation) and/or spectra from the Sloan
Digital Sky Survey that put them in the SDSS DR3 QSO Catalog
\citep{Schneider:2005}.  Eight additional variables with similar
characteristics are listed in Table \ref{qso} as probable QSOs.  

The prototype of the QSOs behind Draco, V203, was found by B\&S
and given in their Table VII, and the light curve spanning over six years
was shown in their Fig. 8.  They did not understand its nature,
but their light curve was the best study of variability of a QSO
at that time.  Of course, the redshift of QSOs and their
characteristic variability was not discovered for two more years
\citep{Schmidt:1963, Matthews:1963}.


\section{Draco and the Oosterhoff dichotomy}

\citet{Oosterhoff:1939} found that five RR Lyrae-rich globular clusters
could be divided into two groups, now known as Oosterhoff groups,
on the basis of the properties of their RR Lyrae stars.  Subsequent
investigations found that almost all of the Milky Way globular
clusters that contain significant numbers of RR Lyrae stars could be placed
into one or the other of the Oosterhoff groups. The RRL in Oosterhoff group I
clusters have $\langle P_{ab}\rangle \sim  0.55^{d}$ and $\langle P_{c}
\rangle \sim 0.32^{d}$.  RRL in Oosterhoff II clusters have $\langle
P_{ab}\rangle \sim 0.64^{d}$ and $\langle P_{c}\rangle \sim 0.37^{d}$.
Oosterhoff II clusters are also relatively richer in RRc variables than
Oosterhoff I clusters, and they are more metal-poor than Oosterhoff I clusters
\citep[see, for example][]{Smith:1995}.  However, not all systems show the
Oosterhoff dichotomy. In contrast to the Milky Way globular clusters, dwarf
spheroidal systems often have Oosterhoff intermediate properties \citep[for
recent discussions, see][]{Pritzl:2002a,Catelan:2004,Catelan:2005}. 

\begin{figure}
\figurenum{14}
\includegraphics[scale=0.7,angle=-90]{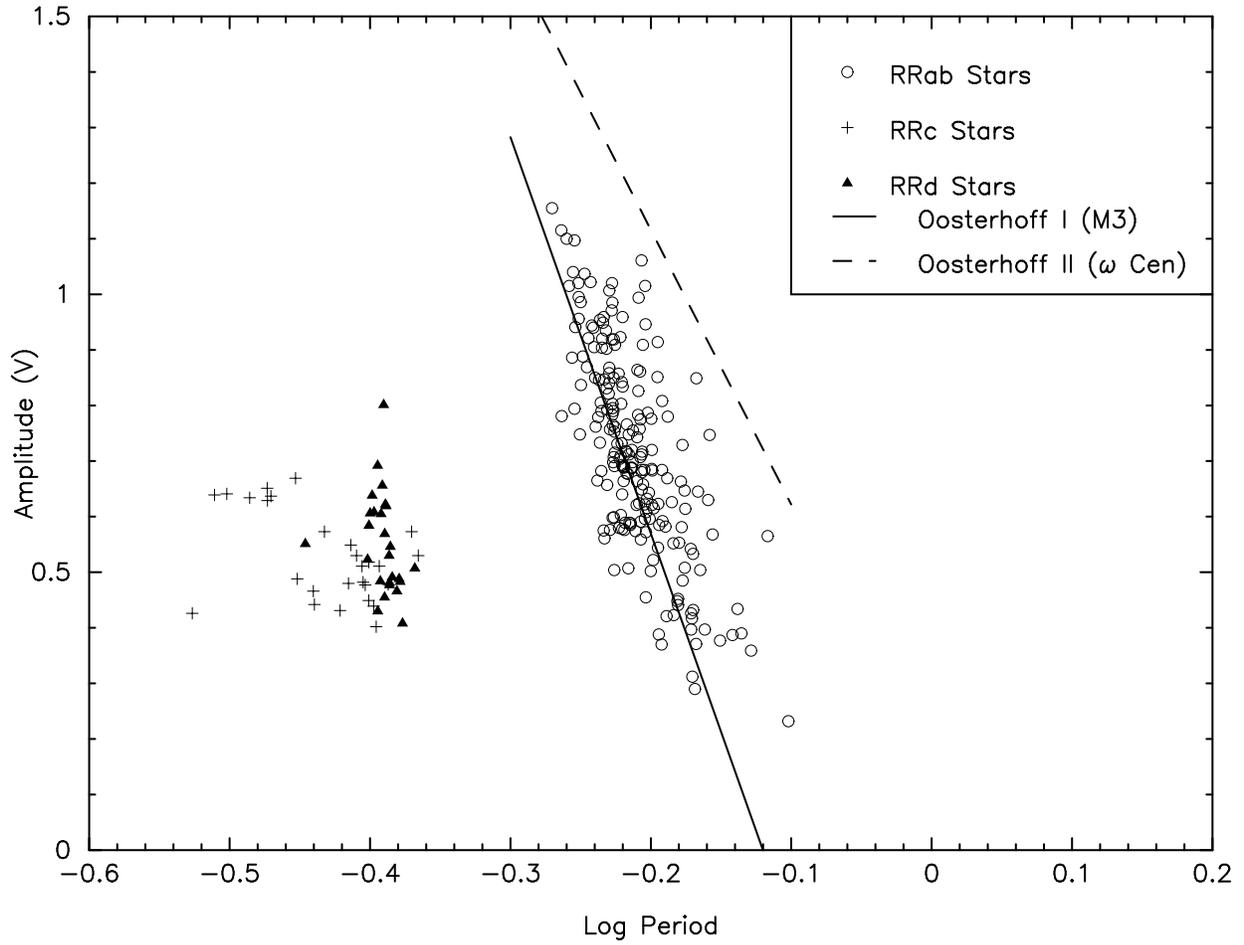}
\caption{Period-amplitude diagram of the Draco RRL stars.
  \citet{Clement:2000}'s relations for Oo I and Oo II are plotted to
  guide the eye for the Oosterhoff classification.}
\label{peramp}
\end{figure}

The mean period of RRab stars in Draco found here, $0.615^{d}$, seems to confirm
its Oosterhoff intermediate nature.  However, a detailed inspection of the
the properties of its RRL suggests a complicated picture.  The Draco
period-amplitude (Bailey) diagram (Figure \ref{peramp}) is consistent with an
Oosterhoff intermediate classification.  Many of the RRab stars occupy
positions in this diagram close to that of the \citet{Clement:2001} Oosterhoff
I line, but many also fall to the right of that line, in the Oosterhoff
intermediate zone.  This result is qualitatively true if the trend lines of
\citet{Cacciari:2005} are used instead of those of \citet{Clement:2000}.  The
\citet{Cacciari:2005} lines are based on the period-amplitude distribution of
RRab, some of which are evolved along the horizontal branch, of M3.
In the Milky Way, a metallicity of $[Fe/H] = -2.1$ would be typical of globular
clusters of Oosterhoff type II.  However, the ratio of RRcd to RRab stars
in Draco, 0.26, is typical of Oosterhoff type I clusters.
In contrast, the RRcd stars in Draco appear dominated by stars having
Oosterhoff type II characteristics.  This is particularly true of the
RRd stars -- all but one of which fall in the Petersen diagram in the
region occupied by double-mode stars observed in Oosterhoff II clusters 
such as M15.  In summary, RRab stars in the period-amplitude diagram
and the mean RRab period suggest an Oosterhoff intermediate class.  However,
the RRcd population has attributes usually associated with an Oosterhoff type
II system.  The mean period of the RRcd stars, 0.39 days, is long and the
location of the RRd stars in the Petersen diagram suggests a mainly Oosterhoff
II class.  Figure \ref{rrlhist} shows a sharp fall off in the number of RRcd
stars as one goes to shorter periods.  This in part reflects an overall
falloff in the number of HB stars as one goes from red to blue across the
instability strip.  There is a hint of a bimodal distribution in the RRc
periods, but its significance is uncertain because of the small number of RRc
stars toward the short period end of the distribution.  Conclusions as to the
Oosterhoff classification of the RRL stars are probably surer when based upon
the more numerous RRab variable stars.

It is plausible that the discordant Oosterhoff properties of the Draco
RRL are in some way related to the overall distribution of stars across its
horizontal branch.  Draco has a HB redder than found among ordinary
Oosterhoff II clusters, or among Milky Way globular clusters having
$[Fe/H] = -2$ \citep[see for example][]{Catelan:2005}. It has been
proposed \citep{Lee:1990, Clement:1999, Clement:2001}
that many RRL in Oosterhoff type II systems have evolved into the blue part of
the instability strip from ZAHB positions.  The paucity of blue HB stars 
in Draco makes it unlikely that the majority of its RRL
have evolved from ZAHB positions on the blue HB, and thus perhaps it
is not surprising that Draco is not an Oosterhoff type II
system despite having a low value of [Fe/H]. There may, however, be 
problems with the hypothesis that Oosterhoff II RRL are evolved
blue HB stars.  Even in the cases of ordinary Oosterhoff
type II globular clusters, it has been argued that, according to
conventional stellar evolutionary theory, the HB stars evolving from
the blue HB would not spend sufficient time in the instability strip
to produce the observed numbers of RRL \citep{Renzini:1988, Rood:1989,
  Pritzl:2002b}. Thus, the exact role played by Draco's redder HB in producing
its confusing Oosterhoff properties remains uncertain.  

According to the $\Lambda$-cold dark matter hierarchical model, the
outer halo has been assembled partly due to the accretion of objects
like the Local Group dwarf galaxy population.  However, almost no globular
clusters of the Galaxy have Oosterhoff intermediate properties.  Nor does the
field RRL population of the halo resemble that of Draco \citep[see, for
example][]{Kinemuchi:2006}.  Thus, it seems likely that systems exactly like
Draco cannot have been a main contributor to building the halo.  In addition,
\citet{Shetrone:2001a} and \citet{Pritzl:2005} found that the patterns of
elemental abundances in the dwarf spheroidal galaxies were distinct from those
in globular clusters and halo field stars.  However, \citet{Bellazzini:2002}
argue that objects like Draco could still be considered as a building block if
we consider that the accretion may have occurred early in the star
formation history of the dwarf galaxy or during an early episode of gas
stripping by the Galaxy.  Our findings with Draco at least imply that
objects like this dwarf galaxy could not have contributed to the
formation of the outer halo, even if it was accreted before the old
population was formed in the dSph galaxy.


\section{Summary}
We have presented the latest census of variable stars of the Draco
dwarf spheroidal galaxy.  We have found 81 new RRab stars, 8 new RRc stars,
and 16 probable new RRd stars, thus bringing to 214 RRab, 30 RRc and
26 RRd the total number of RRL stars of the different types known in Draco.
We have increased the number of anomalous Cepheids to nine from five.  Draco
cannot be clearly classified as an Oo I or an Oo II type system.  Based upon
the mean period of its RRab stars and their location in the period-amplitude
diagram, Draco is Oosterhoff intermediate.  Objects exactly like Draco are
thus not likely to be important building blocks in forming the Galactic halo.
We note, however, that the properties of the RRd stars in Draco are, with a
single exception, similar to those of RRd stars in Oo II clusters.

The anomalous Cepheids in the Draco dSph galaxy show a possible dual
P-L relation stemming from the pulsational modes of the stars.  However, with
so few stars populating the first-overtone relation, we cannot say with
certainty that two P-L relation alternative is the only one capable of
describing the relationships of luminosity and period for Draco AC stars.
In addition to the pulsating variable stars, we find two field
eclipsing binary stars, 30 background QSOs, and 12 long period variable
stars.

\acknowledgments

This research used the facilities of the Canadian Astronomy Data
Centre operated by the National Research Council of Canada with the
support of the Canadian Space Agency.  The identification of QSOs is
based partly on spectra obtained with the Hydra multifiber
spectrograph on the WIYN telescope at Kitt Peak National Observatory,
National Optical Astronomical Observatories.  NOAO is operated by the
Association of Universities for Research in Astronomy, Inc., under
cooperative agreement with the National Science Foundation.  This
research has made use of the USNOFS Image and Catalogue Archive
operated by the United States Naval Observatory, Flagstaff Station
(http://www.nofs.navy.mil/data/fchpix/). 

H.A.S. acknowledges support from NSF grant AST 0607249.  K.K. acknowledges
support from NSF grant AST-0307778.  The authors would like to thank the
referee, Gisella Clementini, for very useful comments and detailed suggestions
for the overall improvement of this paper.

\bibliography{apj-jour,dracoref2}

\begin{thebibliography}{75}
\expandafter\ifx\csname natexlab\endcsname\relax\def\natexlab#1{#1}\fi

\bibitem[{{Aaronson} {et~al.}(1983){Aaronson}, {Hodge}, \&
  {Olszewski}}]{Aaronson:1983}
{Aaronson}, M., {Hodge}, P.~W., \& {Olszewski}, E.~W. 1983, \apj, 267, 271

\bibitem[{{Aaronson} {et~al.}(1982){Aaronson}, {Liebert}, \&
  {Stocke}}]{Aaronson:1982}
{Aaronson}, M., {Liebert}, J., \& {Stocke}, J. 1982, \apj, 254, 507

\bibitem[{{Aparicio} {et~al.}(2001){Aparicio}, {Carrera}, \&
  {Mart{\'{\i}}nez-Delgado}}]{Aparicio:2001}
{Aparicio}, A., {Carrera}, R., \& {Mart{\'{\i}}nez-Delgado}, D. 2001, \aj, 122,
  2524

\bibitem[{{Armandroff} {et~al.}(1995){Armandroff}, {Olszewski}, \&
  {Pryor}}]{Armandroff:1995}
{Armandroff}, T.~E., {Olszewski}, E.~W., \& {Pryor}, C. 1995, \aj, 110, 2131

\bibitem[{{Azzopardi} {et~al.}(1986){Azzopardi}, {Lequeux}, \&
  {Westerlund}}]{Azzopardi:1986}
{Azzopardi}, M., {Lequeux}, J., \& {Westerlund}, B.~E. 1986, \aap, 161, 232

\bibitem[{{Baade} \& {Swope}(1961)}]{Baade:1961}
{Baade}, W., \& {Swope}, H.~H. 1961, \aj, 66, 300

\bibitem[{{Bellazzini} {et~al.}(2002){Bellazzini}, {Ferraro}, {Origlia},
  {Pancino}, {Monaco}, \& {Oliva}}]{Bellazzini:2002}
{Bellazzini}, M., {Ferraro}, F.~R., {Origlia}, L., {Pancino}, E., {Monaco}, L.,
  \& {Oliva}, E. 2002, \aj, 124, 3222

\bibitem[{{Bonanos} {et~al.}(2004){Bonanos}, {Stanek}, {Szentgyorgyi},
  {Sasselov}, \& {Bakos}}]{Bonanos:2004}
{Bonanos}, A.~Z., {Stanek}, K.~Z., {Szentgyorgyi}, A.~H., {Sasselov}, D.~D., \&
  {Bakos}, G.~{\'A}. 2004, \aj, 127, 861

\bibitem[{{Bono} {et~al.}(1997){Bono}, {Caputo}, {Santolamazza}, {Cassisi}, \&
  {Piersimoni}}]{Bono:1997}
{Bono}, G., {Caputo}, F., {Santolamazza}, P., {Cassisi}, S., \& {Piersimoni},
  A. 1997, \aj, 113, 2209+

\bibitem[{{Cacciari} \& {Clementini}(2003)}]{Cacciari:2003}
{Cacciari}, C., \& {Clementini}, G. 2003, in Lecture Notes in Physics, Berlin
  Springer Verlag, Vol. 635, Stellar Candles for the Extragalactic Distance
  Scale, ed. D.~{Alloin} \& W.~{Gieren}, 105--122

\bibitem[{{Cacciari} {et~al.}(2005){Cacciari}, {Corwin}, \&
  {Carney}}]{Cacciari:2005}
{Cacciari}, C., {Corwin}, T.~M., \& {Carney}, B.~W. 2005, \aj, 129, 267

\bibitem[{{Cardelli} {et~al.}(1989){Cardelli}, {Clayton}, \&
  {Mathis}}]{Cardelli:1989}
{Cardelli}, J.~A., {Clayton}, G.~C., \& {Mathis}, J.~S. 1989, \apj, 345, 245

\bibitem[{{Catelan}(2004)}]{Catelan:2004}
{Catelan}, M. 2004, in Astronomical Society of the Pacific Conference Series,
  113--+

\bibitem[{{Catelan}(2005)}]{Catelan:2005}
{Catelan}, M. 2005, ArXiv Astrophysics e-prints

\bibitem[{{Clement} {et~al.}(2001){Clement}, {Muzzin}, {Dufton}, {Ponnampalam},
  {Wang}, {Burford}, {Richardson}, {Rosebery}, {Rowe}, \&
  {Hogg}}]{Clement:2001}
{Clement}, C.~M., {Muzzin}, A., {Dufton}, Q., {Ponnampalam}, T., {Wang}, J.,
  {Burford}, J., {Richardson}, A., {Rosebery}, T., {Rowe}, J., \& {Hogg}, H.~S.
  2001, \aj, 122, 2587

\bibitem[{{Clement} \& {Rowe}(2000)}]{Clement:2000}
{Clement}, C.~M., \& {Rowe}, J. 2000, \aj, 120, 2579

\bibitem[{{Clement} \& {Shelton}(1999)}]{Clement:1999}
{Clement}, C.~M., \& {Shelton}, I. 1999, \apjl, 515, L85

\bibitem[{{Dall'Ora} {et~al.}(2003){Dall'Ora}, {Ripepi}, {Caputo},
  {Castellani}, {Bono}, {Smith}, {Brocato}, {Buonanno}, {Castellani}, {Corsi},
  {Marconi}, {Monelli}, {Nonino}, {Pulone}, \& {Walker}}]{Dall'Ora:2003}
{Dall'Ora}, M., {Ripepi}, V., {Caputo}, F., {Castellani}, V., {Bono}, G.,
  {Smith}, H.~A., {Brocato}, E., {Buonanno}, R., {Castellani}, M., {Corsi},
  C.~E., {Marconi}, M., {Monelli}, M., {Nonino}, M., {Pulone}, L., \& {Walker},
  A.~R. 2003, \aj, 126, 197

\bibitem[{{Faria} {et~al.}(2007){Faria}, {Feltzing}, {Lundstr{\"o}m},
  {Gilmore}, {Wahlgren}, {Ardeberg}, \& {Linde}}]{Faria:2007}
{Faria}, D., {Feltzing}, S., {Lundstr{\"o}m}, I., {Gilmore}, G., {Wahlgren},
  G.~M., {Ardeberg}, A., \& {Linde}, P. 2007, \aap, 465, 357

\bibitem[{{Ferraz-Mello}(1981)}]{Ferraz:1981}
{Ferraz-Mello}, S. 1981, \aj, 86, 619+

\bibitem[{{Foster}(1995)}]{Foster:1995}
{Foster}, G. 1995, \aj, 109, 1889

\bibitem[{{Goranskij}(1982)}]{Goranskij:1982}
{Goranskij}, V.~P. 1982, Astronomicheskij Tsirkulyar, 1216, 5+

\bibitem[{{Harris}(1987)}]{Harris:1987}
{Harris}, H.~C. 1987, in LNP Vol. 274: Stellar Pulsation, ed. A.~N. {Cox},
  W.~M. {Sparks}, \& S.~G. {Starrfield}, 274--283

\bibitem[{{Hilker}(2000)}]{Hilker:2000}
{Hilker}, M. 2000, \aap, 355, 994

\bibitem[{{Honeycutt}(1992)}]{Honeycutt:1992}
{Honeycutt}, R.~K. 1992, \pasp, 104, 435

\bibitem[{{Jurcsik}(1995)}]{Jurcsik:1995}
{Jurcsik}, J. 1995, Acta Astronomica, 45, 653

\bibitem[{{Jurcsik} \& {Kovacs}(1996)}]{Jurcsik:1996}
{Jurcsik}, J., \& {Kovacs}, G. 1996, \aap, 312, 111

\bibitem[{{Kinemuchi} {et~al.}(2006){Kinemuchi}, {Smith}, {Wo{\'z}niak}, \&
  {McKay}}]{Kinemuchi:2006}
{Kinemuchi}, K., {Smith}, H.~A., {Wo{\'z}niak}, P.~R., \& {McKay}, T.~A. 2006,
  \aj, 132, 1202

\bibitem[{{Kinman} {et~al.}(1994){Kinman}, {Suntzeff}, \&
  {Kraft}}]{Kinman:1994}
{Kinman}, T.~D., {Suntzeff}, N.~B., \& {Kraft}, R.~P. 1994, \aj, 108, 1722

\bibitem[{{Kolenberg} {et~al.}(2006){Kolenberg}, {Smith}, {Gazeas},
  {Elmasl{\i}}, {Breger}, {Guggenberger}, {van Cauteren}, {Lampens}, {Reegen},
  {Niarchos}, {Albayrak}, {Selam}, {{\"O}zavc{\i}}, \& {Aksu}}]{Kolenberg:2006}
{Kolenberg}, K., {Smith}, H.~A., {Gazeas}, K.~D., {Elmasl{\i}}, A., {Breger},
  M., {Guggenberger}, E., {van Cauteren}, P., {Lampens}, P., {Reegen}, P.,
  {Niarchos}, P.~G., {Albayrak}, B., {Selam}, S.~O., {{\"O}zavc{\i}}, I., \&
  {Aksu}, O. 2006, \aap, 459, 577

\bibitem[{{Kovacs} \& {Kanbur}(1998)}]{Kovacs:1998}
{Kovacs}, G., \& {Kanbur}, S.~M. 1998, \mnras, 295, 834

\bibitem[{{Landolt}(1992)}]{Landolt:1992}
{Landolt}, A.~U. 1992, \aj, 104, 340

\bibitem[{{Lee} {et~al.}(1990){Lee}, {Demarque}, \& {Zinn}}]{Lee:1990}
{Lee}, Y., {Demarque}, P., \& {Zinn}, R. 1990, \apj, 350, 155

\bibitem[{{Lenz} \& {Breger}(2005)}]{Lenz:2005}
{Lenz}, P., \& {Breger}, M. 2005, Communications in Asteroseismology, 146, 53

\bibitem[{{Margon} {et~al.}(2002){Margon}, {Anderson}, {Harris}, {Strauss},
  {Knapp}, {Fan}, {Schneider}, {Vanden Berk}, {Schlegel}, {Deutsch},
  {Ivezi{\'c}}, {Hall}, {Williams}, {Davidsen}, {Brinkmann}, {Csabai}, {Hayes},
  {Hennessy}, {Kinney}, {Kleinman}, {Lamb}, {Long}, {Neilsen}, {Nichol},
  {Nitta}, {Snedden}, \& {York}}]{Margon:2002}
{Margon}, B., {Anderson}, S.~F., {Harris}, H.~C., {Strauss}, M.~A., {Knapp},
  G.~R., {Fan}, X., {Schneider}, D.~P., {Vanden Berk}, D.~E., {Schlegel},
  D.~J., {Deutsch}, E.~W., {Ivezi{\'c}}, {\v Z}., {Hall}, P.~B., {Williams},
  B.~F., {Davidsen}, A.~F., {Brinkmann}, J., {Csabai}, I., {Hayes}, J.~J.~E.,
  {Hennessy}, G., {Kinney}, E.~K., {Kleinman}, S.~J., {Lamb}, D.~Q., {Long},
  D., {Neilsen}, E.~H., {Nichol}, R., {Nitta}, A., {Snedden}, S.~A., \& {York},
  D.~G. 2002, \aj, 124, 1651

\bibitem[{{Matthews} \& {Sandage}(1963)}]{Matthews:1963}
{Matthews}, T.~A., \& {Sandage}, A.~R. 1963, \apj, 138, 30

\bibitem[{{Momany} {et~al.}(2007){Momany}, {Held}, {Saviane}, {Zaggia},
  {Rizzi}, \& {Gullieuszik}}]{Momany:2007}
{Momany}, Y., {Held}, E.~V., {Saviane}, I., {Zaggia}, S., {Rizzi}, L., \&
  {Gullieuszik}, M. 2007, \aap, 468, 973

\bibitem[{{Nemec} {et~al.}(1988a){Nemec}, {Mendes de Oliveira}, \&
  {Wehlau}}]{Nemec:1988a}
{Nemec}, J., {Mendes de Oliveira}, C., \& {Wehlau}, A. 1988a, in ASP Conf. Ser.
  4: The Extragalactic Distance Scale, 180+

\bibitem[{{Nemec}(1985a)}]{Nemec:1985a}
{Nemec}, J.~M. 1985a, \aj, 90, 204

\bibitem[{{Nemec}(1985b)}]{Nemec:1985b}
---. 1985b, \aj, 90, 240

\bibitem[{{Nemec} {et~al.}(1994){Nemec}, {Nemec}, \& {Lutz}}]{Nemec:1994}
{Nemec}, J.~M., {Nemec}, A.~F.~L., \& {Lutz}, T.~E. 1994, \aj, 108, 222

\bibitem[{{Norris} \& {Zinn}(1975)}]{Norris:1975}
{Norris}, J., \& {Zinn}, R. 1975, \apj, 202, 335

\bibitem[{{Olszewski} {et~al.}(1995){Olszewski}, {Aaronson}, \&
  {Hill}}]{Olszewski:1995}
{Olszewski}, E.~W., {Aaronson}, M., \& {Hill}, J.~M. 1995, \aj, 110, 2120

\bibitem[{{Oosterhoff}(1939)}]{Oosterhoff:1939}
{Oosterhoff}, P.~T. 1939, The Observatory, 62, 104

\bibitem[{{Petersen}(1973)}]{Petersen:1973}
{Petersen}, J.~O. 1973, \aap, 27, 89

\bibitem[{{Pritzl} {et~al.}(2002a){Pritzl}, {Armandroff}, {Jacoby}, \& {Da
  Costa}}]{Pritzl:2002a}
{Pritzl}, B.~J., {Armandroff}, T.~E., {Jacoby}, G.~H., \& {Da Costa}, G.~S.
  2002a, \aj, 124, 1464

\bibitem[{{Pritzl} {et~al.}(2002b){Pritzl}, {Smith}, {Catelan}, \&
  {Sweigart}}]{Pritzl:2002b}
{Pritzl}, B.~J., {Smith}, H.~A., {Catelan}, M., \& {Sweigart}, A.~V. 2002b,
  \aj, 124, 949

\bibitem[{{Pritzl} {et~al.}(2005){Pritzl}, {Venn}, \& {Irwin}}]{Pritzl:2005}
{Pritzl}, B.~J., {Venn}, K.~A., \& {Irwin}, M. 2005, \aj, 130, 2140

\bibitem[{{Purdue} {et~al.}(1995){Purdue}, {Silbermann}, {Gay}, \&
  {Smith}}]{Purdue:1995}
{Purdue}, P., {Silbermann}, N.~A., {Gay}, P., \& {Smith}, H.~A. 1995, \aj, 110,
  1712

\bibitem[{{Reimann}(1994)}]{Reimann:1994}
{Reimann}, J.~D. 1994, Ph.D.~Thesis

\bibitem[{{Renzini} \& {Fusi Pecci}(1988)}]{Renzini:1988}
{Renzini}, A., \& {Fusi Pecci}, F. 1988, \araa, 26, 199

\bibitem[{{Rood} \& {Crocker}(1989)}]{Rood:1989}
{Rood}, R.~T., \& {Crocker}, D.~A. 1989, in IAU Colloq. 111: The Use of
  pulsating stars in fundamental problems of astronomy, ed. E.~G. {Schmidt},
  103--119

\bibitem[{{Sandage}(1958)}]{Sandage:1958}
{Sandage}, A. 1958, Ricerche Astronomiche, 5, 41

\bibitem[{{Sandage} {et~al.}(1981){Sandage}, {Katem}, \&
  {Sandage}}]{Sandage:1981}
{Sandage}, A., {Katem}, B., \& {Sandage}, M. 1981, \apjs, 46, 41

\bibitem[{{Schlegel} {et~al.}(1998){Schlegel}, {Finkbeiner}, \&
  {Davis}}]{Schlegel:1998}
{Schlegel}, D.~J., {Finkbeiner}, D.~P., \& {Davis}, M. 1998, \apj, 500, 525

\bibitem[{{Schmidt}(1963)}]{Schmidt:1963}
{Schmidt}, M. 1963, \nat, 197, 1040

\bibitem[{{Schneider} {et~al.}(2005){Schneider}, {Hall}, {Richards}, {Vanden
  Berk}, {Anderson}, {Fan}, {Jester}, {Stoughton}, {Strauss}, {SubbaRao},
  {Brandt}, {Gunn}, {Yanny}, {Bahcall}, {Barentine}, {Blanton}, {Boroski},
  {Brewington}, {Brinkmann}, {Brunner}, {Csabai}, {Doi}, {Eisenstein},
  {Frieman}, {Fukugita}, {Gray}, {Harvanek}, {Heckman}, {Ivezi{\'c}}, {Kent},
  {Kleinman}, {Knapp}, {Kron}, {Krzesinski}, {Long}, {Loveday}, {Lupton},
  {Margon}, {Munn}, {Neilsen}, {Newberg}, {Newman}, {Nichol}, {Nitta}, {Pier},
  {Rockosi}, {Saxe}, {Schlegel}, {Snedden}, {Szalay}, {Thakar}, {Uomoto},
  {Voges}, \& {York}}]{Schneider:2005}
{Schneider}, D.~P., {Hall}, P.~B., {Richards}, G.~T., {Vanden Berk}, D.~E.,
  {Anderson}, S.~F., {Fan}, X., {Jester}, S., {Stoughton}, C., {Strauss},
  M.~A., {SubbaRao}, M., {Brandt}, W.~N., {Gunn}, J.~E., {Yanny}, B.,
  {Bahcall}, N.~A., {Barentine}, J.~C., {Blanton}, M.~R., {Boroski}, W.~N.,
  {Brewington}, H.~J., {Brinkmann}, J., {Brunner}, R., {Csabai}, I., {Doi}, M.,
  {Eisenstein}, D.~J., {Frieman}, J.~A., {Fukugita}, M., {Gray}, J.,
  {Harvanek}, M., {Heckman}, T.~M., {Ivezi{\'c}}, {\v Z}., {Kent}, S.,
  {Kleinman}, S.~J., {Knapp}, G.~R., {Kron}, R.~G., {Krzesinski}, J., {Long},
  D.~C., {Loveday}, J., {Lupton}, R.~H., {Margon}, B., {Munn}, J.~A.,
  {Neilsen}, E.~H., {Newberg}, H.~J., {Newman}, P.~R., {Nichol}, R.~C.,
  {Nitta}, A., {Pier}, J.~R., {Rockosi}, C.~M., {Saxe}, D.~H., {Schlegel},
  D.~J., {Snedden}, S.~A., {Szalay}, A.~S., {Thakar}, A.~R., {Uomoto}, A.,
  {Voges}, W., \& {York}, D.~G. 2005, \aj, 130, 367

\bibitem[{{Shetrone} {et~al.}(2001b){Shetrone}, {C{\^o}t{\'e}}, \&
  {Sargent}}]{Shetrone:2001b}
{Shetrone}, M.~D., {C{\^o}t{\'e}}, P., \& {Sargent}, W.~L.~W. 2001b, \apj, 548,
  592

\bibitem[{{Shetrone} {et~al.}(2001a){Shetrone}, {C{\^o}t{\'e}}, \&
  {Stetson}}]{Shetrone:2001a}
{Shetrone}, M.~D., {C{\^o}t{\'e}}, P., \& {Stetson}, P.~B. 2001a, \pasp, 113,
  1122

\bibitem[{{Simon}(1979)}]{Simon:1979}
{Simon}, N.~R. 1979, \aap, 74, 30

\bibitem[{{Simon} \& {Teays}(1982)}]{Simon:1982}
{Simon}, N.~R., \& {Teays}, T.~J. 1982, \apj, 261, 586

\bibitem[{{Smith}(1995)}]{Smith:1995}
{Smith}, H.~A. 1995, {RR Lyrae stars} (Cambridge Astrophysics Series,
  Cambridge, New York: Cambridge University Press, |c1995)

\bibitem[{{Stellingwerf}(1978)}]{Stellingwerf:1978}
{Stellingwerf}, R.~F. 1978, \apj, 224, 953

\bibitem[{{Stetson}(1980)}]{Stetson:1980}
{Stetson}, P.~B. 1980, \aj, 85, 387

\bibitem[{{Stetson}(1987)}]{Stetson:1987}
---. 1987, \pasp, 99, 191

\bibitem[{{Stetson}(1994)}]{Stetson:1994}
---. 1994, \pasp, 106, 250

\bibitem[{{Stothers}(2006)}]{Stothers:2006}
{Stothers}, R.~B. 2006, \apj, 652, 643

\bibitem[{{Walker}(1994)}]{Walker:1994}
{Walker}, A.~R. 1994, \aj, 108, 555

\bibitem[{{Walker} \& {Nemec}(1996)}]{Walker:1996}
{Walker}, A.~R., \& {Nemec}, J.~M. 1996, \aj, 112, 2026

\bibitem[{{Wallerstein} \& {Cox}(1984)}]{Wallerstein:1984}
{Wallerstein}, G., \& {Cox}, A.~N. 1984, \pasp, 96, 677

\bibitem[{{Wehlau} {et~al.}(1986){Wehlau}, {Bohlender}, \&
  {Demers}}]{Wehlau:1986}
{Wehlau}, A., {Bohlender}, D., \& {Demers}, S. 1986, \pasp, 98, 872

\bibitem[{{Welch} \& {Stetson}(1993)}]{Welch:1993}
{Welch}, D.~L., \& {Stetson}, P.~B. 1993, \aj, 105, 1813

\bibitem[{{Zinn}(1978)}]{Zinn:1978}
{Zinn}, R. 1978, \apj, 225, 790

\bibitem[{{Zinn} \& {Dahn}(1976)}]{Zinn:1976b}
{Zinn}, R., \& {Dahn}, C.~C. 1976, \aj, 81, 527

\bibitem[{{Zinn} \& {Searle}(1976)}]{Zinn:1976a}
{Zinn}, R., \& {Searle}, L. 1976, \apj, 209, 734

\end{thebibliography}

\clearpage


\clearpage

\begin{deluxetable}{cc|c}
\tablecaption{Observations of Draco dwarf galaxy in Heliocentric Julian date
  (2449000.0 + days).}
\tablenum{1}
\tablewidth{0pc}
\tablehead{\colhead{Year} & \colhead{USNO} & \colhead{WIRO}}
\startdata
1993 & --  & 183 - 187 \\
1994 & 424 - 427 & -- \\
 & 453 - 456 & -- \\
 & 463 - 464 & -- \\
 & 482 & -- \\
 & 501 & -- \\
 & 512 & 511 - 513 \\
 & 541 - 544 & -- \\
 & 560 & -- \\
 & 597 - 600 & -- \\
 & 629 - 630 & -- \\
1995 & 804 - 805 & -- \\
 & 833 - 836 & -- \\
 & 892 - 893 & -- \\
 & 923 - 926 & -- \\
\enddata
\label{jdates}
\end{deluxetable}
\begin{deluxetable}{cccl}

\tablecaption{Stars in the Draco field not observed or found not to be
  variable.}

\tablenum{2}
\tablewidth{0pc}

\tablehead{\colhead{B\&S id} & \colhead{RA} & \colhead{DEC} & \colhead{Comments} \\ 
\colhead{} & \colhead{(2000.0)} & \colhead{(2000.0)} & \colhead{} }

\startdata
V333 & 17:17:24.0 & 58:02:32.0 & West of USNO fields \\
V5 & 17:20:20.4 & 58:00:55.0 & In the gap between Quad 1 \& 4 (P = 0.57006d) \\
V117 & 17:20:21.1 & 58:05:43.0 & In the gap between Quad 1 \& 4, not measured by B\&S 1961 \\
V10 & 17:20:23.9 & 57:58:44.0 & In the gap between Quad 1 \& 4,\\ 
& & & near bright star, not measured by B\&S 1961. \\
V205 & 17:23:19.6 & 57:57:55.0 & East of USNO fields \\
 \\
Non-variable stars \\
V168 & 17:19:27.24 & 58:00:35.0 & Outside central field.  Not measured by B\&S 1961 \\
V195 & 17:20:27.61 & 57:52:58.7 & Near bright star.  Not measured by B\&S 1961 \\
V111 & 17:20:28.60 & 57:52:58.9 & Near bright star.  Not measured by B\&S 1961 \\
\enddata

\label{omitstars}
\end{deluxetable}
\begin{deluxetable}{cccccc}
\tablecaption{Transformation coefficients for USNO dataset}
\tablenum{3}
\tablewidth{0pc}
\tablehead{\colhead{Filter} & \colhead{Obs. date} &  \colhead{$C_{0}$} &
  \colhead{$C_{1}$} & \colhead{$C_{2}$} & \colhead{$\sigma_{stds}$}  }
\startdata
USNO V & 08 Jul 1994 & 4.938 & 0.000 & 0.161 & 0.008 \\
       & 24 Jun 1995 & 4.885 & 0.010 & 0.170 & 0.015 \\
       & 25 Jun 1995 & 4.885 & 0.010 & 0.177 & 0.014 \\
       & 23 Jun 1998 & 3.720 & 0.014 & 0.126 & 0.016 \\
USNO I & 08 Jul 1994 & 5.480 & 0.029 & 0.062 & 0.008 \\
       & 24 Jun 1995 & 5.390 & 0.039 & 0.076 & 0.020 \\
       & 25 Jun 1995 & 5.390 & 0.039 & 0.079 & 0.019 \\
       & 23 Jun 1998 & 4.408 & 0.024 & 0.054 & 0.010 \\
\enddata
\label{usnotrans}
\end{deluxetable}

\begin{deluxetable}{c|cc}
\tablecaption{Transformation coefficients for WIRO dataset\tablenotemark{a}}
\tablenum{4}
\tablewidth{0pc}
\tablehead{\colhead{Field} & \colhead{$\alpha_{V}$} & \colhead{$\alpha_{I}$} }
\startdata
WIRO Field 1 & $-3.453 \pm 0.002$ & $-1.734 \pm 0.003$ \\
WIRO Field 2 & $-3.421 \pm 0.002$ & $-1.715 \pm 0.002$ \\
WIRO Field 3 & $-3.426 \pm 0.002$ & $-1.769 \pm 0.003$ \\
\enddata
\tablenotetext{a}{The $\beta$ coefficient is $0.081 \pm 0.01$ and $\gamma =
  -0.09 \pm 0.01$ in Equation \ref{vtran}.}
\label{trans}
\end{deluxetable}

\begin{deluxetable}{cccccccl}

\tablecaption{Properties of Draco RRL stars.}

\tablenum{5}
\tablewidth{0pc}

\tablehead{\colhead{ID} & \colhead{RA} & \colhead{DEC} & \colhead{Period} & \colhead{Amp} & \colhead{$\langle V \rangle$\tablenotemark{a}} & \colhead{$\langle I \rangle$\tablenotemark{a}} & \colhead{Type\tablenotemark{b}}   \\ 
\colhead{} & \colhead{(2000.0)} & \colhead{(2000.0)} & \colhead{(days)} & \colhead{} & \colhead{} & \colhead{} & \colhead{} } 

\startdata
 1 & 17:20:13.59 & 58:05:24.2 & 0.39363 & 0.48 & 20.18 & 19.60 & c,Bl? \\
 2 & 17:19:42.54 & 58:03:26.8 & 0.59259 & 0.80 & 19.96 & 19.41 & ab \\
 3 & 17:20:14.88 & 58:01:46.8 & 0.64876 & 0.78 & 20.00 & 19.39 & ab;B04 \\
 4 & 17:20:29.95 & 58:00:57.7 & 0.62625 & 0.61 & 20.18 & 19.63 & ab;B04 \\
 6 & 17:20:18.95 & 58:00:37.9 & 0.69485 & 0.75 & 19.94 & 19.34 & ab;B04 \\
 7 & 17:20:09.56 & 57:59:57.4 & 0.61896 & 0.62 & 20.19 & 19.57 & ab;B04 \\
 8 & 17:20:15.23 & 57:59:17.3 & 0.56957 & 0.92 & 20.10 & 19.55 & ab;B04 \\
 9 & 17:19:35.72 & 57:58:32.2 & 0.68418 & 0.50 & 20.02 & 19.39 & ab;B04 \\
 11 & 17:20:41.93 & 57:58:27.4 & 0.41100 & 0.48 & 20.08 & 19.52 & d;B04 \\
 12 & 17:20:41.86 & 57:57:50.0 & 0.57638 & 0.76 & 20.17 & 19.61 & ab;B04 \\
 13 & 17:20:47.19 & 57:57:59.5 & 0.53657 & 1.16 & 20.10 & 19.64 & ab;B04 \\
 14 & 17:19:30.51 & 57:56:33.8 & 0.61839 & 0.99 & 20.01 & 19.45 & ab;B04 \\
 15 & 17:19:23.29 & 57:55:55.9 & 0.57803 & 0.67 & 20.23 & 19.63 & ab,Bl?;B04 \\
 16 & 17:19:56.68 & 57:56:00.6 & 0.62489 & 0.60 & 20.13 & 19.52 & ab \\
 17 & 17:20:40.33 & 57:56:04.1 & 0.59852 & 0.86 & 20.15 & 19.66 & ab;B04 \\
 18 & 17:19:34.07 & 57:55:35.7 & 0.54963 & 1.10 & 20.19 & 19.73 & ab,Bl;B04 \\
 19 & 17:19:43.94 & 57:55:09.6 & 0.63183 & 0.69 & 20.14 & 19.53 & ab;B04 \\
 20 & 17:19:48.26 & 57:54:51.4 & 0.61997 & 0.86 & 20.16 & 19.49 & ab;B04 \\
 21 & 17:20:32.52 & 57:55:09.6 & 0.56061 & 0.99 & 20.30 & 19.88 & ab;B04 \\
 22 & 17:19:11.96 & 57:54:37.2 & 0.58047 & 0.73 & 20.16 & 19.65 & ab,Bl?;B04 \\
 23 & 17:19:22.90 & 57:54:11.8 & 0.61791 & 0.78 & 20.11 & 19.55 & ab,Bl?;B04 \\
 24 & 17:19:58.92 & 57:54:14.7 & 0.63211 & 0.62 & 20.19 & 19.62 & ab \\
 25 & 17:20:49.78 & 57:54:05.1 & 0.56174 & 0.86 & 20.25 & 19.74 & ab;B04 \\
 26 & 17:21:16.56 & 57:53:31.9 & 0.60392 & 0.69 & 20.14 & 19.47 & ab,Bl?;B04 \\
 27 & 17:20:06.01 & 57:53:49.0 & 0.76413 & 0.56 & 19.95 & 19.32 & ab;B04 \\
 28 & 17:19:24.53 & 57:53:36.4 & 0.62593 & 0.57 & 20.10 & 19.47 & ab;B04 \\
 29 & 17:20:16.66 & 57:53:12.1 & 0.56919 & 0.89 & 20.09 & 19.33 & ab;B04 \\
 30 & 17:20:53.28 & 57:53:14.2 & 0.62977 & 0.60 & 20.19 & 19.58 & ab;B04 \\
 32 & 17:19:56.27 & 57:52:47.8 & 0.55184 & 1.01 & 20.10 & 19.55 & ab,Bl? \\
 33 & 17:19:52.43 & 57:52:26.5 & 0.61087 & 0.69 & 20.11 & 19.49 & ab \\
 34 & 17:20:16.96 & 57:52:40.2 & 0.54511 & 0.78 & 20.24 & 19.70 & ab;B04 \\
 35 & 17:20:38.39 & 57:52:35.9 & 0.57488 & 0.91 & 20.17 & 19.59 & ab,Bl?;B04 \\
 36 & 17:20:17.63 & 57:52:00.7 & 0.62547 & 0.95 & 20.01 & 19.42 & ab;B04 \\
 37 & 17:20:08.44 & 57:52:03.3 & 0.55452 & 0.89 & 20.16 & 19.64 & ab;B04 \\
 38 & 17:19:47.67 & 57:51:41.6 & 0.62148 & 0.71 & 20.16 & 19.57 & ab,Bl? \\
 39 & 17:19:55.37 & 57:51:37.6 & 0.57422 & 0.94 & 20.17 & 19.62 & ab,Bl? \\
 40 & 17:20:36.85 & 57:52:12.9 & 0.61640 & 0.62 & 20.13 & 19.56 & ab;B04 \\
 41 & 17:20:42.87 & 57:51:28.9 & 0.57880 & 0.78 & 20.13 & 19.55 & ab,Bl?;B04 \\
 42 & 17:19:21.80 & 57:51:21.4 & 0.69306 & 0.63 & 20.01 & 19.34 & ab \\
 43 & 17:20:05.77 & 57:51:07.8 & 0.60368 & 0.69 & 20.10 & 19.54 & ab \\
 44 & 17:20:16.89 & 57:50:18.2 & 0.38436 & 0.48 & 20.12 & 19.63 & c \\
 45 & 17:19:34.99 & 57:50:47.8 & 0.58048 & 0.95 & 20.13 & 19.55 & ab \\
 46 & 17:19:36.43 & 57:49:30.0 & 0.33633 & 0.65 & 20.19 & 19.73 & c;B04 \\
 47 & 17:19:28.93 & 57:49:17.2 & 0.63242 & 0.68 & 20.16 & 19.52 & ab;B04 \\
 48 & 17:19:49.91 & 57:49:04.6 & 0.58167 & 0.68 & 20.04 & 19.55 & ab,Bl?;B04 \\
 49 & 17:20:08.55 & 57:47:28.5 & 0.62026 & 0.71 & 20.10 & 19.52 & ab;B04 \\
 50 & 17:19:38.37 & 57:47:24.7 & 0.37904 & 0.43 & 20.08 & 19.65 & c;B04 \\
 51 & 17:21:15.59 & 58:05:21.9 & 0.60658 & 0.77 & 20.10 & 19.53 & ab \\
 52 & 17:20:59.26 & 58:05:22.9 & 0.60143 & 0.73 & 20.08 & 19.47 & ab \\
 53 & 17:20:41.85 & 58:03:45.2 & 0.64255 & 0.68 & 20.01 & 19.42 & ab,Bl? \\
 54 & 17:19:03.26 & 58:03:07.7 & 0.63874 & 0.62 & 20.14 & 19.57 & ab \\
 55 & 17:19:59.50 & 58:03:34.4 & 0.60100 & 0.80 & 20.03 & 19.50 & ab \\
 56 & 17:21:14.16 & 58:03:17.2 & 0.59356 & 0.85 & 20.06 & 19.53 & ab \\
 57 & 17:20:08.17 & 58:02:31.7 & 0.60485 & 0.59 & 20.03 & 19.38 & ab \\
 58 & 17:20:06.14 & 58:02:06.6 & 0.60427 & 0.58 & 20.16 & 19.54 & ab;B04 \\
 59 & 17:20:02.26 & 58:01:20.6 & 0.58929 & 0.86 & 20.15 & 19.59 & ab \\
 60 & 17:19:24.78 & 57:58:47.4 & 0.60930 & 0.59 & 20.08 & 19.45 & ab;B04 \\
 61 & 17:21:30.12 & 57:58:41.1 & 0.38941 & 0.53 & 20.02 & 19.50 & c \\
 62 & 17:20:57.20 & 57:58:21.3 & 0.60347 & 0.66 & 20.14 & 19.55 & ab;B04 \\
 63 & 17:20:46.55 & 57:57:41.8 & 0.61063 & 0.59 & 20.12 & 19.52 & ab;B04 \\
 64 & 17:19:31.78 & 57:57:05.1 & 0.59874 & 0.58 & 20.15 & 19.52 & ab,Bl?;B04 \\
 65 & 17:19:59.22 & 57:56:46.5 & 0.58911 & 0.84 & 20.10 & 19.59 & ab \\
 66 & 17:19:05.65 & 57:55:20.2 & 0.64745 & 0.42 & 20.18 & 19.57 & ab;B04 \\
 67 & 17:21:31.31 & 57:55:04.7 & 0.58752 & 0.79 & 20.15 & 19.62 & ab,Bl \\
 68 & 17:20:09.24 & 57:54:38.5 & 0.61683 & 0.66 & 20.03 & 19.43 & ab,Bl?;B04 \\
 69 & 17:21:26.95 & 57:54:19.6 & 0.59496 & 0.76 & 20.11 & 19.52 & ab \\
 70 & 17:21:07.10 & 57:54:08.9 & 0.62508 & 1.01 & 20.02 & 19.45 & ab;B04 \\
 71 & 17:20:15.59 & 57:54:18.0 & 0.62146 & 1.06 & 20.32 & 19.67 & ab,Bl;B04 \\
 72 & 17:20:12.44 & 57:54:11.4 & 0.40715 & 0.99 & 20.12 & 19.64 & d;B04 \\
 73 & 17:19:17.57 & 57:53:32.7 & 0.58470 & 0.85 & 20.17 & 19.61 & ab;B04 \\
 74 & 17:20:00.32 & 57:53:26.4 & 0.59173 & 0.97 & 20.09 & 19.57 & ab \\
 75 & 17:20:56.66 & 57:53:52.5 & 0.60288 & 0.69 & 20.19 & 19.61 & ab,Bl?;B04 \\
 76 & 17:18:58.17 & 57:52:56.7 & 0.58336 & 0.95 & 20.03 & 19.52 & ab;B04 \\
 77 & 17:19:45.55 & 57:52:41.0 & 0.63960 & 0.58 & 20.23 & 19.65 & ab,Bl?;B04 \\
 78 & 17:19:54.55 & 57:52:56.1 & 0.59315 & 0.70 & 20.12 & 19.56 & ab,Bl? \\
 79 & 17:19:54.93 & 57:52:28.0 & 0.61144 & 0.69 & 20.18 & 19.56 & ab,Bl? \\
 80 & 17:21:02.69 & 57:52:50.6 & 0.60234 & 0.96 & 20.14 & 19.71 & ab;B04 \\
 81 & 17:20:21.07 & 57:52:19.0 & 0.73202 & 0.39 & 19.95 & 19.37 & ab;B04 \\
 82 & 17:20:41.68 & 57:49:52.0 & 0.59222 & 0.78 & 20.14 & 19.57 & ab \\
 83 & 17:20:48.80 & 57:50:01.8 & 0.40078 & 0.61 & 20.11 & 19.54 & d \\
 84 & 17:19:53.42 & 57:48:45.6 & 0.59195 & 1.02 & 19.97 & 19.49 & ab;B04 \\
 85 & 17:19:51.29 & 57:48:43.7 & 0.61164 & 0.72 & 20.08 & 19.50 & ab;B04 \\
 86 & 17:19:46.23 & 57:47:44.4 & 0.62896 & 0.64 & 20.13 & 19.57 & ab,Bl?;B04 \\
 87 & 17:21:08.69 & 57:47:46.6 & 0.61526 & 0.57 & 20.22 & 19.62 & ab;B04 \\
 88 & 17:20:10.69 & 57:45:59.0 & 0.60195 & 0.64 & 20.19 & 19.65 & ab;B04 \\
 89 & 17:20:09.22 & 57:45:38.1 & 0.60867 & 0.71 & 20.10 & 19.54 & ab,Bl?;B04 \\
 90 & 17:19:45.89 & 58:05:18.7 & 0.30851 & 0.64 & 20.14 & 19.80 & c \\
 92 & 17:19:39.52 & 58:02:46.8 & 0.56437 & 0.89 & 20.17 & 19.62 & ab,Bl? \\
 93 & 17:19:49.13 & 58:02:12.5 & 0.58668 & 0.90 & 20.10 & 19.58 & ab \\
 94 & 17:20:52.12 & 58:01:25.7 & 0.56032 & 1.02 & 20.10 & 19.54 & ab;B04 \\
 95 & 17:21:07.01 & 57:59:42.4 & 0.61719 & 0.86 & 20.01 & 19.43 & ab,Bl?;B04 \\
 96 & 17:21:03.54 & 57:59:50.7 & 0.58401 & 0.57 & 20.06 & 19.45 & ab,Bl?;B04 \\
 97 & 17:20:59.14 & 58:00:05.8 & 0.31477 & 0.64 & 20.01 & 19.61 & c;B04 \\
 98 & 17:20:07.12 & 57:59:49.4 & 0.62786 & 0.79 & 20.00 & 19.46 & ab;B04 \\
 100 & 17:19:29.46 & 57:58:25.7 & 0.74363 & 0.36 & 19.96 & 19.36 & ab;B04 \\
 101 & 17:19:39.07 & 57:58:04.0 & 0.61953 & 0.76 & 20.14 & 19.57 & ab;B04 \\
 102 & 17:19:19.50 & 57:57:38.9 & 0.58258 & 0.92 & 20.08 & 19.54 & ab;B04 \\
 103 & 17:20:28.25 & 57:57:00.8 & 0.60638 & 0.68 & 20.17 & 19.66 & ab;B04 \\
 104 & 17:19:45.84 & 57:56:26.1 & 0.59182 & 0.92 & 20.11 & 19.55 & ab;B04 \\
 105 & 17:21:31.28 & 57:55:10.1 & 0.61174 & 0.69 & 20.21 & 19.61 & ab,Bl? \\
 106 & 17:20:26.69 & 57:54:27.5 & 0.62048 & 0.59 & 20.19 & 19.65 & ab,Bl;B04 \\
 107 & 17:19:26.62 & 57:53:34.2 & 0.58120 & 0.81 & 20.10 & 19.55 & ab;B04 \\
 108 & 17:19:56.89 & 57:53:50.1 & 0.65976 & 0.44 & 20.16 & 19.52 & ab,Bl? \\
 109 & 17:18:58.72 & 57:52:57.1 & 0.59473 & 0.71 & 20.13 & 19.52 & ab;B04 \\
 110 & 17:19:18.40 & 57:52:45.6 & 0.36924 & 0.53 & 20.16 & 19.63 & c;B04 \\
 112 & 17:21:06.42 & 57:51:52.9 & 0.42845 & 0.51 & 20.06 & 19.48 & d;B04 \\
 113 & 17:20:03.11 & 57:49:51.9 & 0.36274 & 0.47 & 20.10 & 19.62 & c \\
 114 & 17:19:55.40 & 57:49:00.7 & 0.64636 & 0.58 & 19.88 & 19.29 & ab;B04 \\
 115 & 17:19:02.18 & 57:47:54.5 & 0.59372 & 0.71 & 20.14 & 19.62 & ab,Bl?;B04 \\
 116 & 17:19:27.08 & 57:46:54.1 & 0.57984 & 0.90 & 20.16 & 19.64 & ab;B04 \\
 118 & 17:20:59.32 & 58:01:26.4 & 0.58940 & 0.87 & 19.98 & 19.43 & ab,Bl;B04 \\
 119 & 17:20:25.90 & 58:00:02.2 & 0.66464 & 0.48 & 20.10 & 19.53 & ab;B04 \\
 120 & 17:18:58.80 & 57:58:05.6 & 0.40051 & 0.44 & 20.01 & 19.51 & c;B04 \\
 121 & 17:19:39.88 & 57:57:53.6 & 0.33647 & 0.63 & 20.02 & 19.55 & c;B04 \\
 122 & 17:19:57.18 & 57:58:19.8 & 0.63837 & 0.54 & 20.17 & 19.59 & ab \\
 123 & 17:20:11.53 & 57:58:02.6 & 0.58474 & 0.56 & 20.15 & 19.61 & ab,Bl;B04 \\
 124 & 17:20:33.07 & 57:57:30.7 & 0.55673 & 1.10 & 20.13 & 19.62 & ab;B04 \\
 125 & 17:20:53.49 & 57:57:04.6 & 0.68171 & 0.65 & 20.04 & 19.44 & ab;B04 \\
 126 & 17:20:08.90 & 57:56:22.9 & 0.59283 & 0.77 & 20.07 & 19.49 & ab;B04 \\
 127 & 17:19:44.97 & 57:54:18.1 & 0.66454 & 0.73 & 20.14 & 19.57 & ab;B04 \\
 128 & 17:21:14.05 & 57:54:35.5 & 0.63400 & 0.61 & 20.16 & 19.54 & ab;B04 \\
 129 & 17:19:41.43 & 57:53:27.6 & 0.59012 & 0.58 & 20.23 & 19.66 & ab;B04 \\
 130 & 17:21:19.05 & 57:53:24.5 & 0.57159 & 1.02 & 20.16 & 19.57 & ab,Bl;B04 \\
 131 & 17:21:19.47 & 57:52:35.7 & 0.40621 & 0.66 & 20.07 & 19.57 & d;B04 \\
 132 & 17:20:24.20 & 57:51:41.3 & 0.63331 & 0.52 & 20.10 & 19.55 & ab;B04 \\
 133 & 17:20:51.20 & 57:51:47.8 & 0.61001 & 0.58 & 20.08 & 19.49 & ab;B04 \\
 135 & 17:20:30.16 & 57:50:40.1 & 0.63139 & 0.78 & 20.07 & 19.46 & ab \\
 136 & 17:20:44.12 & 57:50:27.0 & 0.55487 & 1.19 & 20.14 & 19.56 & ab \\
 137 & 17:19:23.01 & 57:49:58.5 & 0.60245 & 0.83 & 20.19 & 19.59 & ab;B04 \\
 138 & 17:20:35.88 & 57:49:18.7 & 0.40773 & 0.46 & 19.91 & 19.30 & d \\
 139 & 17:21:21.90 & 57:49:26.8 & 0.33841 & 0.64 & 20.09 & 19.70 & c \\
 140 & 17:20:17.09 & 57:46:41.0 & 0.62578 & 0.51 & 20.08 & 19.53 & ab;B04 \\
 142 & 17:19:13.30 & 58:04:54.5 & 0.63813 & 0.91 & 20.05 & 19.51 & ab \\
 143 & 17:19:31.77 & 57:59:27.1 & 0.40324 & 0.43 & 20.02 & 19.14 & d;B04 \\
 144 & 17:19:52.18 & 57:59:09.4 & 0.58887 & 0.82 & 20.16 & 19.59 & ab,Bl?;B04 \\
 145 & 17:20:57.87 & 57:58:48.5 & 0.39738 & 0.52 & 20.09 & 19.61 & c;B04 \\
 146 & 17:21:26.73 & 57:57:25.9 & 0.58186 & 0.79 & 19.95 & 19.38 & ab,Bl \\
 147 & 17:19:48.75 & 57:56:57.0 & 0.58732 & 0.66 & 20.09 & 19.47 & ab,Bl;B04 \\
 148 & 17:19:56.80 & 57:54:59.0 & 0.67413 & 0.40 & 20.18 & 19.60 & ab,Bl? \\
 149 & 17:19:22.46 & 57:54:04.3 & 0.67536 & 0.31 & 20.21 & 19.61 & ab;B04 \\
 150 & 17:20:31.38 & 57:53:02.4 & 0.67633 & 0.43 & 20.05 & 19.31 & ab;B04 \\
 151 & 17:20:53.13 & 57:53:03.7 & 0.62067 & 0.56 & 20.14 & 19.51 & ab;B04 \\
 152 & 17:20:02.61 & 57:51:30.6 & 0.62690 & 0.63 & 20.13 & 19.56 & ab,Bl? \\
 153 & 17:20:17.46 & 57:46:01.8 & 0.40215 & 0.40 & 20.03 & 19.51 & c;B04 \\
 154 & 17:20:50.94 & 57:45:17.0 & 0.63200 & 0.72 & 20.05 & 19.46 & ab;B04 \\
 155 & 17:20:03.72 & 58:05:21.6 & 0.41989 & 0.41 & 20.02 & 19.54 & d \\
 156 & 17:19:55.09 & 58:01:09.9 & 0.40871 & 0.62 & 19.96 & 19.54 & d \\
 158 & 17:20:31.15 & 57:57:37.2 & 0.65465 & 0.55 & 20.10 & 19.53 & ab,Bl;B04 \\
 159 & 17:19:05.65 & 57:55:38.8 & 0.65295 & 0.63 & 20.06 & 19.46 & ab;B04 \\
 160 & 17:20:08.92 & 57:55:29.3 & 0.64320 & 0.59 & 20.16 & 19.59 & ab,Bl;B04 \\
 161 & 17:20:40.59 & 57:54:52.1 & 0.62158 & 0.65 & 20.20 & 19.63 & ab;B04 \\
 162 & 17:20:37.99 & 57:55:31.1 & 0.62171 & 0.68 & 20.11 & 19.37 & ab;B04 \\
 163 & 17:20:58.94 & 57:53:44.2 & 0.56060 & 0.96 & 20.23 & 19.71 & ab,Bl;B04 \\
 164 & 17:20:45.14 & 57:51:27.7 & 0.62464 & 0.62 & 20.10 & 19.50 & ab,Bl?;B04 \\
 165 & 17:20:08.64 & 57:50:07.1 & 0.35802 & 0.55 & 20.21 & 19.64 & d \\
 166 & 17:21:20.23 & 57:49:18.2 & 0.36339 & 0.44 & 20.10 & 19.62 & c \\
 167 & 17:20:07.68 & 58:01:40.2 & 0.66756 & 0.61 & 20.10 & 19.51 & ab;B04 \\
 169 & 17:20:42.33 & 57:58:52.4 & 0.40317 & 0.69 & 20.12 & 19.60 & d;B04 \\
 170 & 17:20:52.01 & 57:55:32.0 & 0.40370 & 0.42 & 20.21 & 19.84 & c;B04 \\
 171 & 17:20:15.40 & 57:53:28.1 & 0.59963 & 0.70 & 20.10 & 19.57 & ab,Bl?;B04 \\
 172 & 17:20:36.84 & 57:48:20.6 & 0.66282 & 0.66 & 19.96 & 19.37 & ab;B04 \\
 173 & 17:20:59.52 & 57:55:42.3 & 0.36946 & 0.57 & 20.04 & 19.31 & c;B04 \\
 174 & 17:19:42.91 & 57:55:27.1 & 0.67612 & 0.53 & 20.00 & 19.45 & ab;B04 \\
 175 & 17:20:58.47 & 57:53:32.0 & 0.56243 & 0.99 & 20.11 & 19.61 & ab;B04 \\
 176 & 17:20:48.99 & 57:51:03.2 & 0.60211 & 0.58 & 19.68 & 18.96 & ab, blended? \\
 177 & 17:20:52.79 & 57:50:41.8 & 0.59242 & 0.98 & 20.11 & 19.49 & ab \\
 178 & 17:20:57.24 & 57:50:01.2 & 0.59386 & 0.92 & 20.09 & 19.49 & ab \\
 179 & 17:21:10.77 & 57:47:36.2 & 0.39293 & 0.51 & 20.04 & 19.57 & c;B04 \\
 180 & 17:20:46.73 & 58:03:03.2 & 0.65982 & 0.45 & 20.04 & 19.38 & ab \\
 181 & 17:21:12.74 & 58:01:31.0 & 0.38572 & 0.55 & 20.10 & 19.66 & c;B04 \\
 182 & 17:19:28.21 & 58:00:43.5 & 0.41022 & 0.48 & 20.04 & 19.57 & c;B04 \\
 183 & 17:21:07.35 & 57:58:00.6 & 0.59506 & 0.91 & 20.11 & 19.57 & ab;B04 \\
 184 & 17:20:13.73 & 57:57:24.9 & 0.59430 & 0.69 & 20.08 & 19.54 & ab,Bl? \\
 185 & 17:20:47.28 & 57:55:23.0 & 0.59385 & 0.60 & 20.25 & 19.62 & ab,Bl?;B04 \\
 186 & 17:20:06.57 & 57:49:45.0 & 0.59717 & 0.73 & 20.24 & 19.66 & ab,Bl? \\
 187 & 17:19:26.12 & 57:48:51.3 & 0.68939 & 0.40 & 20.10 & 19.49 & ab;B04 \\
 188 & 17:19:42.62 & 57:53:30.0 & 0.67368 & 0.54 & 20.10 & 19.47 & ab;B04 \\
 189 & 17:21:13.02 & 57:53:50.8 & 0.59440 & 0.75 & 20.14 & 19.53 & ab,Bl?;B04 \\
 190 & 17:20:42.51 & 57:51:53.1 & 0.39652 & 0.52 & 20.07 & 19.54 & d;B04 \\
 191 & 17:19:17.45 & 57:48:42.9 & 0.39729 & 0.45 & 20.04 & 19.57 & c;B04 \\
 192 & 17:20:13.14 & 57:55:26.4 & 0.66098 & 0.55 & 20.22 & 19.63 & ab,Bl;B04 \\
 193 & 17:20:21.58 & 57:54:31.0 & 0.67818 & 0.29 & 20.11 & 19.53 & ab,Bl;B04 \\
 196 & 17:19:51.33 & 57:53:20.9 & 0.58936 & 1.01 & 20.16 & 19.65 & ab;B04 \\
 197 & 17:21:25.64 & 57:50:30.4 & 0.59265 & 0.60 & 20.12 & 19.51 & ab,Bl \\
 198 & 17:20:51.82 & 57:56:35.9 & 0.67956 & 0.37 & 20.07 & 19.46 & ab;B04 \\
 199 & 17:21:18.63 & 57:47:43.8 & 0.66690 & 0.65 & 20.09 & 19.52 & ab \\
 200 & 17:21:17.02 & 58:05:11.9 & 0.41700 & 0.32 & 20.03 & 19.52 & c \\
 201 & 17:20:36.57 & 58:03:57.7 & 0.65900 & 0.45 & 20.11 & 19.50 & ab \\
 202 & 17:21:21.04 & 57:50:44.8 & 0.64213 & 0.37 & 20.17 & 19.50 & ab,Bl? \\
 207 & 17:23:05.77 & 57:59:13.4 & 0.56817 & 0.87 & 20.11 & 19.56 & ab \\
 213 & 17:22:41.11 & 57:58:03.8 & 0.62290 & 0.66 & 20.15 & 19.54 & ab \\
 216 & 17:22:30.47 & 57:42:24.1 & 0.59338 & 0.79 & 20.11 & 19.55 & ab \\
 217 & 17:22:22.09 & 57:58:02.7 & 0.41166 & 0.55 & 20.06 & 19.50 & d \\
 218 & 17:22:20.80 & 57:55:54.4 & 0.60842 & 0.75 & 20.11 & 19.53 & ab \\
 219 & 17:22:19.42 & 57:53:10.9 & 0.60534 & 0.63 & 20.21 & 19.62 & ab \\
 220 & 17:22:19.44 & 57:51:43.8 & 0.62203 & 0.72 & 20.12 & 19.52 & ab \\
 221 & 17:22:16.13 & 57:50:03.7 & 0.40788 & 0.57 & 20.13 & 19.54 & d \\
 223 & 17:22:09.97 & 57:46:40.5 & 0.60209 & 0.69 & 20.19 & 19.60 & ab \\
 225 & 17:22:03.62 & 57:43:59.2 & 0.57588 & 0.85 & 20.13 & 19.65 & ab,Bl? \\
 226 & 17:22:00.30 & 57:58:01.1 & 0.29750 & 0.43 & 20.05 & 19.66 & c \\
 227 & 17:22:00.12 & 57:58:23.4 & 0.60069 & 0.60 & 20.13 & 19.55 & ab \\
 228 & 17:21:48.28 & 58:11:29.1 & 0.41605 & 0.47 & 20.10 & 19.55 & d \\
 232 & 17:21:47.30 & 57:53:06.5 & 0.41082 & 0.53 & 20.05 & 19.47 & d \\
 233 & 17:21:46.39 & 58:07:32.8 & 0.59427 & 0.50 & 20.16 & 19.64 & ab,Bl \\
 234 & 17:21:45.75 & 57:51:49.5 & 0.65512 & 0.42 & 20.13 & 19.49 & ab,Bl? \\
 235 & 17:21:45.39 & 57:41:45.7 & 0.39954 & 0.64 & 20.15 & 19.71 & d \\
 236 & 17:21:42.85 & 57:37:02.0 & 0.40420 & 0.51 & 20.04 & 19.56 & c \\
 237 & 17:21:41.89 & 57:54:29.7 & 0.61271 & 0.76 & 20.20 & 19.67 & ab \\
 238 & 17:21:41.81 & 57:38:11.9 & 0.56264 & 0.84 & 20.07 & 19.51 & ab \\
 239 & 17:21:41.23 & 57:55:43.5 & 0.59004 & 0.93 & 20.14 & 19.59 & ab,Bl? \\
 242 & 17:21:40.90 & 57:59:39.4 & 0.35332 & 0.49 & 20.11 & 19.70 & c \\
 243 & 17:21:41.15 & 57:54:23.9 & 0.72130 & 0.39 & 20.00 & 19.37 & ab \\
 244 & 17:21:40.66 & 57:54:36.2 & 0.56161 & 0.67 & 20.17 & 19.66 & ab \\
 245 & 17:21:40.34 & 57:51:32.2 & 0.41106 & 0.48 & 20.04 & 19.45 & d \\
 246 & 17:21:38.70 & 57:52:42.0 & 0.63086 & 0.50 & 20.19 & 19.56 & ab,Bl? \\
 247 & 17:21:36.46 & 57:55:58.8 & 0.41762 & 0.49 & 20.09 & 19.55 & d \\
 248 & 17:21:35.07 & 57:53:04.9 & 0.41828 & 0.48 & 20.06 & 19.51 & d \\
 249 & 17:21:30.07 & 58:07:48.5 & 0.60027 & 0.92 & 20.09 & 19.60 & ab \\
 250 & 17:21:30.37 & 57:48:52.8 & 0.40488 & 0.48 & 20.09 & 19.59 & d \\
 252 & 17:21:29.08 & 58:03:16.1 & 0.58232 & 0.90 & 20.10 & 19.54 & ab \\
 253 & 17:21:26.82 & 58:06:58.0 & 0.59247 & 0.80 & 20.21 & 19.68 & ab,Bl \\
 258 & 17:21:09.54 & 58:09:14.7 & 0.58614 & 0.94 & 20.03 & 19.47 & ab \\
 260 & 17:21:00.07 & 58:06:20.3 & 0.55743 & 0.94 & 20.08 & 19.56 & ab \\
 261 & 17:20:58.51 & 58:09:18.0 & 0.55670 & 0.79 & 20.10 & 19.55 & ab \\
 262 & 17:20:55.08 & 57:43:35.2 & 0.61697 & 0.74 & 20.09 & 19.53 & ab \\
 265 & 17:20:51.99 & 58:15:02.1 & 0.58396 & 0.96 & 20.05 & 19.60 & ab,Bl \\
 267 & 17:20:46.52 & 57:48:18.9 & 0.67495 & 0.42 & 20.04 & 19.41 & ab,Bl \\
 269 & 17:20:43.61 & 58:08:30.9 & 0.54478 & 1.11 & 20.05 & 19.54 & ab \\
 270 & 17:20:42.46 & 57:39:55.8 & 0.56602 & 1.04 & 19.95 & 19.41 & ab \\
 273 & 17:20:38.99 & 57:57:32.4 & 0.79061 & 0.23 & 19.97 & 19.36 & ab \\
 275 & 17:20:29.33 & 57:58:07.6 & 0.65198 & 0.32 & 20.19 & 19.64 & ab \\
 276 & 17:20:19.11 & 58:16:21.8 & 0.63478 & 0.43 & 17.57 & 16.88 & ab;field \\
 277 & 17:20:14.32 & 57:44:02.0 & 0.64281 & 0.81 & 20.02 & 19.49 & ab,Bl \\
 278 & 17:20:08.35 & 57:35:01.8 & 0.61812 & 0.83 & 20.10 & 19.60 & ab \\
 279 & 17:20:00.79 & 57:44:11.7 & 0.60644 & 0.72 & 19.99 & 19.43 & ab \\
 281 & 17:20:00.66 & 57:42:20.3 & 0.69818 & 0.58 & 19.97 & 19.39 & ab \\
 284 & 17:19:57.88 & 57:41:57.3 & 0.60760 & 0.68 & 20.02 & 19.45 & ab \\
 285 & 17:19:44.74 & 57:57:37.3 & 0.65136 & 0.57 & 19.84 & 19.27 & ab;B04 \\
 286 & 17:19:43.55 & 58:06:02.9 & 0.60153 & 0.84 & 20.11 & 19.55 & ab,Bl \\
 289 & 17:19:29.26 & 57:41:59.5 & 0.39742 & 0.58 & 20.13 & 19.64 & d \\
 290 & 17:19:23.22 & 58:08:20.3 & 0.70680 & 0.38 & 20.11 & 19.47 & ab \\
 291 & 17:19:21.05 & 57:36:41.4 & 0.72744 & 0.43 & 19.92 & 19.34 & ab,Bl? \\
 294 & 17:19:08.98 & 57:34:06.7 & 0.40548 & 0.60 & 20.11 & 19.69 & d,Bl? \\
 295 & 17:19:07.75 & 57:44:32.7 & 0.42625 & 0.57 & 20.04 & 19.61 & c \\
 297 & 17:19:04.96 & 58:03:30.3 & 0.66699 & 0.51 & 20.12 & 19.48 & ab,Bl? \\
 298 & 17:19:04.47 & 58:06:17.1 & 0.63785 & 0.85 & 20.15 & 19.68 & ab,Bl? \\
 301 & 17:19:00.43 & 57:37:29.9 & 0.41287 & 0.49 & 20.04 & 19.61 & d \\
 303 & 17:18:51.86 & 57:47:28.1 & 0.62029 & 0.78 & 20.19 & 19.68 & ab \\
 304 & 17:18:49.62 & 57:53:56.1 & 0.66365 & 0.58 & 20.08 & 19.51 & ab;B04 \\
 305 & 17:18:47.75 & 58:03:31.9 & 0.64080 & 0.31 & 20.12 & 19.46 & ab \\
 306 & 17:18:47.01 & 58:14:08.9 & 0.39823 & 0.61 & 20.04 & 19.54 & d \\
 307 & 17:18:45.25 & 57:52:22.6 & 0.39504 & 0.48 & 20.02 & 19.57 & c;B04 \\
 308 & 17:18:38.43 & 57:52:38.5 & 0.60803 & 0.51 & 20.03 & 19.49 & ab;B04 \\
 309 & 17:18:35.08 & 57:56:54.8 & 0.61993 & 0.63 & 20.00 & 19.24 & ab;B04 \\
 310 & 17:18:33.34 & 57:51:59.9 & 0.64819 & 0.67 & 20.03 & 19.45 & ab \\
 313 & 17:18:29.82 & 58:11:52.7 & 0.57270 & 0.94 & 20.06 & 19.52 & ab \\
 314 & 17:18:27.57 & 57:52:13.3 & 0.43096 & 0.53 & 19.93 & 19.44 & c \\
 315 & 17:18:24.82 & 57:59:48.3 & 0.57953 & 0.85 & 20.04 & 19.53 & ab \\
 316 & 17:18:24.92 & 57:43:31.7 & 0.32679 & 0.63 & 20.06 & 19.74 & c \\
 317 & 17:18:19.34 & 57:58:42.7 & 0.58685 & 0.83 & 20.02 & 19.48 & ab,Bl \\
 318 & 17:18:19.52 & 57:55:49.9 & 0.40829 & 0.62 & 20.16 & 19.61 & d \\
 319 & 17:18:19.52 & 57:46:21.1 & 0.62276 & 0.91 & 19.99 & 19.45 & ab,Bl \\
 321 & 17:18:12.87 & 58:02:56.0 & 0.59716 & 0.61 & 17.93 & 17.37 & ab;field \\
 323 & 17:18:09.26 & 57:37:05.7 & 0.55540 & 1.04 & 20.00 & 19.53 & ab \\
 324 & 17:18:04.66 & 58:02:35.7 & 0.60035 & 0.71 & 20.02 & 19.47 & ab \\
 325 & 17:18:02.69 & 57:48:14.2 & 0.62424 & 0.68 & 19.99 & 19.43 & ab \\
 326 & 17:17:59.11 & 58:02:05.6 & 0.62179 & 0.68 & 20.02 & 19.45 & ab \\
 327 & 17:17:55.93 & 57:40:00.0 & 0.61859 & 0.94 & 17.55 & 17.05 & ab;field \\
 330 & 17:17:46.91 & 57:43:27.8 & 0.35233 & 0.67 & 19.99 & 19.54 & c \\
 332 & 17:17:29.12 & 57:34:25.6 & 0.61233 & 0.70 & 20.15 & 19.61 & ab \\
\enddata
\tablenotetext{a}{Intensity-weighted magnitudes}
\tablenotetext{b}{Bl = Blazhko effect}
\tablerefs{B04 = \citet{Bonanos:2004}}
\label{properties}

\end{deluxetable}

\begin{deluxetable}{ccccccc}
\tablecaption{Properties of the Draco RRd stars.}
\tablenum{6}
\tablewidth{0pc}
\tablehead{\colhead{ID} & \colhead{$P_1$} & \colhead{Error} & \colhead{$P_0$} & \colhead{Error} & \colhead{$P_1/P_0$} &\colhead{Error\tablenotemark{a}}}

\startdata 
11 & 0.41100 & .00002	& 0.55114 & .00010 & 0.7457 & .0002\\
72 & 0.40711 & .00002	& 0.54599 & .00006 & 0.7456 & .0001\\
83 & 0.40075 & .00002 & 0.53720 & .00006 & 0.7460 & .0001\\
112 & 0.42844 & .00002 & 0.57446 & .00005 & 0.7458 & .0001\\
131 & 0.40626 & .00002	& 0.54424 & .00006 & 0.7465 & .0001\\
138 & 0.40773 & .00003	& 0.54601 & .00012 & 0.7467 & .0002\\
143 & 0.40317 & .00004	& 0.54042 & .0003  & 0.7460 & .0006*\\
155 & 0.41393 & .00003	& 0.55476 & .00009 & 0.7461 & .0002*\\
156 & 0.40868 & .00002	& 0.54778 & .00006 & 0.7461 & .0001\\
165 & 0.35798 & .00002 & 0.48064 & .00004 & 0.7448 & .0001\\
169 & 0.40316 & .00003	& 0.54059 & .00008 & 0.7458 & .0002\\
190 & 0.39652 & .00001	& 0.53080 & .00006 & 0.7470 & .0001*\\
217 & 0.41166 & .00003	& 0.55149 & .00014 & 0.7465 & .0002*\\
221 & 0.40788 & .00003	& 0.54671 & .00008 & 0.7461 & .0002\\
228 & 0.41606 & .00003	& 0.55784 & .00010 & 0.7458 & .0002\\
232 & 0.41081 & .00001	& 0.55017 & .00007 & 0.7467 & .0001*\\
235 & 0.39954 & .00003	& 0.53560 & .00011 & 0.7460 & .0002\\
245 & 0.41105 & .00001	& 0.55029 & .00010 & 0.7470 & .0002*\\
247 & 0.41760 & .00002	& 0.55946 & .00006 & 0.7464 & .0001\\
248 & 0.41828 & .00002	& 0.56055 & .00005 & 0.7462 & .0001\\
250 & 0.40491 & .00002	& 0.54218 & .00014 & 0.7468 & .0003*\\
289 & 0.39743 & .00002	& 0.53258 & .00008 & 0.7462 & .0002\\
294 & 0.39998 & .00002	& 0.53622 & .00007 & 0.7459 & .0001*\\
301 & 0.41286 & .00002	& 0.55306 & .00007 & 0.7465 & .0002\\
306 & 0.39824 & .00002	& 0.53323 & .00009 & 0.7468 & .0002\\
318 & 0.40264 & .00004	& 0.53995 & .00007 & 0.7457 & .0002*\\
\enddata
\tablenotetext{a}{Stars with an * denote some uncertainty with the
  period solutions due to aliasing.}
\label{rrd}
\end{deluxetable}


\begin{deluxetable}{cccccccccccl}
\tablecaption{Fourier decomposition parameters for RRab stars.}
\tablenum{7}
\tablewidth{0pc}
\tablehead{\colhead{ID} & \colhead{$A_{0}$} & \colhead{$R_{21}$} &
  \colhead{$R_{31}$} & \colhead{$R_{41}$} & \colhead{$\phi_{21}$} &
  \colhead{$\phi_{31}$} & \colhead{$\phi_{41}$} &
  \colhead{$\sigma_{\phi_{31}}$} & \colhead{[Fe/H]} &
  \colhead{$\sigma_{[Fe/H]}$} & \colhead{$D_{M}$ pass?\tablenotemark{a}} \\ 
\colhead{} & \colhead{} & \colhead{} & \colhead{} & \colhead{} & \colhead{} & \colhead{} & \colhead{} & \colhead{} & \colhead{} & \colhead{} & \colhead{}} 

\startdata
2 & 20.9797 & 0.3221 & 0.0858 & 0.1693 & 3.5756 & 1.403 & 5.449 & 0.553 & -2.43 & 0.75 & \\
3 & 21.0260 & 0.3453 & 0.2780 & 0.2168 & 3.9108 & 1.569 & 6.036 & 0.180 & -2.51 & 0.25 & * \\
4 & 21.1916 & 0.4781 & 0.1509 & 0.0581 & 3.8671 & 1.302 & 1.641 & 0.356 & -2.76 & 0.48 & \\
6 & 20.9611 & 0.5028 & 0.2893 & 0.1936 & 3.8149 & 2.042 & 6.092 & 0.134 & -2.11 & 0.18 & \\
7 & 21.2041 & 0.4867 & 0.3420 & 0.0415 & 3.7750 & 1.915 & 5.866 & 0.179 & -1.86 & 0.24 & \\
8 & 21.1371 & 0.5084 & 0.3504 & 0.2033 & 3.8696 & 1.671 & 6.105 & 0.128 & -1.92 & 0.17 & * \\
9 & 21.0360 & 0.4902 & 0.2396 & 0.1688 & 3.7757 & 1.799 & 0.785 & 0.222 & -2.39 & 0.30 & \\
12 & 21.2055 & 0.4099 & 0.2215 & 0.1029 & 3.8709 & 1.743 & 5.930 & 0.238 & -1.86 & 0.32 & * \\
13 & 21.1566 & 0.4181 & 0.2999 & 0.1742 & 3.8595 & 1.408 & 5.677 & 0.108 & -2.10 & 0.15 & * \\
14 & 21.0536 & 0.3410 & 0.3305 & 0.3288 & 3.8314 & 1.857 & 5.995 & 0.173 & -1.93 & 0.23 & * \\
15 & 21.2359 & 0.4504 & 0.2656 & 0.0374 & 3.5087 & 1.533 & 5.400 & 0.278 & -2.16 & 0.38 & * \\
16 & 21.1378 & 0.3367 & 0.1520 & 0.0652 & 4.0760 & 1.823 & 4.995 & 0.453 & -2.02 & 0.61 & \\
17 & 21.1852 & 0.3700 & 0.2613 & 0.0471 & 3.5913 & 1.052 & 6.038 & 0.272 & -2.96 & 0.37 & \\
19 & 21.1497 & 0.4018 & 0.2557 & 0.1974 & 4.1254 & 2.255 & 0.236 & 0.186 & -1.45 & 0.25 & \\
20 & 21.1713 & 0.4051 & 0.3398 & 0.2134 & 3.8324 & 1.714 & 0.185 & 0.159 & -2.15 & 0.22 & \\
21 & 21.3392 & 0.3626 & 0.2600 & 0.2263 & 3.7277 & 1.337 & 5.622 & 0.199 & -2.34 & 0.27 & * \\
22 & 21.1894 & 0.3796 & 0.3104 & 0.1628 & 4.1072 & 1.950 & 0.004 & 0.286 & -1.59 & 0.39 & \\
23 & 21.1370 & 0.5135 & 0.2314 & 0.2189 & 3.7667 & 1.487 & 5.489 & 0.425 & -2.45 & 0.57 & \\
24 & 21.1933 & 0.3830 & 0.2605 & 0.1930 & 4.0358 & 1.462 & 0.398 & 0.177 & -2.57 & 0.24 & \\
25 & 21.2862 & 0.3835 & 0.3411 & 0.3119 & 4.0705 & 1.710 & 5.930 & 0.136 & -1.82 & 0.18 & * \\
27 & 20.9620 & 0.4201 & 0.3533 & 0.0944 & 3.9173 & 2.328 & 0.985 & 0.156 & -2.09 & 0.21 & \\
28 & 21.1103 & 0.4367 & 0.3305 & 0.1880 & 3.9909 & 1.712 & 6.234 & 0.156 & -2.18 & 0.21 & * \\
29 & 21.0918 & 0.5195 & 0.6465 & 0.1672 & 3.4223 & 1.529 & 5.983 & 0.225 & -2.12 & 0.30 & \\
30 & 21.2023 & 0.3684 & 0.2356 & 0.1205 & 3.6366 & 2.060 & 0.933 & 0.183 & -1.71 & 0.25 & \\
32 & 21.0954 & 0.3225 & 0.3235 & 0.3422 & 3.9962 & 2.185 & 5.995 & 0.270 & -1.09 & 0.36 & \\
33 & 21.1307 & 0.3173 & 0.3863 & 0.1106 & 3.7809 & 1.359 & 6.199 & 0.155 & -2.59 & 0.21 & \\
34 & 21.2633 & 0.4045 & 0.2217 & 0.2377 & 4.2836 & 2.324 & 0.540 & 0.206 & -0.86 & 0.28 & \\
36 & 21.0509 & 0.3891 & 0.3177 & 0.2871 & 3.9562 & 1.524 & 5.890 & 0.100 & -2.44 & 0.14 & * \\
37 & 21.2053 & 0.3731 & 0.2829 & 0.1995 & 3.6804 & 1.418 & 5.892 & 0.155 & -2.19 & 0.21 & * \\
40 & 21.1469 & 0.3599 & 0.2782 & 0.1971 & 3.8310 & 1.481 & 6.267 & 0.176 & -2.45 & 0.24 & \\
42 & 21.0392 & 0.3932 & 0.3089 & 0.1697 & 3.6708 & 1.710 & 6.045 & 0.150 & -2.56 & 0.20 & * \\
43 & 21.1188 & 0.3645 & 0.3324 & 0.2243 & 3.8744 & 1.791 & 6.075 & 0.151 & -1.94 & 0.20 & * \\
45 & 21.1690 & 0.4129 & 0.3552 & 0.2593 & 3.8733 & 1.391 & 5.584 & 0.133 & -2.38 & 0.18 & * \\
47 & 21.1743 & 0.3426 & 0.2996 & 0.1977 & 3.6999 & 1.895 & 6.150 & 0.178 & -1.96 & 0.24 & * \\
49 & 21.1237 & 0.3706 & 0.2392 & 0.1690 & 3.6302 & 1.731 & 5.878 & 0.196 & -2.12 & 0.26 & * \\
51 & 21.1322 & 0.3698 & 0.2734 & 0.1621 & 3.9393 & 1.626 & 5.700 & 0.161 & -2.19 & 0.22 & * \\
52 & 21.1344 & 0.5317 & 0.3492 & 0.0578 & 4.3725 & 2.834 & 1.716 & 0.122 & -0.46 & 0.17 & \\
53 & 21.0354 & 0.5495 & 0.3102 & 0.1556 & 4.0508 & 2.151 & 0.046 & 0.237 & -1.65 & 0.32 & \\
54 & 21.1544 & 0.4462 & 0.2972 & 0.2325 & 3.7941 & 2.109 & 0.717 & 0.221 & -1.69 & 0.30 & \\
55 & 21.0562 & 0.3807 & 0.2762 & 0.2315 & 3.9886 & 1.590 & 5.568 & 0.178 & -2.21 & 0.24 & * \\
56 & 21.1826 & 0.7219 & 0.7131 & 0.6050 & 4.0904 & 2.675 & 1.197 & 0.117 & -0.64 & 0.16 & \\
57 & 21.0546 & 0.4037 & 0.2996 & 0.1998 & 3.7945 & 1.673 & 6.110 & 0.203 & -2.12 & 0.28 & * \\
58 & 21.1601 & 0.4425 & 0.3006 & 0.1511 & 3.8126 & 2.012 & 6.026 & 0.299 & -1.63 & 0.40 & * \\
59 & 21.1954 & 0.3742 & 0.3120 & 0.1503 & 3.8208 & 1.539 & 5.581 & 0.111 & -2.22 & 0.15 & * \\
60 & 21.0932 & 0.3856 & 0.3667 & 0.0409 & 4.5573 & 2.133 & 1.630 & 0.292 & -1.49 & 0.39 & \\
62 & 21.1604 & 0.2769 & 0.2547 & 0.1082 & 3.6509 & 1.119 & 0.357 & 0.249 & -2.89 & 0.34 & \\
63 & 21.1308 & 0.4094 & 0.3034 & 0.1444 & 3.8708 & 1.520 & 0.566 & 0.239 & -2.37 & 0.32 & \\
64 & 21.1639 & 0.4044 & 0.2060 & 0.1087 & 3.5979 & 1.046 & 4.927 & 0.336 & -2.97 & 0.46 & \\
65 & 21.1341 & 0.4056 & 0.3401 & 0.1892 & 3.7348 & 1.263 & 5.416 & 0.196 & -2.61 & 0.27 & * \\
66 & 21.1917 & 0.4033 & 0.2293 & 0.1271 & 3.8670 & 1.352 & 1.749 & 0.227 & -2.81 & 0.31 & \\
69 & 21.1418 & 0.4391 & 0.3548 & 0.2247 & 3.8962 & 1.606 & 5.866 & 0.129 & -2.15 & 0.18 & * \\
70 & 21.0688 & 0.4295 & 0.3121 & 0.2121 & 3.6864 & 1.545 & 5.568 & 0.161 & -2.41 & 0.22 & * \\
74 & 21.1333 & 0.3436 & 0.3541 & 0.1874 & 3.8226 & 1.401 & 5.798 & 0.103 & -2.43 & 0.14 & \\
76 & 21.0668 & 0.5111 & 0.4381 & 0.3828 & 3.8159 & 1.663 & 5.946 & 0.135 & -2.01 & 0.18 & * \\
80 & 21.1615 & 0.4577 & 0.3232 & 0.2484 & 4.0232 & 1.726 & 5.872 & 0.195 & -2.03 & 0.26 & * \\
81 & 20.9603 & 0.4094 & 0.3117 & 0.1080 & 3.6554 & 2.259 & 2.538 & 0.239 & -2.01 & 0.32 & \\
82 & 21.1601 & 0.4776 & 0.2322 & 0.1698 & 3.8021 & 1.630 & 5.921 & 0.189 & -2.11 & 0.26 & * \\
84 & 21.0138 & 0.3621 & 0.2656 & 0.2341 & 3.9118 & 1.392 & 5.785 & 0.140 & -2.44 & 0.19 & \\
85 & 21.1101 & 0.3688 & 0.3734 & 0.1821 & 3.8898 & 1.411 & 5.390 & 0.175 & -2.53 & 0.24 & * \\
86 & 21.1539 & 0.4207 & 0.2335 & 0.0733 & 3.7654 & 1.913 & 6.089 & 0.337 & -1.91 & 0.45 & \\
87 & 21.2338 & 0.3278 & 0.4763 & 0.1339 & 4.5433 & 2.335 & 1.365 & 0.200 & -1.24 & 0.27 & \\
88 & 21.1992 & 0.4304 & 0.3728 & 0.2675 & 3.6537 & 1.639 & 5.915 & 0.181 & -2.15 & 0.25 & * \\
89 & 21.1122 & 0.4373 & 0.2633 & 0.2134 & 3.9794 & 1.640 & 5.929 & 0.271 & -2.18 & 0.37 & * \\
92 & 21.2012 & 0.4593 & 0.2218 & 0.2490 & 4.0976 & 1.786 & 5.667 & 0.245 & -1.73 & 0.33 & \\
93 & 21.1188 & 0.3776 & 0.3039 & 0.2448 & 4.0668 & 1.558 & 5.937 & 0.132 & -2.17 & 0.18 & * \\
94 & 21.1455 & 0.4514 & 0.2628 & 0.2586 & 3.7857 & 1.543 & 6.005 & 0.125 & -2.05 & 0.17 & \\
95 & 21.0371 & 0.3973 & 0.2391 & 0.1905 & 3.8537 & 1.621 & 5.608 & 0.176 & -2.26 & 0.24 & * \\
98 & 21.0444 & 0.4845 & 0.3347 & 0.1981 & 3.8040 & 1.665 & 0.389 & 0.189 & -2.26 & 0.26 & \\
100 & 20.9654 & 0.3373 & 0.1896 & 0.0863 & 4.1915 & 1.908 & 1.242 & 0.299 & -2.57 & 0.40 & \\
101 & 21.1938 & 0.3419 & 0.4871 & 0.3298 & 3.9467 & 1.824 & 5.786 & 0.151 & -1.99 & 0.20 & * \\
102 & 21.0912 & 0.3846 & 0.1889 & 0.2116 & 3.5606 & 1.948 & 0.430 & 0.187 & -1.60 & 0.25 & \\
103 & 21.1933 & 0.4730 & 0.3462 & 0.2441 & 3.8644 & 1.724 & 5.684 & 0.177 & -2.05 & 0.24 & * \\
104 & 21.1413 & 0.3145 & 0.3785 & 0.2147 & 3.9168 & 1.419 & 5.685 & 0.111 & -2.40 & 0.16 & * \\
105 & 21.2349 & 0.3906 & 0.3318 & 0.1716 & 3.8405 & 1.935 & 6.225 & 0.147 & -1.79 & 0.20 & * \\
107 & 21.1280 & 0.4623 & 0.3735 & 0.2258 & 3.5859 & 1.489 & 5.910 & 0.163 & -2.24 & 0.22 & * \\
114 & 20.8940 & 0.3156 & 0.1729 & 0.1486 & 3.9526 & 1.759 & 6.230 & 0.237 & -2.23 & 0.32 & * \\
115 & 21.0860 & 1.6991 & 1.7997 & 0.9826 & 2.0955 & 0.667 & 6.012 & 0.324 & -3.47 & 0.44 & \\
116 & 21.1802 & 2.2240 & 2.8269 & 0.6754 & 0.1569 & 1.101 & 4.885 & 3.001 & -3.35 & 4.04 & \\
119 & 21.1065 & 0.3107 & 0.2153 & 0.1698 & 4.1220 & 1.915 & 0.517 & 0.278 & -2.11 & 0.38 & \\
122 & 21.1773 & 0.3763 & 0.3510 & 0.1807 & 3.8666 & 1.947 & 0.367 & 0.369 & -1.92 & 0.50 & \\
124 & 21.1814 & 0.4343 & 0.2887 & 0.2845 & 3.7266 & 1.512 & 5.377 & 0.118 & -2.07 & 0.16 & \\
125 & 21.0576 & 0.3962 & 0.2374 & 0.1322 & 3.9208 & 1.807 & 0.055 & 0.143 & -2.36 & 0.19 & \\
126 & 21.1108 & 0.3502 & 0.3280 & 0.2460 & 3.6596 & 1.565 & 5.892 & 0.147 & -2.20 & 0.20 & * \\
127 & 21.1686 & 0.3964 & 0.2482 & 0.1976 & 3.9177 & 1.577 & 6.208 & 0.142 & -2.59 & 0.20 & \\
128 & 21.1830 & 0.3977 & 0.3906 & 0.2593 & 3.6900 & 1.872 & 5.872 & 0.172 & -2.00 & 0.23 & * \\
129 & 21.2547 & 0.3290 & 0.1382 & 0.2019 & 4.1716 & 1.255 & 5.564 & 0.417 & -2.62 & 0.56 & \\
132 & 21.0987 & 0.1121 & 0.3401 & 0.1890 & 3.5035 & 2.589 & 6.112 & 0.166 & -0.98 & 0.23 & \\
133 & 21.0932 & 0.4467 & 0.2818 & 0.1231 & 3.8670 & 1.863 & 0.030 & 0.196 & -1.88 & 0.26 & \\
135 & 21.0948 & 0.4023 & 0.3453 & 0.1891 & 3.8332 & 1.757 & 5.891 & 0.117 & -2.15 & 0.16 & * \\
136 & 21.1900 & 0.3180 & 0.2411 & 0.2928 & 3.9378 & 1.784 & 5.756 & 0.424 & -1.68 & 0.57 & \\
137 & 21.2138 & 0.3758 & 0.2625 & 0.1822 & 3.9266 & 1.589 & 5.710 & 0.168 & -2.22 & 0.23 & * \\
140 & 21.0901 & 0.2879 & 0.1967 & 0.2204 & 2.3131 & 1.191 & 4.132 & 1.184 & -2.92 & 1.59 & \\
142 & 21.0906 & 0.4665 & 0.3385 & 0.2569 & 3.7860 & 1.634 & 5.772 & 0.102 & -2.36 & 0.14 & * \\
144 & 21.3136 & 0.8180 & 0.2715 & 0.5360 & 2.4477 & 2.581 & 3.875 & 0.333 & -0.74 & 0.45 & \\
149 & 21.2231 & 0.3700 & 0.1774 & 0.1194 & 3.9287 & 3.235 & 0.367 & 0.394 & -0.31 & 0.54 & \\
150 & 21.0500 & 0.2377 & 0.1382 & 0.1382 & 4.3527 & 2.121 & 5.571 & 0.726 & -1.89 & 0.98 & \\
151 & 21.1498 & 0.3626 & 0.3154 & 0.0990 & 3.9220 & 1.622 & 0.008 & 0.209 & -2.28 & 0.28 & \\
152 & 21.1560 & 0.3985 & 0.3967 & 0.2306 & 3.6581 & 1.786 & 6.260 & 0.119 & -2.08 & 0.16 & * \\
154 & 21.0598 & 0.3671 & 0.3369 & 0.1177 & 3.8109 & 1.313 & 5.592 & 0.158 & -2.78 & 0.22 & * \\
159 & 21.0791 & 0.4384 & 0.2648 & 0.2246 & 4.0971 & 1.953 & 6.277 & 0.135 & -1.99 & 0.18 & * \\
161 & 21.2265 & 0.3989 & 0.3224 & 0.2057 & 3.7740 & 1.729 & 6.173 & 0.143 & -2.13 & 0.19 & * \\
162 & 21.1348 & 0.4299 & 0.2925 & 0.1395 & 3.9167 & 1.702 & 5.922 & 0.162 & -2.17 & 0.22 & * \\
163 & 21.2684 & 0.4072 & 0.3391 & 0.2046 & 4.0556 & 2.067 & 6.148 & 0.314 & -1.31 & 0.42 & * \\
164 & 21.1045 & 0.3637 & 0.3045 & 0.2021 & 3.6531 & 1.378 & 5.967 & 0.316 & -2.64 & 0.43 & * \\
167 & 21.1274 & 0.3788 & 0.2584 & 0.2988 & 3.9965 & 1.796 & 0.072 & 0.203 & -2.30 & 0.28 & \\
171 & 21.1111 & 0.4379 & 0.2636 & 0.1877 & 3.8213 & 2.009 & 6.264 & 0.255 & -1.61 & 0.34 & * \\
172 & 20.9743 & 0.3988 & 0.3147 & 0.1148 & 3.7654 & 1.898 & 0.471 & 0.124 & -2.13 & 0.17 & \\
174 & 20.9688 & 0.5443 & 0.2898 & 0.2588 & 4.2496 & 2.078 & 6.195 & 0.163 & -1.95 & 0.22 & * \\
175 & 21.1595 & 0.4584 & 0.2483 & 0.1598 & 3.8483 & 1.593 & 5.813 & 0.151 & -1.99 & 0.20 & * \\
176 & 20.7044 & 0.4697 & 0.2381 & 0.1348 & 3.7171 & 1.321 & 5.613 & 0.214 & -2.60 & 0.29 & * \\
177 & 21.0706 & 0.9609 & 0.5962 & 0.8721 & 5.1753 & 5.931 & 0.622 & 0.377 & 3.97 & 0.56 & \\
178 & 21.1075 & 0.3318 & 0.3033 & 0.2832 & 3.8723 & 1.613 & 0.042 & 0.133 & -2.14 & 0.18 & \\
180 & 21.0667 & 0.4906 & 0.3366 & 0.1060 & 3.9508 & 2.642 & 1.553 & 0.367 & -1.06 & 0.50 & \\
183 & 21.1632 & 0.4588 & 0.3604 & 0.2331 & 3.7826 & 1.490 & 5.939 & 0.153 & -2.32 & 0.21 & * \\
185 & 21.2703 & 0.4016 & 0.2601 & 0.1861 & 3.8097 & 1.540 & 5.831 & 0.242 & -2.24 & 0.33 & * \\
187 & 21.1778 & 0.6475 & 0.4987 & 0.3929 & 2.0980 & 2.169 & 3.850 & 0.154 & -1.89 & 0.21 & \\
196 & 21.2136 & 0.4445 & 0.3840 & 0.1823 & 3.8577 & 1.506 & 5.976 & 0.137 & -2.27 & 0.19 & * \\
198 & 21.0808 & 0.3530 & 0.2225 & 0.0692 & 3.6715 & 1.804 & 1.183 & 0.255 & -2.35 & 0.34 & \\
199 & 21.1064 & 0.4746 & 0.3464 & 0.1720 & 4.0304 & 2.074 & 0.341 & 0.148 & -1.90 & 0.20 & \\
201 & 21.1134 & 0.3684 & 0.4381 & 0.2510 & 4.2339 & 1.746 & 0.049 & 0.216 & -2.32 & 0.29 & \\
207 & 21.1579 & 0.3866 & 0.3150 & 0.2058 & 3.6818 & 1.372 & 5.550 & 0.217 & -2.33 & 0.29 & * \\
213 & 21.1666 & 0.4452 & 0.2605 & 0.1771 & 3.9542 & 2.009 & 5.510 & 0.276 & -1.75 & 0.37 & \\
216 & 21.1330 & 0.4461 & 0.3912 & 0.2201 & 3.3530 & 1.542 & 6.217 & 0.105 & -2.24 & 0.14 & \\
218 & 21.1418 & 0.4827 & 0.3112 & 0.2325 & 3.8122 & 1.236 & 4.587 & 0.253 & -2.75 & 0.34 & \\
219 & 21.2374 & 0.4351 & 0.3167 & 0.1716 & 3.9298 & 1.522 & 5.540 & 0.164 & -2.33 & 0.22 & * \\
220 & 21.1473 & 0.4628 & 0.2067 & 0.1128 & 3.5244 & 1.388 & 5.629 & 0.286 & -2.62 & 0.39 & * \\
223 & 21.2273 & 0.3467 & 0.3532 & 0.1648 & 3.9499 & 1.824 & 0.257 & 0.109 & -1.89 & 0.15 & \\
225 & 21.1668 & 0.3397 & 0.3088 & 0.1798 & 3.8777 & 1.429 & 5.352 & 0.268 & -2.30 & 0.36 & * \\
227 & 21.1436 & 0.4977 & 0.4236 & 0.2166 & 4.1281 & 1.515 & 0.083 & 0.167 & -2.32 & 0.23 & \\
237 & 21.2335 & 0.4387 & 0.3462 & 0.1824 & 3.8619 & 1.578 & 5.537 & 0.127 & -2.30 & 0.17 & * \\
238 & 21.0892 & 0.3513 & 0.2782 & 0.1435 & 3.6480 & 1.495 & 5.621 & 0.130 & -2.13 & 0.18 & * \\
243 & 21.0025 & 0.2285 & 0.1968 & 0.1412 & 4.1500 & 2.210 & 1.336 & 0.314 & -2.02 & 0.42 & \\
244 & 21.2050 & 0.2709 & 0.1472 & 0.1502 & 3.9090 & 1.318 & 5.572 & 0.206 & -2.37 & 0.28 & * \\
249 & 21.1367 & 0.3947 & 0.3224 & 0.2133 & 3.8850 & 1.689 & 5.687 & 0.086 & -2.07 & 0.12 & * \\
252 & 21.1382 & 0.4386 & 0.3597 & 0.2268 & 3.8585 & 1.453 & 5.479 & 0.133 & -2.30 & 0.18 & * \\
253 & 21.2784 & 0.2496 & 0.1984 & 0.5095 & 4.7083 & 4.776 & 4.581 & 0.357 & 2.34 & 0.51 & \\
258 & 21.0697 & 0.3559 & 0.3024 & 0.1948 & 3.7039 & 1.365 & 5.571 & 0.124 & -2.45 & 0.17 & * \\
260 & 21.1105 & 0.5214 & 0.3487 & 0.2697 & 3.9380 & 1.576 & 5.989 & 0.085 & -1.98 & 0.12 & \\
261 & 21.1359 & 0.4000 & 0.2921 & 0.1263 & 3.7906 & 1.831 & 0.182 & 0.167 & -1.62 & 0.22 & \\
262 & 21.1091 & 0.3866 & 0.3278 & 0.1781 & 3.5986 & 1.408 & 6.122 & 0.092 & -2.56 & 0.13 & \\
265 & 21.0829 & 0.4074 & 0.2783 & 0.2220 & 3.7921 & 1.488 & 5.868 & 0.403 & -2.26 & 0.54 & * \\
269 & 21.0926 & 0.4347 & 0.3432 & 0.2381 & 3.7428 & 1.413 & 5.332 & 0.127 & -2.14 & 0.17 & * \\
270 & 20.9785 & 0.3460 & 0.3707 & 0.2696 & 3.7410 & 1.306 & 5.605 & 0.085 & -2.41 & 0.12 & * \\
273 & 20.9756 & 0.0759 & 0.0814 & 0.1433 & 5.2331 & 2.817 & 1.126 & 0.859 & -1.55 & 1.16 & \\
278 & 21.1312 & 0.3491 & 0.3719 & 0.2490 & 3.7717 & 1.668 & 5.850 & 0.143 & -2.20 & 0.19 & * \\
279 & 21.0067 & 0.4360 & 0.2030 & 0.1760 & 3.6151 & 1.206 & 5.978 & 0.223 & -2.79 & 0.31 & \\
281 & 20.8779 & 0.2710 & 0.6362 & 0.3831 & 1.8806 & 4.044 & 0.350 & 0.089 & 0.70 & 0.17 & \\
284 & 21.0144 & 0.1990 & 0.3959 & 0.2275 & 4.0643 & 1.113 & 5.740 & 0.228 & -2.92 & 0.31 & \\
285 & 20.8499 & 0.3775 & 0.2681 & 0.1193 & 4.1360 & 1.897 & 0.200 & 0.281 & -2.06 & 0.38 & \\ 
290 & 21.1048 & 0.4284 & 0.1804 & 0.1608 & 3.8519 & 1.824 & 1.108 & 0.491 & -2.48 & 0.66 & \\
291 & 20.7762 & 0.7227 & 0.5494 & 0.4493 & 3.8448 & 1.340 & 4.926 & 0.102 & -3.28 & 0.15 & \\
298 & 21.1788 & 0.3668 & 0.2674 & 0.2986 & 3.8061 & 0.882 & 4.773 & 0.257 & -3.42 & 0.35 &  \\
303 & 21.2060 & 0.3931 & 0.3419 & 0.1996 & 4.3250 & 2.005 & 0.412 & 0.102 & -1.74 & 0.14 & \\
304 & 21.1033 & 0.4119 & 0.2785 & 0.1459 & 3.9121 & 1.802 & 5.678 & 0.234 & -2.27 & 0.32 & * \\
308 & 21.0490 & 0.3422 & 0.0718 & 0.2947 & 3.0191 & 0.890 & 4.789 & 0.953 & -3.24 & 1.28 & \\
309 & 21.0466 & 0.3957 & 0.2182 & 0.2359 & 3.2974 & 1.333 & 5.126 & 0.208 & -2.68 & 0.28 & \\
310 & 21.0316 & 0.4385 & 0.3614 & 0.3383 & 4.1751 & 2.114 & 0.496 & 0.170 & -1.74 & 0.23 & \\
313 & 21.0945 & 0.3778 & 0.2171 & 0.1559 & 3.8761 & 1.283 & 4.963 & 0.184 & -2.48 & 0.25 & * \\
315 & 21.0593 & 0.4135 & 0.1561 & 0.1555 & 3.5577 & 1.364 & 0.402 & 0.316 & -2.41 & 0.43 & \\
323 & 21.0327 & 0.3434 & 0.2483 & 0.2011 & 3.9195 & 1.239 & 5.380 & 0.153 & -2.45 & 0.21 & *\\
324 & 21.0505 & 0.4123 & 0.2956 & 0.1909 & 3.7772 & 1.367 & 5.935 & 0.132 & -2.52 & 0.18 & \\
325 & 21.0045 & 0.2638 & 0.3611 & 0.1385 & 4.5856 & 2.350 & 0.748 & 0.269 & -1.27 & 0.36 & \\
326 & 21.0456 & 0.4710 & 0.2458 & 0.2248 & 3.4030 & 1.104 & 5.710 & 0.160 & -3.02 & 0.22 & \\
332 & 21.1832 & 0.2445 & 0.3541 & 0.4182 & 3.7120 & 0.430 & 5.168 & 0.443 & -3.91 & 0.60 & \\

\enddata
\tablenotetext{a}{Stars with an * have passed the $D_{M} < 5.0$ criterion.}

\label{met}
\end{deluxetable}

\begin{deluxetable}{cccccccc}

\tablecaption{Table of parameters for Draco anomalous Cepheids.}
\tablenum{8}

\tablewidth{0pc}

\tablehead{\colhead{ID\tablenotemark{a}} & \colhead{RA} & \colhead{DEC} & \colhead{Period} & \colhead{Amp} & \colhead{$\langle V \rangle$} & \colhead{$\langle I \rangle$} & \colhead{$M_{V}$} \\ 
\colhead{} & \colhead{(2000.0)} & \colhead{(2000.0)} & \colhead{(days)} & \colhead{} & \colhead{} & \colhead{} & \colhead{} } 

\startdata
31 & 17:20:25.15 & 57:52:53.3 & 0.61763 & 0.71 & 19.57 & 18.78 & -0.01  \\
134* & 17:19:06.37 & 57:49:48.2 & 0.59228 & 0.85 & 18.78 & 18.40 & -0.80  \\
141* & 17:20:17.82 & 57:57:07.8 & 0.90087 & 0.67 & 19.20 & 18.63 & -0.38  \\
157* & 17:19:08.08 & 57:58:35.2 & 0.93649 & 1.04 & 18.85 & 18.41 & -0.74  \\
194* & 17:19:36.06 & 57:54:15.6 & 1.59027 & 0.48 & 18.11 & 17.53 & -1.47  \\
204* & 17:22:00.74 & 57:50:21.2 & 0.45413 & 0.75 & 19.23 & 18.77 & -0.35  \\
230 & 17:21:47.77 & 57:53:18.9 & 0.60816 & 0.38 & 19.25 & 18.54 & -0.33  \\
282 & 17:19:42.55 & 57:54:49.8 & 0.55187 & 0.60 & 19.51 & 18.90 & -0.05  \\
312 & 17:18:30.56 & 57:56:04.8 & 0.90735 & 0.90 & 19.15 & 18.59 & -0.43  \\
\enddata
\tablenotetext{a}{Stars denoted with an asterisk are previously known
  anomalous Cepheids \citep{Norris:1975, Zinn:1976a, Nemec:1988a}}
\label{tableac}
\end{deluxetable}

\begin{deluxetable}{lllcccccl}

\tablecaption{Long period, semi-irregular red variable stars and carbon stars.}
\tablenum{9}
\tablewidth{0pc}
\tablehead{\colhead{ID} & \colhead{RA} & \colhead{DEC} & \colhead{V} &
  \colhead{I} & \colhead{$V-I$} & \colhead{$\sigma_{V}$} &
  \colhead{$\sigma_{I}$} & \colhead{Comment} \\ 
  \colhead{} & \colhead{(2000.0)} & \colhead{(2000.0)} & \colhead{} &
  \colhead{} & \colhead{} & \colhead{} & \colhead{} & \colhead{} }

\startdata
302 & 17:18:52.12 & 58:04:13.2 & 17.26 & 15.88 & 1.38 & 0.04 & 0.03 & 1   \\
293 & 17:19:10.82 & 57:59:17.7 & 17.14 & 15.76 & 1.39 & 0.03 & 0.01 & 1   \\
292 & 17:19:17.52 & 58:01:07.4 & 16.84 & 15.35 & 1.48 & 0.06 & 0.03 & 1,2 var.vel.  \\
288 & 17:19:42.39 & 57:58:38.0 & 17.25 & 15.99 & 1.26 & 0.07 & 0.08 & 1 carbon star (C5) \\
283 & 17:19:57.66 & 57:50:05.7 & 17.15 & 15.66 & 1.49 & 0.08 & 0.04 & 1,2 carbon star (C1), var.vel.  \\
280 & 17:20:00.70 & 57:53:46.8 & 17.30 & 15.99 & 1.31 & 0.05 & 0.04 & 1,2,3 carbon star (C2)  \\
274 & 17:20:32.85 & 57:51:44.2 & 16.91 & 15.38 & 1.53 & 0.06 & 0.05 & 1,2,3   \\
272 & 17:20:40.26 & 57:57:33.1 & 16.44 & 14.90 & 1.55 & 0.02 & 0.02 & 1,3  \\
271 & 17:20:41.85 & 58:00:25.1 & 16.93 & 15.44 & 1.48 & 0.05 & 0.02 & 1,2  \\
268 & 17:20:43.69 & 57:48:44.3 & 16.51 & 15.03 & 1.48 & 0.04 & 0.03 & 1,2  var.vel  \\
263 & 17:20:53.01 & 57:55:58.0 & 17.23 & 15.70 & 1.53 & 0.02 & 0.02 & 1,2,3  \\
259 & 17:21:02.23 & 58:15:38.7 & 17.15 & 15.75 & 1.41 & 0.04 & 0.03 & 1  \\
 & Eclipsing  binaries &  \\
296 & 17:19:06.16 & 57:41:21.1 & 19.54 & 18.16 & 1.38 & 0.15 & 0.12 & P=0.2435d  \\
256 & 17:21:18.30 & 58:14:29.9 & 18.54 & 17.28 & 1.26 & 0.03 & 0.10 & P=0.1253d or 0.2300d \\

\enddata

\tablerefs{1 = Draco RV member \citep{Armandroff:1995}; 2 = Draco RV member \citep{Olszewski:1995}; 3 = Draco proper motion member \citep{Stetson:1980}}
\label{longper}

\end{deluxetable}

\begin{deluxetable}{ccccccccl}
\tablecaption{QSOs found in our Draco survey.  The last column lists where the spectra, if available, were obtained.}
\tablenum{10}
\tablewidth{0pc}
\tablehead{\colhead{ID} & \colhead{RA} & \colhead{DEC} & \colhead{$V$} & \colhead{$I$} & \colhead{$V-I$} & \colhead{$\sigma_{V}$} & \colhead{$\sigma_{I}$} & \colhead{Comment}  \\ 
\colhead{} & \colhead{(2000.0)} & \colhead{(2000.0)} & \colhead{} & \colhead{} & \colhead{} & \colhead{} & \colhead{} & \colhead{} } 

\startdata
331 & 17:17:35.09 & 57:56:26.2 & 19.66 & 18.98 & 0.67 & 0.07 & 0.06 & WIYN  \\
329 & 17:17:50.11 & 58:11:08.1 & 19.96 & 19.13 & 0.83 & 0.08 & 0.06 & probable QSO  \\
328 & 17:17:50.57 & 58:15:14.7 & 17.71 & 16.92 & 0.79 & 0.06 & 0.06 & SDSS  \\
322 & 17:18:09.35 & 58:07:16.3 & 19.86 & 19.44 & 0.42 & 0.07 & 0.07 & WIYN  \\
320 & 17:18:19.40 & 57:39:35.2 & 20.18 & 19.30 & 0.87 & 0.15 & 0.06 & WIYN  \\
311 & 17:18:31.90 & 58:08:44.3 & 19.17 & 18.85 & 0.32 & 0.14 & 0.04 & WIYN  \\
300 & 17:19:01.71 & 58:00:29.1 & 19.31 & 18.60 & 0.71 & 0.06 & 0.02 & WIYN, SDSS  \\
299 & 17:19:04.18 & 58:03:29.4 & 20.36 & 19.88 & 0.48 & 0.11 & 0.07 & WIYN  \\
203 & 17:19:34.43 & 57:58:49.8 & 19.51 & 18.80 & 0.71 & 0.20 & 0.10 & WIYN\tablenotemark{a}
\\
287 & 17:19:43.77 & 58:11:12.4 & 19.87 & 18.85 & 1.01 & 0.06 & 0.04 & SDSS  \\
266 & 17:20:51.96 & 57:41:59.9 & 19.52 & 18.99 & 0.53 & 0.07 & 0.04 & probable QSO \\
264 & 17:20:52.31 & 57:55:13.4 & 19.86 & 19.43 & 0.40 & 0.08 & 0.06 & WIYN  \\
257 & 17:21:18.12 & 57:33:30.5 & 20.25 & 19.19 & 1.05 & 0.17 & 0.08 & probable QSO  \\
255 & 17:21:22.85 & 57:50:29.5 & 17.90 & 17.26 & 0.64 & 0.02 & 0.01 & WIYN, SDSS  \\
254 & 17:21:25.49 & 58:15:28.2 & 20.34 & 19.65 & 0.69 & 0.08 & 0.05 & SDSS  \\
251 & 17:21:30.06 & 57:40:15.8 & 19.69 & 19.15 & 0.54 & 0.06 & 0.05 & WIYN  \\
240 & 17:21:41.47 & 57:52:35.6 & 20.65 & 19.83 & 0.82 & 0.09 & 0.08 & probable QSO \\
241 & 17:21:41.57 & 57:33:18.9 & 18.88 & 18.24 & 0.64 & 0.06 & 0.17 & WIYN, SDSS  \\
231 & 17:21:47.52 & 58:15:07.8 & 19.96 & 18.46 & 1.50 & 0.14 & 0.04 & WIYN  \\
229 & 17:21:48.30 & 57:58:05.8 & 20.70 & 19.75 & 0.95 & 0.11 & 0.07 & SDSS  \\
224 & 17:22:07.34 & 58:14:25.0 & 20.79 & 20.25 & 0.54 & 0.25 & 0.14 & probable QSO \\
222 & 17:22:11.66 & 57:56:52.2 & 20.74 & 19.77 & 0.98 & 0.16 & 0.09 & WIYN, SDSS \\
215 & 17:22:36.06 & 57:37:05.0 & 20.10 & 19.90 & 0.21 & 0.19 & 0.08 & WIYN  \\
214 & 17:22:41.04 & 57:45:27.1 & 20.78 & 19.94 & 0.83 & 0.10 & 0.07 & probable QSO  \\
212 & 17:22:44.68 & 57:41:24.2 & 20.40 & 19.47 & 0.93 & 0.15 & 0.08 & probable QSO \\
211 & 17:22:51.01 & 57:41:18.5 & 19.80 & 19.27 & 0.53 & 0.14 & 0.06 & WIYN  \\
210 & 17:22:56.95 & 58:11:10.8 & 19.76 & 19.01 & 0.75 & 0.14 & 0.08 & WIYN, SDSS  \\
209 & 17:23:01.83 & 58:04:06.7 & 20.52 & 19.10 & 1.42 & 0.14 & 0.07 & probably QSO\tablenotemark{b}  \\
208 & 17:23:02.20 & 58:04:14.5 & 20.17 & 19.45 & 0.71 & 0.08 & 0.06 & WIYN  \\
206 & 17:23:14.18 & 58:14:07.4 & 19.99 & 19.57 & 0.42 & 0.17 & 0.09 & WIYN  \\
\enddata
\tablenotetext{a}{B\&S V203}
\tablenotetext{b}{only 8\arcsec from QSO at RA=17:23:02.20, DEC=+58:04:14.5.}
\label{qso}
\end{deluxetable}

\end{document}